\documentclass[10pt, letterpaper, twocolumn, journal, keeplastbox]{IEEEtran}
\usepackage{cite, graphicx,amssymb,color,pict2e, multirow, rotating} 
\usepackage{flushend}
\usepackage{amsmath}
\usepackage{epsfig}
\usepackage{amssymb}
\usepackage{hyperref}
\usepackage{color}
\usepackage{psfrag}
\usepackage{caption}
\usepackage{subfig}
\usepackage[usenames,dvipsnames]{xcolor}
\usepackage{textcomp}
\usepackage{array}
\usepackage{tikz}
\usepackage{longtable}
\usepackage{float}
\usetikzlibrary{matrix,decorations.pathreplacing}

\usepackage{tabularx,ragged2e}

\usepackage{algorithmic}
\usepackage{algorithm}
\usepackage{url}
\usepackage{xcolor}
\usepackage{xpatch}

\begin{document}

\title{Holistic Network Virtualization and Pervasive Network Intelligence for 6G}


\author{Xuemin~(Sherman)~Shen,~\IEEEmembership{Fellow,~IEEE,} 
Jie~Gao,~\IEEEmembership{Senior~Member,~IEEE,}
Wen~Wu,~\IEEEmembership{Member,~IEEE,}
Mushu~Li,~\IEEEmembership{Member,~IEEE,}
Conghao~Zhou,~\IEEEmembership{Student~Member,~IEEE,}
and~Weihua~Zhuang,~\IEEEmembership{Fellow,~IEEE} 
\thanks{
This work was supported by research grants from the Natural Sciences and Engineering Research Council (NSERC) of Canada.

Xuemin (Sherman) Shen, Mushu Li, Conghao Zhou, and Weihua Zhuang are with the Department of Electrical and Computer Engineering, University of Waterloo, Waterloo, ON, Canada, N2L 3G1 (email: \{sshen, m475li, c89zhou,  wzhuang\}@uwaterloo.ca). 
    
Jie Gao is with the Department of Electrical and Computer Engineering, Marquette University, Milwaukee, WI, USA 53233 (email: j.gao@marquette.edu).

Wen Wu is with the Frontier Research Center, Peng Cheng Laboratory, Shenzhen, Guangdong, China, 518055 (email: wuw02@pcl.ac.cn).

The corresponding author of this paper is Weihua Zhuang.
}
}%

\IEEEspecialpapernotice{(Invited Paper)}

\maketitle

\begin{abstract}
In this tutorial paper, we look into the evolution and prospect of network architecture and propose a novel conceptual  architecture for the 6th generation (6G) networks. The proposed architecture has two key elements, i.e., holistic network virtualization and pervasive artificial intelligence (AI). The holistic network virtualization consists of network slicing and digital twin, from the aspects of service provision and service demand, respectively, to incorporate service-centric and user-centric networking. The pervasive network intelligence integrates AI into future networks from the perspectives of networking for AI and AI for networking, respectively. Building on holistic network virtualization and pervasive network intelligence, the proposed architecture can facilitate three types of interplay, i.e., the interplay between digital twin and network slicing paradigms, between model-driven and data-driven methods for network management, and between virtualization and AI, to maximize the flexibility, scalability, adaptivity, and intelligence for 6G networks. We also identify challenges and open issues related to the proposed architecture. By providing our vision, we aim to inspire further discussions and developments on the potential architecture of 6G.

\end{abstract}










\section{Introduction}
\label{sec:introduction}
\subsection{Background}

With the ongoing worldwide deployment of the 5th generation (5G) networks, the technical community in wireless communications and networking has started looking into the 6th generation (6G) networks for 2030 and beyond. While the exact concepts and techniques that define 6G are not determined yet, visions, requirements, use cases, and candidate techniques are discussed in an increasing amount of works, e.g., \cite{J_You2021towards, M_WSaad_2020, M_MGiordani_2020}. Among these  discussions, some preliminary consensus regarding 6G emerges. For instance, in terms of main requirements of 6G, the urgency of improving security~\cite{MNet_XShen_2020} and energy efficiency~\cite{J_ASodhro_2020} is understood unanimously. For use cases of 6G, the combination of enhanced mobile broadband (eMBB), ultra-reliable and low-latency communications (uRLLC), and massive machine-type communications (mMTC) have been brought up, despite the different terminologies used in different works \cite{M_ZZhang_2019}, \cite{M_BZong2019}. As to candidate techniques, commonly mentioned examples include the integration of satellite, aerial, terrestrial, and underwater networks~\cite{M_NZhang}, \cite{M_SChen_2020}, (sub)terahertz and visible light communications~\cite{M_ECStrinati2019}, artificial intelligence (AI) empowered networks~\cite{J_XShen_2020, M_KLetaief2019, yang2020artificial}, to name a few.              

One consensus deserving special attention is that 6G may need a brand-new network architecture. Driven by cost effectiveness and efficiency, the evolution of network architecture follows the evolving services provided by the networks. For instance, to introduce data service, a packet-switched core network component emerged in the 3G architecture as a complement to its circuit-switched counterpart for voice service~\cite{M_MSayed_2002}. Then, to accommodate the exponential growth of data traffic, 4G introduced a redesigned and simplified network architecture for a flat all-Internet protocol (IP) network with increased data rate and reduced latency~\cite{M_BBjerke_2011}. In the era of 5G, as networks become more heterogeneous than ever while services become diversified, various network architecture innovations have been proposed towards  flexible service-oriented networking, including software defined networking (SDN) \cite{S_DKreutz_2015}, cloud radio access network (C-RAN) \cite{S_AChecko_2015}, and network slicing~\cite{J_MBagaa2018}, \cite{J_WWu2020}. Therefore, envisioned to support unprecedentedly diverse services with exceedingly stringent quality of service (QoS) or quality of experience (QoE) requirements, 6G will most likely need ground-breaking innovations in network architecture.    

While conceiving an architecture for 6G, it is difficult to overlook two key elements, i.e., virtualization and AI. Network virtualization already plays an important role in the architecture of 5G~\cite{M_KQu_2020}. The virtualization of resources, functions, and networks enables resource sharing, software implementation of network functions, and service-orientated networking, respectively, and thereby increases resource utilization while reducing the cost for deploying and operating networks. Virtualization reflects a trend of softwarization for flexible, scalable, and adaptive network management~\cite{S_IAfolabi2018}. Therefore, it is foreseeable that virtualization will remain crucial in the architecture of 6G. As for the second key element, i.e., AI, a growing number of research teams worldwide are investigating AI-driven networks, and high expectation is placed on AI for empowering 6G~\cite{J_You2021towards, O_Ali6gWhite_2020}. In comparison with  heuristic or mathematical model based approaches for communications and networking, AI based approaches can handle complicated networking  problems and obtain accurate results, provided that sufficient data are available for training. This advantage suits the increasingly heterogeneous and dynamic networks, where mathematical models may not exist or cannot accurately characterize the considered problems. Therefore, it is not difficult to predict the significance of AI in 6G.         

\subsection{Architectural Innovations Required for 6G}

Recognizing the importance of virtualization and AI, we further look into their limitations in the state-of-the-art to comprehend the architectural innovations required for 6G. Existing virtualization techniques mostly deal with \textit{service provision} in communication networks. For instance, network slicing highlights available network resources, service provision capability, and QoS satisfaction for various services~\cite{J_HZhang2016}. While such virtualization, with a focus on service provision, enables 5G to handle diverse coexisting services, it may not suffice for 6G since the characteristics of end user \textit{service demand} can be the key to achieving user-centric networking. Therefore, in the future, virtualization should focus on both the service provision capability of a network and the service demand of end users in the network. This will lead to the virtualization of end users in addition to the virtualization of networks. As for AI, existing research on AI mostly addresses specific functions (e.g., routing~\cite{S_Fadlullah2017}),  layers (e.g., physical layer~\cite{J_AToma2020}), network segments (e.g., access networks~\cite{M_SHan2020}), or applications (e.g., autonomous driving~\cite{J_MLi2021}) of a network. Meanwhile, how to integrate AI into the network architecture across different layers or network segments needs further investigation. The scope and extent of AI-driven networks are yet to be determined.         

As virtualization extends to cover both service provision and service demand while AI pervades every corner of the network, close connections between the two elements are foreseeable and can dominate the architectural needs of 6G. The first connection is through network and end user \textit{data}~\cite{li2021slicing}. Virtualization facilitates the characterization of network service provision capability, service performance, resource utilization and, in future networks, end user service demand. As a result, a vast amount of data will be generated, which can be exploited to characterize the network and end users. Such data, if properly managed, can empower both AI-driven networking and AI applications (e.g., object detection)~\cite{O_PChemouil2020}. The second connection is through network \textit{control}. AI can be used to make decisions pertinent to virtualization, including service admission, slice establishment, dynamic virtual network function orchestration, and resource scheduling. In the future, AI can also help control data collection for the virtualization of end users and extract features of virtualized end users. Thus, AI has the potential to improve the efficacy and adaptivity of virtualization. The third connection is through network \textit{resources}. A main motivation of network virtualization is to coordinate resource sharing among different services and thereby improve network resource utilization and service satisfaction. AI-driven networking can target efficient utilization of network resources. As both virtualization and AI consume computing, communication, and storage resources, they may compete for network resources. However, AI has a potential to increase the efficiency of virtualization through intelligent network planning and operations, while virtualization may increase the efficiency of AI through proper data provision and management. As a result, they should work together to enhance network resource utilization and service quality.

A rudiment of the above connections through data and control can be observed in the existing architecture of 5G. For instance, the 3rd Generation Partnership Project (3GPP) introduces a network data analytics function (NWDAF) for 5G in Release 15~\cite{3gpp.29.520} and enablers for network automation (eNA) in Release 16~\cite{3gpp.21.916}. The architecture design provides a framework for the NWDAF to collect data from other network functions (such as policy control and network slice selection functions) and provide analytics (such as data traffic statistics and predictions) back to these network functions. In 6G, the scope and level of both data collection and analytics will expand significantly. Most likely, network data analysis, instead of being limited to one or two specific functions, will be AI-driven and available everywhere in a network. Similarly, the data available for network management, instead of being limited in type, content, or format, should provide information of the network and end users as needed. Such expectations can be fulfilled by extending the roles of virtualization and AI in the network architecture.  

\subsection{Our Vision}

Our vision of network architecture for 6G is based on the importance of virtualization and AI, their limitations in existing networks, and the essential connections between them. Specifically, we aim to design a network architecture that i) supports virtualization of the network and  end users from the perspectives of service provision and service demand, respectively, ii) integrates  AI  in various network functions, layers, segments, and applications under a unified architecture, and, more importantly, iii) facilitates the interplay between virtualization and AI, enabling their coexistence, integration, and mutual enhancement. To consolidate the vision, we raise the following three key questions: 
\begin{itemize}
	\item How to further advance virtualization beyond network slicing?    
	\item How to enable AI into every facet of a network?     
	\item How to effectively integrate virtualization and AI through network architecture design? 
\end{itemize}

In pursuit of answering the preceding questions, we develop the ideas of holistic network virtualization and pervasive network intelligence for 6G network architecture. \textit{Holistic network virtualization} advances virtualization toward 6G by incorporating network slicing and digital twin paradigms. The former enables service-centric network management, and the latter adds a user-centric perspective to virtualization for future networks. \textit{Pervasive network intelligence} enables generic integration of AI into a network from the perspectives of AI for networking and networking for AI. The former emphasizes the role of AI in network management, while the latter leverages network design to support AI applications. In this tutorial paper, for both holistic network virtualization and pervasive network intelligence, we survey existing studies, present our network architecture designs, and illustrate their benefits. Unifying these two components, we further introduce an overall conceptual network architecture, which fulfills our vision of unprecedentedly flexible, scalable, adaptive, and intelligent networks for 6G.           

 \begin{figure*}[tt]
	\centering
	\renewcommand{\figurename}{Fig.}
	\includegraphics[width=1\textwidth]{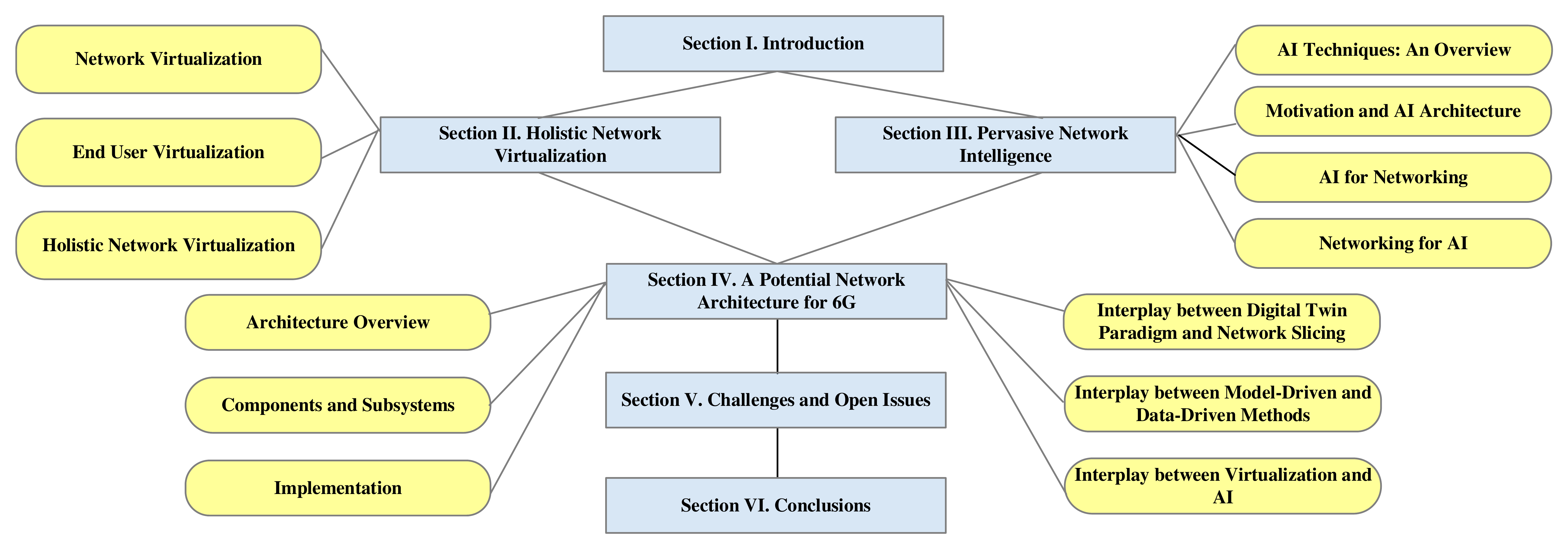}
	\caption{The structure of this paper.}
	\label{fig:structure}
\end{figure*}

This tutorial paper can provide useful information and benefit readers from three aspects. First, for readers who are interested in the historical and current developments of virtualization and AI techniques, we survey the literature and provide a review of both in the context of communication networks. Second, for readers who are exploring future directions in virtualization and AI, we propose original ideas for advancing them toward 6G. Specifically, we illustrate designs and ideas, such as incorporating digital twins for holistic network virtualization, connected AI for network management, AI slices with training and inference separation, and hybrid data-model driven methods, throughout this paper. Last, after introducing our vision of holistic network virtualization and pervasive network intelligence, we present open issues and challenges to inspire further research.

There are a few surveys on virtualization and AI in the literature~\cite{S_IAfolabi2018, J_RMinerva_ProcIEEE_2020,boutaba2018comprehensive, zhang2019deep}. Regarding virtualization, Minerva \emph{et al.} present existing digital twin based applications in the context of IoT~\cite{J_RMinerva_ProcIEEE_2020}, and another survey introduces the key enabling technologies and design principles of network slicing~\cite{S_IAfolabi2018}. Regarding AI, Boutaba \emph{et al.} undertake a comprehensive survey on AI applications in various areas of networking~\cite{boutaba2018comprehensive}, and another survey focuses on  deep learning (DL) based applications in wireless networking~\cite{zhang2019deep}. In comparison, this tutorial paper focuses on the vision of 6G. Specifically, after introducing state-of-the-art virtualization and AI techniques, we propose original designs, including holistic network virtualization and pervasive network intelligence, to establish a novel conceptual architecture for 6G networks.

\subsection{Structure of the Paper}

The structure of this tutorial paper is shown in Fig.~\ref{fig:structure}.

Section~\ref{sec:HNV} illustrates our vision of 6G networks from the aspect of holistic network virtualization. We review existing network virtualization concepts and techniques in Subsection~\ref{ss:NetVirtual}. Then, we introduce end user virtualization with a focus on digital twins in Subsection~\ref{ss:EUVirtual}. Lastly, we present our idea of holistic network virtualization, highlighting a six-layer virtualization architecture, in Subsection~\ref{ss:HNV}.

Section~\ref{sec:PI} illustrates our vision of 6G networks from the aspect of pervasive network intelligence. Subsection~\ref{ss:AIOv} presents an overview of representative AI techniques that are potentially useful for 6G networks. Subsection~\ref{ss:Moti} introduces the motivation for pervasive network intelligence and presents a four-level AI architecture. Subsections~\ref{ss:AIforNet} and \ref{ss:NetforAI} summarize the existing research and present our ideas on AI for networking and networking for AI, respectively.

Section~\ref{sec:PNA} integrates holistic network virtualization and pervasive network intelligence and presents our overall vision for 6G. Subsection~\ref{ss:related6Garc} reviews related studies on architectures for 6G. Subsection~\ref{ss:ArcOv} introduces a conceptual architecture for 6G networks that incorporates holistic network virtualization and pervasive network intelligence. Subsections~\ref{ss:comp}~and~\ref{ss:Implement} discuss the components, subsystems, and potential implementation of the proposed architecture. Subsections~\ref{ss:IntDTNS}~to~\ref{ss:IntVTAI} elaborate on three types of interplay enabled by the proposed architecture, i.e., the interplay between digital twin and network slicing, between data-driven and model-driven methods, and between virtualization and AI, respectively.

Section~\ref{sec:Open} identifies key challenges and open issues related to the proposed network architecture, and Section~\ref{sec:conclusions} concludes this research. 

Table~\ref{t:acronym} lists the acronyms used in this paper.

\begin{table}
\footnotesize 
\centering
\captionsetup{justification=centering,singlelinecheck=false}
\caption{List of Acronyms}\label{t:acronym}
\begin{tabular}{l|l}
\hline\hline
3GPP & 3rd Generation Partnership Project \\
5G & 5th Generation \\
6G & 6th Generation \\
AI & Artificial Intelligence \\
AL & AI Level \\
AP & Access Point \\
API & Application Programming Interface \\
ARQ & Automatic Repeat-Request \\
BS & Base Station \\
C-RAN & Cloud Radio Access Network \\
DL & Deep Learning \\
DNN & Deep Neural Network \\
DRL & Deep Reinforcement Learning \\
eMBB & Enhanced Mobile Broadband \\
FL & Federated Learning \\
IoT & Internet of Things \\
IP & Internet Protocol \\
ITU & International Telecommunication Union \\
LSTM & Long Short-Term Memory \\
LTE & Long Term Evolution \\
mMTC & Massive Machine-Type Communications \\
MIMO & Multiple-Input Multiple-Output\\
ML & Machine Learning \\
NFV & Network Function Virtualization \\
NN & Neural Network \\
NWDAF & Network Data Analytics Function \\
QoE & Quality of Experience \\
QoS & Quality of Service \\
RAN & Radio Access Network \\
SBS & Small Base Station \\
SDN & Software Defined Networking \\
SNR & Signal-to-Noise Ratio \\
UAV & Unmanned Aerial Vehicle \\
uRLLC & Ultra-Reliable and Low-Latency Communications \\
VL & Virtualization Layer \\ 
VM & Virtual Machine \\
WSN & Wireless Sensor Network \\ 
\hline\hline
\end{tabular}
\end{table}

\section{Holistic Network Virtualization}
\label{sec:HNV}
In this section, we first review virtualization techniques in existing networks and their benefits. Then, we introduce the idea of holistic network virtualization.  

\subsection{Network Virtualization}\label{ss:NetVirtual}

The concept and techniques of network virtualization have been evolving over more than three decades~\cite{M_Mosharaf_COMM_2009}. Early research on network virtualization includes virtual local area networks motivated by facilitating different types of operations (services) in distributed systems~\cite{rossi1986proposal}, as well as providing flexible network control and improving link utilization~\cite{J_KSato_TCOM_1990}.  Another example of network virtualization is virtual private networks, which establish efficient and secure communication links to connect geographically dispersed end users. Over time, the desire for programmable network management extends to the objective of enhancing network architecture.    

The advancement in cloud computing has propelled recent development in network virtualization, including network function virtualization (NFV) and network slicing. With NFV, software instances running on virtual machines at general computing servers replace customized and proprietary hardware for various network functions~\cite{J_KQu_TCom_2020}. At the network core, NFV applies to functions such as switching, firewall, deep packet inspection, and session border controller~\cite{C_chioC_2012}. At radio access networks, NFV applies to frame generation, modulation, carrier allocation, etc.~\cite{J_AGDalla-Costa_JSAC_2020}. The realization of NFV becomes an enabler for network slicing, which is a key network architecture innovation in 5G. Network slicing emphasizes a service-oriented perspective in network management by creating multiple end-to-end virtual networks, i.e.,  slices, for different services on top of shared physical network infrastructure. With network slicing, network resources are first reserved for respective services in network planning stages and later allocated to individual users in network operation stages~\cite{J_XShen_2020}. The creation, adjustment, and termination of slices are based on the varying spatiotemporal distribution of service demands to provide a high level of flexibility and adaptivity in network management~\cite{J_HZhang2016}.\footnote{SDN and C-RAN are also closely related to network virtualization since virtualization significantly simplifies and expedites their realization in modern wireless networks.}

Virtualization can be applied on different levels and scales in a network. Existing techniques include virtualization at \emph{node, link, resource}, and \emph{network} levels. Virtual nodes are abstractions of substrate nodes in a network such as servers, routers, and switches, and typical examples of node virtualization are storage and computing server virtualization \cite{J_GZhang2007}, \cite{J_FXu_2014}. Virtual links are the logical channels that interconnect virtual nodes. Virtual resources are abstractions of computing, memory, storage, and communication resources in a network~\cite{J_PBellavista_2019}, while physical resources at different locations can form virtual resource pools~\cite{J_KQu_TCom_2020}. For instance, the virtualization of a network function is the execution of a network control or service function by running software, supported with necessary resources. A virtual network is the combination of virtual nodes and links with proper virtual resource allocation for a service request to meet its QoS requirements, supported by necessary networking protocols. Besides the aforementioned works, more representative research works on node, link, resource, and network virtualization are summarized in Table~\ref{Table:NetVirtual}.

\begin{table*}[hbt!]
	\centering
\scriptsize{
	\caption{Some Representative Works on Node, Link, Resource, and Network Virtualization}\label{Table:NetVirtual}
\begin{tabularx}{0.92\textwidth}{|p{1.5cm}|l|p{2.5cm}|p{3cm}|X|}
\hline
\textbf{Type}    &  \textbf{Work}  & \textbf{Scenario} &  \textbf{Research Focus} & \textbf{Objective} \\ \hline
\multirow{10}{*}{\parbox{2cm}{Node \\Virtualization}} & \cite{J_XYuan2021} & Edge computing & Virtual edge node placement & Low-cost placement and fast response to user requests \\ \cline{2-5}
 & \cite{J_SYang2017} & Cloud computing & Virtual machine (VM) placement  & Reliable VM placement and routing \\ \cline{2-5}
 & \cite{J_MNagy_2018} & IP network & Virtual node/router as IP overlay & Practical IP-level resilience to link failures \\\cline{2-5}
 & \cite{M_Khan2015} & Wireless sensor network (WSN) & Architecture for sensor virtualization in WSN & Multiple applications share the same WSN  \\\cline{2-5}
 & \cite{J_SZaidi_2019} & C-RAN & Clustering of access points & Forming user-specific virtual base stations given QoS requirements \\
 \hline
\multirow{6.5}{*}{\parbox{2cm}{Link \\Virtualization}} & \cite{J_HDu2013} & WSN & Virtual backbone construction  & Enabling low-complexity backbone construction with performance guarantee  \\ \cline{2-5}
& \cite{J_FHosseini_2019} & Generic & Virtual link embedding & Reducing congestion probability given bandwidth demands \\ \cline{2-5}
&\cite{J_STomovic2019} & Internet service provider (ISP) network with SDN & Virtual link provision &  Maximizing network throughput subject to QoS and robustness constraints \\\cline{2-5}
\hline
\multirow{13}{*}{\parbox{2cm}{Resource \\Virtualization}} & \cite{J_CPapagianni_2013} & \multirow{2}*{Cloud computing}  & Composite virtual resource mapping  & Efficient mapping of computing and networking resources to substrate resources within
networked clouds  \\ \cline{2-5}
& \cite{J_TWood2015} & Cloud computing & VM migration & Low-cost transferring of VM storage and 
memory during VM migration over wide area networks \\ \cline{2-5}
&\cite{L_MKali_2016} & Radio access network (RAN) & Radio resource virtualization &  Maximizing throughput with fairness among multiple mobile network operators \\\cline{2-5}
&\cite{J_XLu_2019} & RAN & Radio resource virtualization &  Delay-bounded QoS provisioning through radio resource virtualization \\\cline{2-5}
& \cite{J_SZhang2020} & Vehicular network & Resource sharing among slices  & Reusing communication and caching resources to support applications with different QoS requirements \\ \cline{2-5}
\hline
\multirow{13}{*}{\parbox{2cm}{Network \\Virtualization}}  & \cite{J_OAlhussein_2020} & 5G core network with SDN & Network function chain embedding & Minimizing embedding cost subject to network resource constraints \\ \cline{2-5}
& \cite{J_NZhang_2017} & Core network with SDN & Network function chain embedding &  Minimizing total flow in the network subject to network resource constraints \\ \cline{2-5}
& \cite{J_JTang2019}  & C-RAN & Slice request admission &  Maximizing the revenue of the C-RAN operator subject to network resource constraints \\ \cline{2-5}
& \cite{J_QYe2018}  &  Heterogeneous wireless network &  Dynamic radio resource slicing &  Maximizing network utility through optimal bandwidth slicing and user association \\ \cline{2-5}
& \cite{J_TGuo2019}  & 5G RAN &  Radio resource allocation in RAN slicing &  Satisfying QoS requirements by proper resource mapping and scheduling \\ \cline{2-5}  
& \cite{J_JNi2018} & IoT & Service-oriented authentication & Privacy-preserving slice selection and secure access of service data \\
\hline
\end{tabularx}
}
\end{table*}

Regardless of its level and scale, virtualization in the context of networking typically demonstrates the following characteristics:
\begin{itemize}
	\item Abstraction - Abstraction provides a high-level overview of a network while hiding details of the underlying physical network entities (nodes, links, or networks) or resources~\cite{J_MKessler_2008}. This simplifies network management and facilitates flexible service provision; 
	\item Co-existence - Multiple virtual entities corresponding to a shared physical entity co-exist, or multiple virtual resource pools co-exist on the same physical resource pool~\cite{M_Mosharaf_COMM_2009}. This enables service-oriented virtual networks and improves network resource utilization efficiency;
	\item Isolation - Coexisting virtual entities corresponding to the same physical entity should function independently~\cite{S_JVBelt_2017}. This is necessary for guaranteeing service reliability, security, scalability, and QoS satisfaction.     
\end{itemize}    

Both academia and industry have spent a significant amount of efforts on network virtualization. For virtualizing core networks, some works leverage SDN techniques to separate the control and data planes through different protocols or application programming interface (API), e.g., OpenFlow~\cite{mckeown2008openflow}. Furthermore, network virtualization has been extended to radio access networks (RANs), and several frameworks for RAN virtualization are proposed. A SoftRAN framework enables both centralized and distributed RAN control based on the time sensitiveness of control decisions~\cite{gudipati2013softran}. Another framework, FlexRAN, offers a hierarchical architecture for real-time RAN control and incorporates a flexible  API to separate control and data planes in RANs~\cite{foukas2016flexran}. Initiated by industry, such as AT$\&$T and China Mobile, Open-RAN (O-RAN) is proposed as an open-source and open-interface platform to support RAN virtualization~\cite{ORAN}, which can incorporate AI and provide APIs for data-driven networking~\cite{breen2021powder,johnson2021open,foukas2021concordia}.

The adoption of virtualization techniques renders modern networks programmable, flexible, and scalable, which significantly increases cost effectiveness in network deployment and operation. Due to these benefits, it is foreseeable that advanced virtualization techniques will be essential to 6G. Meanwhile, the existing scope of network virtualization is limited in the sense that virtualization techniques mostly focus on network infrastructure and resources, yet less attention is given to end users. In 6G, end user virtualization will become necessary for two reasons. First, with increasingly diverse end user devices, resource-demanding services, and heterogeneous and dynamic networks, providing QoE guarantee for end users will become more challenging in the era of 6G. Accurate characterization and abstraction of end users, which necessitate end user virtualization, can be a precondition to QoE satisfaction. Second, as AI will be a highlight of 6G, extensive user data are required to fuel AI services and AI-based network management. Given such need for data, end user virtualization can be a competitive approach for collecting, managing, and processing data from end users.

%
%
%
%

%
%
%

\subsection{End User Virtualization}~\label{ss:EUVirtual}

 Until recently, only a few works study end user virtualization in the context of networking. One early example relevant to end user virtualization is network-hosted avatars, i.e., virtual agents, of end users for applications such as file downloading when the users are offline~\cite{J_AConte_2008}. Another example is virtual objects, proposed as a component in Internet of things (IoT) platforms~\cite{S_MNitti_2016}. The motivation is to handle the heterogeneity of physical objects (end users) via virtualization and to facilitate the provision of services to end users.  


As a potential paradigm to enable end user virtualization, digital twin attracts much attention lately. The concept of digital twin was originally conceived by Michael Grieves for product life-cycle management in industry in 2003~\cite{W_MGrieves_2014, khan2021digital}. Later, NASA and U.S. Air Force Vehicles developed a digital twin paradigm for vehicles to forecast their remaining usable life and the mission success probability~\cite{ glaessgen2012digital}. A digital twin is characterized by a full digital representation of a physical object or a process and real-time synchronization between the physical object or process and its corresponding digital replica. Digital twins can contain a large volume of data from physical objects or processes for advanced analytics, and the analytical results can be used to improve the performance of the corresponding physical objects or processes. Exemplary digital twins in general application scenarios, as well as potential requirements for the digital twins to enable big data analytics, are discussed in~\cite{TII_Gao}. Potential implementation of digital twins representing IoT devices in  industrial systems is proposed in~\cite{madni2019leveraging}. Other representative research works on digital twins are summarized in Table~\ref{Table:DigTwin}. 

Most existing research on digital twins in the network field focuses on applications, e.g., distributed clock synchronization~\cite{J_PJia_2020} and computation offloading \cite{J_YDai_2020}. In comparison, the study of digital twins from the perspective of network architecture and network management is limited at the moment.\footnote{Some works focus on distributed networks, e.g., vehicular networks, and adopt digital twins as an approach for network virtualization instead of end user  virtualization~\cite{M_LZhao_2020},\cite{C_JTaylor_2020}.} A digital twin based cloud-centric network architecture is proposed in~\cite{M_Q_Yu_2019}, where digital twins of end users hosted at the network edge play the role of communication assistants or network data loggers. 


\begin{table*}[hbt!]
	\centering
\scriptsize{
	\caption{Some Related Works on Digital Twins}\label{Table:DigTwin}
\begin{tabularx}{1\textwidth}{|l|p{2.3cm}|p{2.3cm}|p{4cm}|X|}
\hline
 \textbf{Work}  & \textbf{Application } & \textbf{Type of Physical Object} & \textbf{Role of Digital Twin} &  \textbf{Target of Using Digital Twin}\\ \hline
 \cite{J_ABarbie_2021} &  Underwater network for ocean observation & Underwater sensors/actuators & Monitoring and testing observation system & Visualizing an ocean observation system and enhancing simulations\\  \hline
 \cite{J_XXu2020} &   Edge computing for internet of vehicles & Vehicles & Collecting and sharing information about vehicles and surroundings & Empowering computing offloading by facilitating data analytics  \\  \hline
 \cite{J_HWang2020} &   5G network slicing & Network slices & Predicting and monitoring slice performance & Assisting autonomous network slicing \\  \hline
 \cite{J_TWang2020} &   Smart factory & Workstations in a conveyor system & Evaluating and validating control strategies
 & Implementing intelligent conveyor systems \\  \hline
\cite{M_OEMarai2020} &   Smart city & Road infrastructure & Monitoring roads and detecting vehicles/persons
 & Supporting smart city applications through gathering and processing data \\  \hline
\cite{J_RMinerval_2021} & IoT  &  Objects with sensing capability & Storing data for detecting events and recognizing behaviors & Facilitating synthetic sensing through situation awareness and explainability \\  \hline
\cite{J_ACastellani2020} &   Industry 4.0 &  Industrial machinery & Generating training dataset and simulations & Achieving accurate anomaly detection with limited labelled data \\  \hline
\cite{J_HElayan2021} &  Smart healthcare &  Patients  & Handling data for analysis and developing AI models & Improving healthcare operations \\  \hline
\cite{J_MSchluse2018} &  Industry 4.0 &  Technical assets (e.g., machine, environment)  & Integrating knowledge from model and data for simulations & Enhancing simulation-based systems engineering \\  \hline
\cite{J_RDong2019} &  Mobile edge computing &   Real-world network environments  & Training learning algorithms and monitoring network environments &  Enabling learning for optimizing user association, resource allocation, and
offloading \\  \hline
\cite{J_CGehrmann2020} &  Industrial 4.0 &  Products, workstations, conveyor belts  &  Data sharing
and control of security-critical processes &  Building a security architecture based on state replication and synchronization \\  \hline
\cite{J_KMAlam2017} &  Cyber-physical systems &   Generic physical devices  &  Monitoring, diagnostics, and prognostics  & Supporting applications such as context aware interaction and driving assistance \\  \hline
\cite{J_LFRivera2021} &  IoT &   Generic physical systems  &  Managing context information and self-adapting & Increasing  autonomy and enhancing cooperation through autonomic digital twins \\  \hline
\cite{zhang2020manufacturing} &  Welding manufacturing &  Human-robot interaction systems  &  Monitoring welding robot and enabling simulations & Visualizing  welder behavior and training welders \\  \hline
\cite{fang2019digital} &   Smart manufacturing &   Job shop scheduling systems  &  Obtaining scheduling data and simulating scheduling strategies & Enabling timely response and reducing scheduling plan deviation \\  \hline
\cite{xu2019digital} &  Smart manufacturing &   Manufacturing systems  &  Predicting and verifying the system performance & Increasing  autonomy and enhancing fault diagnosis \\  \hline
\cite{jain2019digital} &  Smart building &  Photovoltaic energy conversion units  &  Estimating the status of photovoltaic energy conversion unit & Improving the accuracy of fault detection \\  \hline
\cite{IOT_Sun_2021} &  Internet of vehicles &  Vehicles and road side units  &  Monitoring the real-time status of vehicles and road side units & Supporting network resource management  \\  \hline
\cite{TCSS_Zhang_2021} &  Mobile edge caching &   Vehicles  &   Capturing the social characteristics of vehicles & Improving the effectiveness of cache management \\  \hline
\end{tabularx}
}
\end{table*}

Digital twin appears to be an intuitive solution to end user virtualization. Nevertheless, extending the existing network virtualization, represented by network slicing, to end users is not straightforward, given the target of flexible and efficient network management and service provision. For instance, it is trivial to simply use node-level virtualization and to represent end users as virtual data sources or sinks in a virtual network. Moreover, while end users may possess communication and computing resources, resource-level virtualization does not characterize user-specific properties, e.g., location and mobility, or service-specific properties, e.g., data traffic variations, of end users. It is necessary to understand potential \textit{benefits, requirements}, and \textit{implementation} of digital twin based end user virtualization, with a particular focus on the integration of digital twin and existing network virtualization frameworks.  

There are potentially two-fold \textit{benefits} of digital twin based end user virtualization, i.e., extensive end user data and powerful network emulation capability. While the virtualization of network infrastructure and resources characterizes the network status and \textit{service provision} capabilities, digital twins of end users can provide extensive data regarding \textit{service demand} and user QoS/QoE satisfaction. Such data can play a significant role in network management through facilitating well-informed network planning and operation decisions. Moreover, the real-time or near real-time synchronization between end users and their digital twins enables powerful network emulations. For instance, multiple instances of the same virtual network can be created, with real-time end user information, e.g., location and data traffic volume, provided to all instances through synchronized end user digital twins.\footnote{The emulation can apply to a virtual network segment, e.g., the network edge.} Different network planning or operation strategies can be applied and emulated in different instances, while each instance remains synchronized with the real-world network environment through the information provided by the digital twins of end users.



To take part in network virtualization, digital twins of end users should satisfy the following \textit{requirements}:
\begin{itemize}
	\item Flexible: The abstraction of end users into digital twins must be sufficiently flexible to represent heterogeneous physical devices (such as smartphones, vehicles, and industrial sensors) and serve various applications (such as virtual reality gaming, autonomous driving, and industrial automation); 
	\item Compatible: The end user virtualization based on digital twins should complement and enhance the state-of-the-art network virtualization, i.e., network slicing. For instance, digital twins of end users should provide data to support various network slices, while each slice may only have access to a subset of data pertinent to that slice;      
	\item Customizable: The attributes of digital twins should be customized and updated based on the corresponding service, network traffic, resource utilization, etc. For instance, the amount and types of data included in a digital twin should be adaptable rather than fixed. In addition, while the focus of digital twins is placed on end users, digital twins should be able to represent other network entities, e.g., unmanned aerial vehicle (UAV) mounted mobile base stations (BSs). 
\end{itemize}
In addition, network resource consumption from creating and maintaining digital twins should be taken into account.

Noting the aforementioned benefits and requirements, we aim to answer the following key questions with respect to the \textit{implementation} of digital twins: 
\begin{itemize}
	\item Location: Where should digital twins be hosted in a network? 
	\item Affiliation: Should digital twins exist within or outside  network slices? 
	\item Data: What data attributes pertinent to networking should be included in a digital twin? How much historical data should be included for a specific attribute? Should predicted user information be included?
	\item Synchronization: How to determine the frequencies of updating various data entries of a digital twin by acquiring new data from the physical object?     
	\item Control: Who should determine and update digital twin models and based on what information?
\end{itemize}
In the next subsection, we propose a novel conceptual architecture for holistic network virtualization, which integrates digital twins and network slicing, and delve into the above questions.



 \begin{figure*}[tt]
	\centering
	\renewcommand{\figurename}{Fig.}
	\includegraphics[width=1\textwidth]{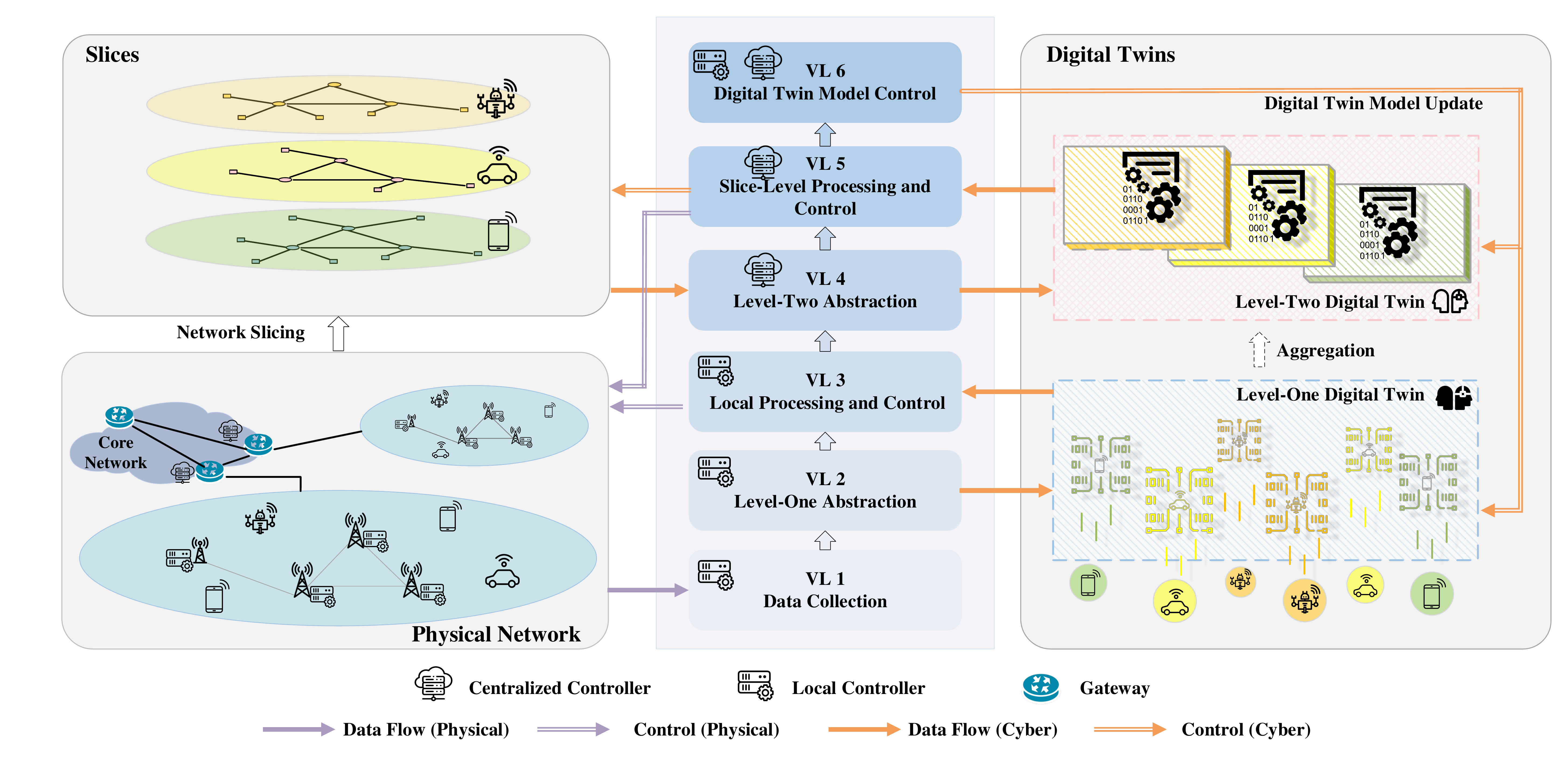}
	\caption{The conceptual six-layer virtualization architecture for holistic network virtualization.}
	\label{fig:digital_twin}
\end{figure*}

\subsection{Holistic Network Virtualization}\label{ss:HNV}

We propose a novel virtualization architecture, i.e., holistic network virtualization, for integrating digital twins into network virtualization, in order to improve network management and service provision capabilities. The proposed virtualization architecture consists of six layers and is illustrated in Fig.~\ref{fig:digital_twin}, in which virtualization layer (VL)~1 is the bottom layer for data collection and VL~6 is the top layer for digital twin model control. The outline of each layer is given as follows:

	\textit{VL~1 -- Data Collection}: Data required for the digital twin representation of selected end users are collected from the corresponding physical entities following prescribed data precision, uploading method, collection frequency,~etc. The data are collected via access points, and the data collection is controlled by local controllers deployed at network edge; 
	
	\textit{VL 2 -- Level-One Abstraction}: Based on the current \textit{digital twin model} from the digital twin model control layer (i.e., VL 6), which determines the content and format of data included in every digital twin, digital twins are formed and updated using data collected by VL~1. The abstraction may include the aggregation of data from different sources, the update of historical data, and the creation of digital twins for new or additional end users. The digital twins created in this layer are \textit{level-one digital twins}, representing individual end users, and hosted at servers connected to local controllers;
	
	\textit{VL 3 -- Local Processing and Control}: The data from level-one digital twins are processed at network edge for predicting behaviors of individual users, such as user data traffic and mobility patterns, and making user-level service decisions, such as computing offloading, content delivery, and link-layer protocol adaption. Local processing may also include emulations of an edge network or a part of it based on level-one digital twins. Local control may include further data aggregation from level-one digital twins for VL~4, the migration of digital twins based on user mobility, and the selection of end users for digital twin representation. Similar to the case of VL~2, the local processing and control occur at servers affiliated with local controllers; 
	
	\textit{VL 4 -- Level-Two Abstraction}: The aggregated data from VL~3 is sorted into service-specific data for respective network slices in VL~4. Additional data that describe slice configuration, slice resource utilization, slice service level agreement satisfaction, etc., are generated for each slice. Then, the aforementioned data are abstracted to form or update the digital twins of various slices. The digital twins created in this layer are \textit{level-two digital twins}, which are associated with virtual networks (slices). The level-two digital twins are hosted at servers connected to the centralized controller of the network;
	
	\textit{VL 5 -- Slice-Level Processing and Control}: The data from level-two digital twins of network slices are processed for service-specific prediction, e.g., spatiotemporal service demand distribution forecast, or slice-level decision making, e.g., planning and operation decisions. Slice-level processing may include emulations of an end-to-end slice or a part of it based on level-two digital twins. Slice-level control may include slice admission, resource reservation, and slice service coverage control. Similar to the case of VL~4, the service-level processing and control occur at servers affiliated with the centralized controller of the network; 
	
	\textit{VL 6 -- Digital Twin Model Control}: This layer determines and updates the models of level-one and level-two digital twins based on  available network resources for digital twins, the performance of network management and service provision decisions derived based on the current digital twins, and the dynamic spatiotemporal service demands. For instance, VL~6 determines data precision, synchronization frequencies for different data attributes, the amount of historical data contained in the digital twins for each attribute, and the inclusion of predicted user information. In addition, this layer decides the subset of data in level-one digital twins that each slice can access. The digital twin model control also occurs at servers affiliated with the centralized controller of the network;

The level-one digital twin model configured by VL~6 may include the following data, which shall be collected by the local controllers from end users at VL~1: (1) connectivity and channel information, such as the AP(s) that an end user is connected to and the channel state information for each connection; (2) service information, such as active service types, data traffic volume of each service, and QoS satisfaction of each service; (3) user information, such as user profile, user location and mobility, network resources allocated to the user, and the local computing and caching capabilities of the user; and (4) additional use case-specific information, such as motion sensor readings for augmented reality interactive gaming or operation log for industrial IoT devices. The level-two digital twin model configured by VL~6 may include the following data, which shall be collected or generated by the centralized controller: (1) slice service demand, such as the number of service requests and the spatiotemporal service request distribution; (2) slice resource configuration, such as the reserved communication, computing, and caching resources for the slice; (3) slice performance, such as the slice service level agreement satisfaction, slice resource utilization, and slice energy consumption; (4) slicing strategy, such as the method or algorithm used for network function deployment, resource reservation, and resource scheduling; (5) additional use case-specific information, such as UAV trajectory configuration for UAV-assisted networks. Note that different end user digital twin models are applicable to different types of end users, and each network slice may have a uniquely defined slice digital twin model. For example, the digital twins of vehicles and industrial IoT devices most likely contain different data, and the digital twin models may differ between slices for industrial IoT and those for smart home or between slices of different network operators. Accordingly, the need for customization necessitates the digital twin model control in VL~6.             

In the conceptual virtualization architecture, VL~1 to VL~3 interface with the local controllers in the network, VL~4 and VL~5 interface with the centralized controller of the network, and VL~6 interfaces with both the local controllers and the centralized controller. This architecture fully exploits the two benefits of digital twins, i.e., providing extensive data for network management and enabling powerful network emulations. It also satisfies the aforementioned requirements for digital twins in terms of flexibility, compatibility, and customization. Last but not least, it answers the key questions regarding the implementation of digital twins raised in Subsection~\ref{ss:EUVirtual}.      

With the architecture design in Fig.~\ref{fig:digital_twin}, digital twins and network slicing are integrated in the idea of holistic network virtualization. Network slicing incorporates existing network virtualization techniques such as NFV. Digital twins enhance network slicing by providing organized and customized end user data to slices and by further abstracting slices into level-two digital twins. The design of two-level digital twins avoids extra resource consumption from creating and maintaining multiple digital twins of the same user for different slices and the resulting burden of synchronizing them. Instead, each slice has access to a subset of data from level-one digital twins pertinent to either the corresponding service or general user information such as location and mobility, and the pertinent data are further aggregated to the level-two digital twins for that slice. In this architecture, network slicing conforms to service-centric network management, while digital twins add a user-centric perspective to the virtualization. Specifically, level-one digital twins characterize end users and their service demands, and level-two digital twins  characterize network service provision capability, network performance, and network resource utilization. Overall, the digital twin paradigm and network slicing jointly support network management and service provision, while the network configures digital twins and network slices as needed, depending on network dynamics. 

\subsection{Holistic Network Virtualization: A Summary}

In this section, we have reviewed the existing scope and techniques of network virtualization, identified the insufficiency of current network virtualization, introduced the idea of holistic network virtualization to incorporate network and end user virtualization, and developed a six-layer virtualization architecture for holistic network virtualization.  

The virtualization of resources, network functions, and networks in 5G is expected to remain important in 6G, since they contribute to flexible and adaptive network management. Meanwhile, the virtualization techniques in 5G, represented by network slicing and NFV, 
mostly focus on network virtualization from the perspective of service provision. In 6G, it will be essential to extend the scope of virtualization and incorporate end user virtualization. 

The digital twin paradigm is a promising solution to end-user virtualization. In 6G, digital twins can be used for characterizing the status and the service demand of individual end users. The study of digital twins in the context of 6G networks is still in an initial stage, and various definitions or implementations exist. In our vision of holistic network virtualization, digital twins are configurable assemblage of data, including both historical and real-time data and both collected and generated data, for describing end users, infrastructure, or network slices. Moreover, the corresponding data collection and processing are also configurable.   

To consolidate holistic network virtualization, we have proposed a six-layer virtualization architecture for 6G. The architecture provides a reference design for systematically integrating digital twins and network slicing and answers important questions related to digital twins in 6G networks, including where are they hosted, what data do they contain, and how to manage them.

\section{Pervasive Network Intelligence}
\label{sec:PI}



Pervasive network intelligence is the second element of our vision for 6G. In this section, we first present an overview of existing AI techniques. Then, we introduce the motivation and propose a four-level architecture for pervasive network intelligence. Next, we elaborate the idea of pervasive network intelligence from the perspectives of AI for networking and networking for AI, and review related works. Rather than surveying specific AI techniques, this section focuses on the architecture and methods of pervasive network intelligence. 

\subsection{{AI Techniques: An Overview}}\label{ss:AIOv}
\begin{table*}[]
\caption{{Common ML algorithms.}}\label{table:AI}
\centering {
\begin{tabular}{|l|l|l|l|}
\hline
& Unsupervised Learning   & Supervised Learning   & Reinforcement Learning  \\ \hline
\begin{tabular}[c]{@{}l@{}}Centralized \\ Learning \\ Algorithms\end{tabular} & 
\begin{tabular}[c]{@{}l@{}} $\bullet$ K-means \cite{hartigan1979algorithm, Parwez}\\ $\bullet$ Mixture models \cite{topchy2004mixture,Shih}\\ $\bullet$ Autoencoders \cite{Bega}\\ $\bullet$ Generative adversarial\\ network \cite{ledig2017photo,Erpek} \end{tabular} &
\begin{tabular}[c]{@{}l@{}}$\bullet$ Support-vector machine \cite{SVM_MAC}\\ $\bullet$ Logistic regression \cite{You_LR}
\\ $\bullet$ Deep neural network \\ \cite{Peng_CNN,zhao2017lstm,Bega}\end{tabular} &
\begin{tabular}[c]{@{}l@{}}$\bullet$ Deep Q-learning \cite{DQN2,DQN1,Yaohua}\\ $\bullet$ Policy gradient\cite{silver2014deterministic,Somuyiwa,Hoang}\\ $\bullet$ Actor-critic \cite{konda2000actor,Cheng_AC}\\ $\bullet$ Deep deterministic policy \\gradient (DDPG) \cite{lillicrap2015continuous,TCCN_MUSHU,J_WWu2020}\end{tabular} \\ \hline
\begin{tabular}[c]{@{}l@{}}Distributed \\ Learning \\ Algorithms\end{tabular} & 
\multicolumn{2}{l|}{\begin{tabular}[c]{@{}l@{}}$\bullet$ Federated learning\cite{konevcny2016federated,Zhaohui}\\ $\bullet$ Split learning\cite{thapa2020advancements}\end{tabular}}                                                                                           
& \begin{tabular}[c]{@{}l@{}} $\bullet$ Multi-agent reinforcement learning\\ \cite{Weisen,Sana,Ruijin,Yunting}       \end{tabular}                                                                                                      \\ \hline
\end{tabular}}
\end{table*}
{The idea of AI is to design intelligent machines or systems to demonstrate human intelligence and perform tasks as humans do or even better~\cite{zhou2019edge}. The  advancement of machine learning (ML) has facilitated the success of AI in both academia and industry. Applications supported by ML techniques, such as computer vision and natural language processing, can achieve beyond human-level accuracy. Lately, for its potential in enabling intelligent networks, AI has received significant attention in the research field of wireless networks. }

{ML techniques can be categorized into three types: unsupervised learning, supervised learning, and reinforcement learning. In terms of learning structures, the techniques can be subdivided into centralized and decentralized techniques. We list common ML techniques used in wireless networks in Table \ref{table:AI}. }

{Unsupervised learning evaluates features and patterns hidden in data for data analysis, such as prediction, without using a labeled dataset. One popular application of unsupervised learning techniques is data clustering, e.g., \textit{k}-means \cite{hartigan1979algorithm} and mixture models \cite{topchy2004mixture}, for solving network planning problems, such as cluster-forming in wireless sensor networks \cite{Parwez} and small-cell deployment \cite{Samarakoon}. Neural networks can be adopted to facilitate novel unsupervised learning algorithms. For example, neural network-based autoencoders can learn the compressed features of input data with a limited number of neurons and can be leveraged for data prediction, such as traffic forecasting \cite{Bega}.}

{Supervised learning exploits the mapping between the input and output data via a given labeled dataset. Supervised learning techniques can derive a mapping function, i.e., a training model, from the input data to the labeled output data in the dataset. Through applying a training model, the output corresponding to a new input can be evaluated, which can be utilized for decision making or prediction. A typical method for supervised learning is using deep neural networks (DNNs). DNNs use layers of artificial neurons to estimate a non-linear correlation between the input and the output data and iteratively improve the estimation accuracy. There have been many successful applications of DNN techniques in communications. For example, convolutional neural networks (CNNs) utilize convolutional and pooling layers to identify the correlation of multi-dimensional input data and have been applied in modulation classification \cite{Peng_CNN}; recurrent neural networks (RNNs) explore the correlation among a sequence of the data and have been widely adopted for traffic prediction \cite{zhao2017lstm} and wireless channel modeling \cite{Zhu_RNN}.}

{Reinforcement learning iteratively learns the optimal policy by interacting with the environment, sensing network states, and evaluating feedback. The goal is to maximize a cumulative reward in a dynamic environment. 
Deep reinforcement learning (DRL), which combines DNN and reinforcement learning techniques, is used extensively in resource management to solve complex decision-making problems.
In DRL, neural networks play the role of approximators to store high-dimensional states or actions, which enables DRL to solve complex problems efficiently. DRL has been widely used for network optimization \cite{Yang_INFOCOM}, resource allocation \cite{J_WWu2020,DRL4,Somuyiwa}, and user association \cite{Cheng_AC,TCCN_MUSHU,Yaohua} in wireless networks.}

{With the development of mobile edge computing, distributed AI has been developed to harvest computing resources at network edge and reduce communication overhead due to data collection and exchange \cite{MEC_survey}. The learning models can be trained and evaluated at network edge in a semi- or fully-distributed manner. Specifically, federated learning (FL), as one of the most popular distributed learning techniques, trains models with data distributed over network edge. A centralized controller aggregates locally-computed learning models and updates parameters in the learning models.  Due to such decentralized model training, FL is capable of preserving privacy and can be applied in privacy-sensitive network management scenarios \cite{Niknam,lim2020federated}. In addition, multi-agent reinforcement learning has been developed to implement reinforcement learning in a distributed manner, which aims to handle scenarios in which network agents cannot obtain sufficient information from each other. Multi-agent reinforcement learning techniques can be used, for example, to solve resource allocation problems in heterogeneous networks \cite{Ruijin,Yunting}.  }

\subsection{Motivation and AI Architecture}\label{ss:Moti}
In 6G, AI is expected to penetrate every facet of the network including end users, the network edge, and the cloud, resulting in \emph{pervasive network intelligence}. Such trend is due to advancements and innovations in the areas of ML, data collection, edge and cloud computing, and programmable network control in recent decades. As such, AI will fundamentally transform modern networks in many aspects and foster a myriad of exciting applications. 



The AI applications can be categorized into management-oriented and service-oriented  applications, which are detailed as follows:
\begin{itemize}
	\item \emph{Management-Oriented AI Applications} - In these applications, AI is used as a tool for network management, such as transmission power allocation in cellular networks~\cite{sun2018learning} and resource reservation in network slices~\cite{J_WWu2020}. AI techniques, such as reinforcement learning, have the potential of handling complicated decision making problems in a dynamic network environment. Resorting to AI techniques, the management-oriented AI applications can analyze a large amount of network data, make real-time network management decisions, and then update network management policies based on the newly analyzed data. Hence, for such applications, the key issue is how to leverage advanced AI techniques to manage and enhance complex networks, which falls into the scope of \emph{AI for networking};
	
	\item \emph{Service-Oriented AI Applications} - In these applications, AI is offered as services for end users. Fuelled by powerful computing servers and well-curated datasets, AI techniques, especially  DL, can outperform traditional techniques in a wide range of applications, such as environmental perception in autonomous driving, audio recognition in intelligent healthcare, and object detection in mobile virtual reality~\cite{zhu2017overview, gunduz2019machine, chen2019deep}. For instance, an AI-based YOLO algorithm can detect objects with a high accuracy~\cite{lin2018architectural,liu2018edge}, and the state-of-the-art DL-based face recognition algorithm can achieve an accuracy of  99\% or higher~\cite{huang2008labeled}. Facilitating service-oriented AI applications in a network consumes a large amount of network resources, including storage  and computing resources for model training/inference, and communication resources for data collection and model uploading. Hence, for such applications, the key issue is how to design and optimize the network to support emerging AI services, which falls into the scope of \emph{networking for AI}. 
\end{itemize}
Note that the scope of AI in 6G includes AI for networking and networking for AI, which is larger than that in 5G, as the latter simply focuses on  applying AI in  communications.


\begin{figure}[t]  
	\centering  
	\includegraphics[width=92mm]{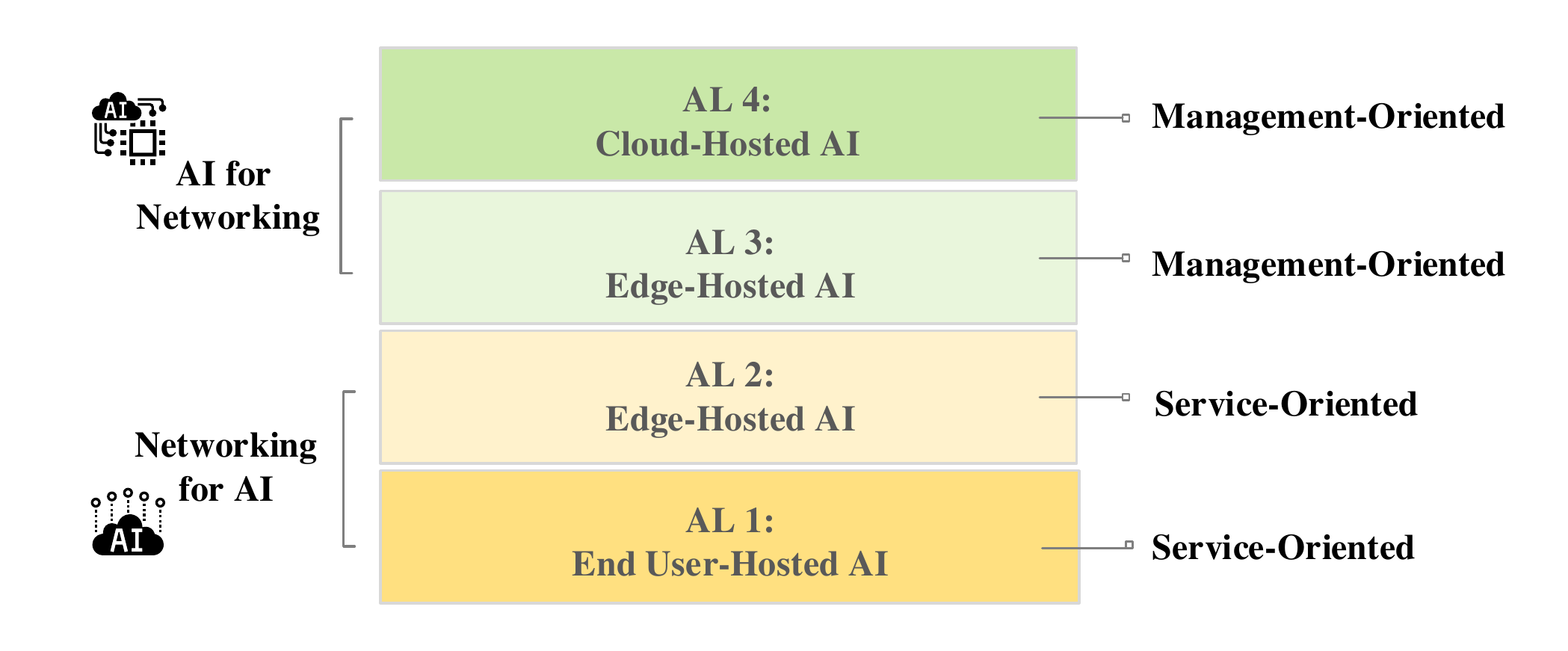}\\
	\caption{An illustration of the four-level AI architecture for pervasive network intelligence.}  
	\label{fig:1} 
\end{figure}

An AI architecture is needed to characterize AI's different functionalities in different network segments. In the literature, there are a few studies on the AI architecture. Edge intelligence (or edge AI) is represented in six levels based on the amount and path length of data offloading~\cite{zhou2019edge}. Moreover, edge intelligence can be categorized into two parts: AI for edge (i.e., to enhance and optimize the network edge with AI techniques) and AI on edge (i.e., to carry out AI models on the network edge)~\cite{deng2020edge, peltonen20206g}. Different from these works on edge intelligence, our work focuses on a broader scope of pervasive network intelligence and categorizes it into multiple levels based on AI's locations and functionalities in the network.

As shown in Fig.~\ref{fig:1}, we propose a four-level AI architecture, in which AI levels (ALs)~1 and 2 focus on service-oriented applications, and ALs~3 and 4 aim at management-oriented applications. We describe each level in detail as follows.

\emph{AL 1 - End User-Hosted Service-Oriented AI}: Utilizing local data and computing resources at end users, end user-hosted service-oriented AI applications are offered as services for end users by processing AI tasks locally, such as next word prediction in mobile keyboards~\cite{hard2018federated}, user traffic demand prediction~\cite{wang2020cellular}, and vehicle trajectory prediction~\cite{deo2018convolutional}. When computing resources of end users are insufficient for computation-intensive AI tasks, partial computation workloads can be offloaded to nearby edge servers for collaborative processing.  

\emph{AL 2 - Edge-Hosted Service-Oriented AI}: Residing at network edge (e.g., Wi-Fi access points and BSs) close to end users, edge-hosted service-oriented AI applications are offered as low-latency services for end users, such as face recognition in video surveillance~\cite{cocskun2017face} and object detection in virtual reality~\cite{erhan2014scalable}. To support edge-hosted service-oriented AI applications, service demand data from end users are collected, stored, and analyzed, and then the analytical results are utilized for service provision.   
	
	
\emph{AL 3 - {Edge-Hosted} Management-Oriented AI}: At this level, AI is hosted at local  controllers at  network edge for network management that is executed in real time, such as spectrum allocation, content caching~\cite{muller2016context}, and computation  offloading~\cite{zhou2020deep}. Specifically, the edge-hosted management-oriented AI is  to allocate network resources to network nodes for supporting services, including AI services at ALs~1 and~2. For instance, the edge-hosted management-oriented AI can be used to perform service migration across edge networks to guarantee service continuity for high-mobility users, e.g., vehicular users. 
	
\emph{AL 4 - Cloud-Hosted Management-Oriented AI}: Cloud-hosted management-oriented AI resides at the centralized controller in the cloud for network management that is executed once every several minutes or hours, such as slice admission control~\cite{sciancalepore2019rl} and virtual network function deployment~\cite{qu2020dynamic}. Since the cloud possesses abundant computing and storage resources, powerful and complex  AI models can be trained and deployed for managing large-scale networks.
	

Next, AI for networking is elaborated in Subsection~\ref{ss:AIforNet} to illustrate AI's role in network management, and networking for AI is discussed in Subsection~\ref{ss:NetforAI} to illustrate AI service provision in 6G networks.

\subsection{AI for Networking}\label{ss:AIforNet}

In this subsection, we discuss how AI techniques can support network management. We first review existing works on AI-based network slicing. Then, we introduce our idea of connected AI solution for AI-based network slicing.

\begin{table*}[h]
	\centering
\scriptsize{
	\caption{Representative Works on AI-based Network Slicing}\label{Table:NetworkSlicing}
\begin{tabularx}{0.92\textwidth}{|p{1.2cm}|l|p{3cm}|p{6cm}|X|} 
\hline
\textbf{Stage}    &  \textbf{Work} & \textbf{Research Focus}  & \textbf{Objective}  &  \textbf{AI Method} \\ \hline
\multirow{13}{*}{{\shortstack[l]{Network\\Planning}}}  & \cite{Bega} & Network capacity prediction  & Forecasting the capacity for individual virtual networks & Deep neural network based autoencoder \\ \cline{2-5}
 & \cite{J_HWang2020} & Virtual representation for network slices & Capturing the relationships among slices and monitoring the end-to-end performance in dynamic network environments & Graph neural networks \\\cline{2-5}
 & \cite{Huynh} & Resource reservation adjustment &  Maximizing the overall reward obtained from the tenants of slices & Deep dueling neural networks \\\cline{2-5}
 & \cite{gnn2} &  Bandwidth allocation &  Jointly maximizing spectrum efficiency and the QoS requirement satisfaction ratio & Generative adversarial network and deep Q network\\\cline{2-5}
 & \cite{Li_CL}& Bandwidth allocation & Jointly maximizing spectrum efficiency and overall service level agreement satisfaction ratio of slices& Long short-term memory and advantage actor-critic\\\cline{2-5}
 & \cite{Chergui} & Traffic prediction and resource provisioning & Minimizing the probability of slice service level agreements violation & Gated recurrent unit\\
 \hline
\multirow{13}{*}{{\shortstack[l]{Network\\ Operation}}}  & \cite{drl2} & Computation offloading & Minimizing average computing time of services and maximizing user computing experience & Deep Q network\\ \cline{2-5}
& \cite{Xiang_TVT} & Slice selection and channel allocation & Minimizing the power consumption of wireless transmission for a sliced fog-RAN & Reinforcement learning \\ \cline{2-5}
& \cite{Xiang_TWC} & Content caching placement and delivery &  Managing caching resources to maximize cache hit ratio while satisifying resource reservation constraints & Deep Q network\\ \cline{2-5}
& \cite{Chen_JSAC} & Inter-slice coordination  & Maximizing long-term payoff from the competition among service providers through resource orchestration & Deep Q network\\ \cline{2-5}
& \cite{Messaoud}& Inter-slice coordination &  Maximizing QoS satisfaction ratio for slices by scheduling transmission power and sharing resources among slices & Multi-agent deep Q learning\\
\hline
\multirow{5.5}{*}{{\shortstack[l]{Two-Stage\\Interplay}}}  & \cite{J_WWu2020} & Computing resource allocation in vehicular networks & Allocating spectrum and computing resources for slices while minimizing computing service delay & Deep deterministic policy gradient  \\ \cline{2-5}
&\cite{Dandachi} &  Cross-slice admission and congestion control & Maximizing operator revenue by resource reservation and adjust reserved resources in real time & State-action-reward-state-action (SARSA) \\
\hline
\end{tabularx}
}
\end{table*}

\subsubsection{AI-Based Network Slicing}\label{ss:AISlicing}

Network slicing includes two stages: network planning stage for resource reservation and network operation stage for resource scheduling~\cite{J_XShen_2020}. In the \textit{network planning} stage, network resources are reserved for network slices on a large time scale (e.g., from several minutes to several hours).  In the \textit{network operation} stage, the reserved resources of each slice are allocated to end users on a small time scale (e.g., from several milliseconds to several seconds). Due to network dynamics such as spatiotemporally changing network traffic, it can be difficult for model-based solutions to attain the optimal network slicing strategies. By contrast, AI techniques can characterize network dynamics by analyzing the collected network data and obtain the optimal network slicing strategies accordingly. Next, we review AI-based network slicing, taking into account the interplay between the planning and operation stages. Representative research works on AI-based network slicing are summarized in Table~\ref{Table:NetworkSlicing}.


On a small time scale, a local controller collects data and provides resource scheduling strategies to allocate resources reserved for each slice to end users. Specifically, the local controller determines resource scheduling strategies based on two factors: the amount of resources reserved for each slice, which is determined by the centralized controller, and the instantaneous user data from level-one digital twins pertinent to that slice, such as service type, user location, and user mobility. The main challenges of determining the optimal resource scheduling strategies are two-fold: a large number of end users and service demand dynamics. AI techniques have potentials to cope with both challenges. First, to schedule resources for a large number of end users, unsupervised learning methods, such as \textit{k}-means~\cite{kmean} and DNN based autoencoders~\cite{Bega}, can be utilized to classify end users according to their service demands. Similar resource scheduling strategies can be applied to end users with similar service demands, which facilitates scalable network management. For instance, end users in close proximity and with similar mobility patterns may experience similar channel statistical behaviors, and the same power control policy can be applicable to them. Second, to deal with network dynamics, reinforcement learning can be applied for generating adaptive resource scheduling strategies~\cite{drl1}.  Reinforcement learning iteratively allocates resources to maximize a long-term reward function and updates the reward function based on feedback from the network environment. Moreover, reinforcement learning can be combined with DNNs, such as recurrent neural networks~\cite{J_MLi2021} and conventional neural networks~\cite{TCCN_MUSHU}, to analyze the spatiotemporal pattern of user data for finding the optimal resource scheduling strategies. 

On a large time scale, local controllers aggregate the collected user-level data to service-level information from level-two digital twins, i.e., slice digital twins. Utilizing information from slice digital twins, the centralized controller reserves network resources for each slice. The challenges of resource reservation are two-fold. First, making proactive resource reservation that can avoid either resource over-provisioning or under-provisioning is challenging with time-varying network traffic. Second, the strategies for resource reservation and scheduling are coupled, which further complicates resource reservation. AI techniques can cope with these challenges as follows. To address the challenge of proactive resource reservation, supervised learning, such as long short-term memory (LSTM) networks, can be used to exploit the features of historic network traffic loads and predict traffic loads in near future~\cite{Li_CL, wang2020cellular}. The centralized controller can then use the predicted traffic loads for proactive resource reservation.
To handle the correlation between resource reservation and scheduling, reinforcement learning can be adopted to reserve resources while considering network operation strategies as a part of the  dynamic network environment~\cite{J_WWu2020,DRL4, Chen_JSAC}. Moreover, an option-based hierarchical reinforcement learning technique can be a potential solution for jointly optimizing resource reservation and network operation policies and addressing network dynamics in both stages. This technique has been used to tackle complex DRL problems by grouping decision variables according to decision time scales~\cite{HDL1} or decision-making agents~\cite{Qian} and then determining the decision variables. Through this novel reinforcement learning technique, the complex correlation between resource reservation and scheduling strategies can be obtained iteratively. To apply this technique in network slicing, the centralized controller can select the resource reservation strategies on a large time scale, and local controllers find optimal resource scheduling strategies on a small time scale, thereby jointly optimizing both the resource reservation and the scheduling strategies. 

\subsubsection{Connected AI Solution for Network Management}\label{ss:nested}
Existing AI applications on network management mostly focus on individual control functions.
For instance, learning-based autoencoders can achieve reliable transmission power control for high-speed data transmission with limited channel state information~\cite{Dorner}, and DNNs can select medium access control protocol parameters with low communication and processing overhead~\cite{JieG_mac1},\cite{Jie_mac2}. 
Although various AI techniques have been proposed for network management, AI models among network control functions are usually isolated. Such isolation may result in inefficient and redundant data processing, which brings up a pressing need for integrating the AI models in AI-based network control functions. 

There are three types of solutions for integrating AI models~\cite{M_ZZhang_2019}. In the first type of solutions, the entire network is viewed as a black box, where a single AI model characterizes the entire network and generates network control policies. Such structure simplifies decision-making processes in network management. However, training the single AI model can be extremely difficult due to high-dimensional input data. Then, the second type of solutions adopts different AI models in a network for different network control functions, and the AI models are generally independent on each other to reduce the complexity of training. However, this approach neglects the correlation and interplay among network functions and thus cannot obtain a global-optimal network management strategy. Moreover, network data may be repetitively processed by different AI models with similar network functions, which degrades network management efficiency. For instance, AI models for user mobility management and computing service migration would repetitively analyze end user mobility. In contrast, the third type of solutions, namely \textit{connected AI}, exploits the correlations among network control functions, connects their AI models, and allows them to jointly make network control decisions. The connected AI solution offers benefits in integrating AI models by highlighting the interplay among them and balancing training complexity and network performance. Therefore, the connected AI solution has great potential in facilitating AI-based network slicing. However, existing research on the connected AI solution is limited. How to apply connected AI solution to network management requires further studies~\cite{M_SHan2020}.

Recent advancements in distributed learning techniques facilitate the development of a connected AI solution for network management. Model partition, investigated in~\cite{Yiping}, can divide a global DNN into multiple sub-neural networks (sub-NNs). The sub-NNs can reside at different network entities, according to the available computing and communication resources, and communicate with each other~\cite{partition1, partition3}. Furthermore, the idea of \textit{nested DNN}, which allows sub-NNs to have their own functionalities while contributing to the global DNN for model inference and training, has been proposed and evaluated in~\cite{cui2021fully} and~\cite{nest1}. Using the above two techniques, each sub-NN can perform a specific network control function. Accordingly, multiple sub-NNs can collaboratively fulfill common control functions, thereby applying the connected AI solution to network management.

\begin{figure}[t]  
  \centering  
    \includegraphics[width=90mm]{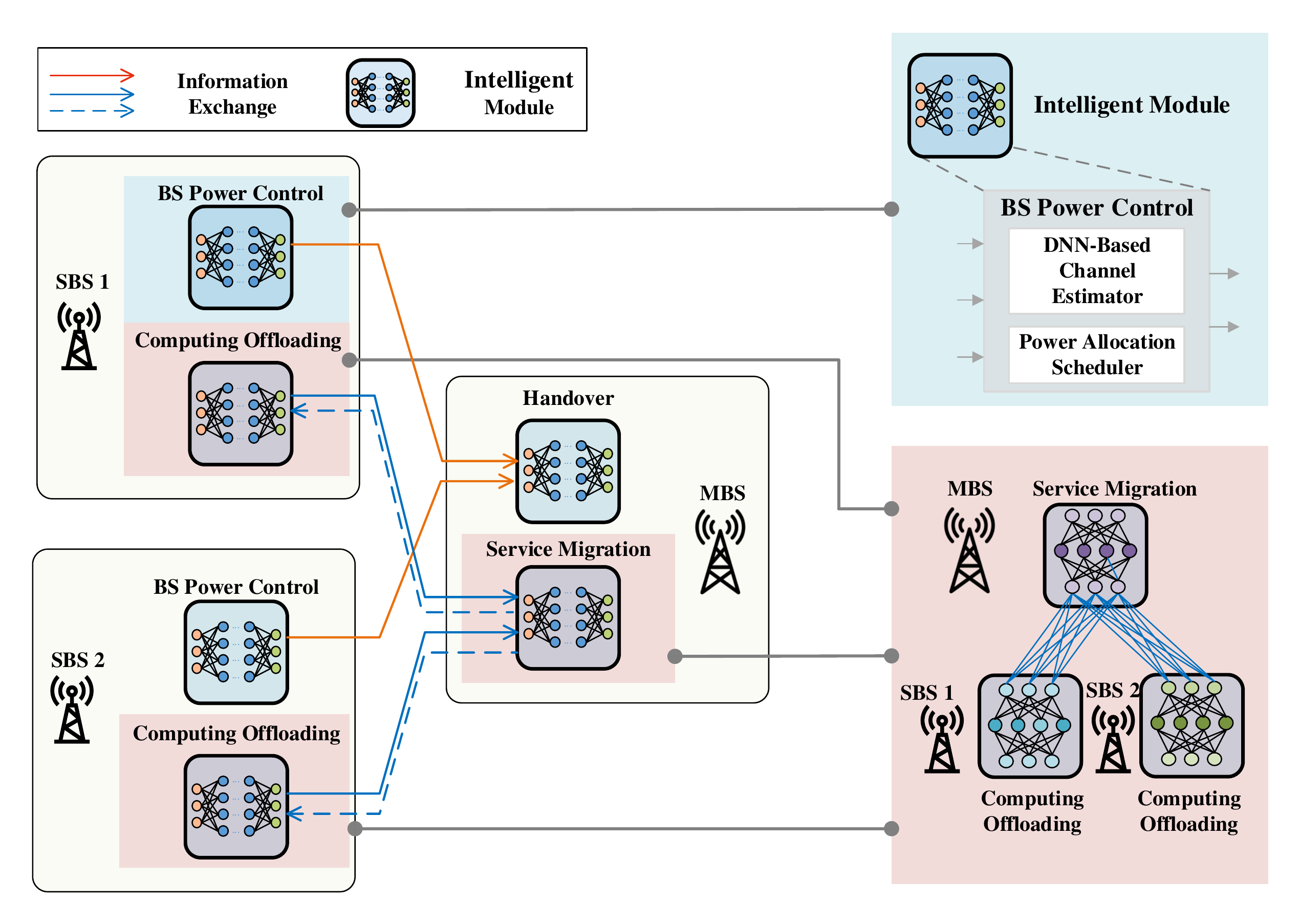}\\
  \caption{The connected AI solution for network management.}  
  \label{fig:3} 
\end{figure}

Based on the above advanced DNN techniques, we present our idea of applying the connected AI solution to network management next. The control functions for network management are encapsulated into \textit{intelligent modules}. An intelligent module can be implemented solely by a DNN or cooperatively by a DNN and conventional model-based techniques. An example is shown in the upper right corner of Fig.~\ref{fig:3}, in which the intelligent module for power control includes a learning-based channel estimator and a model-based power allocation scheme, e.g., water-filling power allocation~\cite{Haykin}. Moreover, the DNN in each intelligent module can play the role of a sub-NN of a global DNN. 
The intelligent modules connect with each other to share information, such as their outputs and gradient information in model training, and aggregated user data.
Via the intelligent modules, control functions can manage the network in a divide-and-conquer manner to avoid the complicated model training required for layer-free AI. With model partition and nested DNN techniques, multiple network control functions can cooperatively make control decisions to achieve globally optimal network management.

Fig.~\ref{fig:3} shows another example of the connected AI design, i.e., supporting mobile edge computing. We explain the design using the case of  vehicular networks as an example. Note that other networks can use the same or a similar design. Small base stations (SBSs), as edge servers, can process computation tasks offloaded by vehicles. Intelligence modules at SBSs provide computing offloading decisions, including the computing tasks to be offloaded, transmit power for offloading, task scheduling, etc., based on network status and computing offloading requests. Due to the high mobility of vehicles and the limited communication coverage of the SBSs, computing tasks are often migrated among the SBSs, referred to as service migration, and migration decisions are determined by a macro base station (MBS). Service migration and computing offloading decisions are highly coupled. For example, the chance of service migration increases when  vehicles offload more computing tasks to an SBS. In addition, service migration requires the collaboration of multiple SBSs. In our idea of connected AI, service migration and computing offloading decisions are jointly determined. Specifically, we split the DNN into multiple sub-NNs by DNN splitting and nested DNN techniques. Some sub-NNs are deployed at the SBSs to provide computing offloading decisions.  These sub-NNs are also connected with a sub-NN deployed at an MBS, which can be leveraged to make migration decisions. In this example, the input of the service migration module includes the output of intelligent modules at the SBSs, e.g., computing offloading decisions and the parameters of sub-NNs, and the output of the service migration module is the service migration policy. In this way, the intelligent modules can cooperate to make decisions for mobile edge computing.



\subsection{Networking for AI}\label{ss:NetforAI}

In addition to managing networks, AI can function as services, namely {AI services}, which reside at ALs~1 and~2 in the proposed AI architecture in Fig.~\ref{fig:1}. \emph{Networking for AI} is to design and optimize networks to facilitate AI services. In this subsection, we first introduce the motivation of networking for AI. Next, existing works are reviewed, and research challenges are presented. Finally, the idea of AI slice is proposed and elaborated. 


\subsubsection{Motivation}\label{subsubsec: Motivation}
Networking for AI is attracting great attention in both academia and industry. In academia, networking for AI calls for extensive interdisciplinary research efforts between networking researchers and AI researchers to develop new communication standards and technologies to cater for  AI services at scale~\cite{chemouil2019special, sorour2020returning, gunduz2019machine,wu2021ai}. In industry, the International Telecommunication Union (ITU) is discussing high-level architectures to integrate, orchestrate, and update AI components for future networks, including IMT-2020 networks~\cite{itu2019architectural, itu2019Unified}. Some 3GPP working groups are studying data collection frameworks in the network for supporting AI services~\cite{3GPP2019Study, wilhelmi2020flexible}. Notably, networking for AI is becoming an indispensable component for facilitating AI services in networks and is expected to be a key enabling technology in 6G.


Networking for AI should take the following factors into consideration:
\begin{itemize}
    \item \emph{Distributed Data} -  With the wide deployment of various IoT devices and small BSs, massive data are generated from many distributed network nodes, e.g., end users and the network edge. In the traditional cloud-based AI paradigm, the cloud  collects massive distributed data for model training, and a well-trained model is deployed at the cloud for model inference. This paradigm suffers from spectrum resource scarcity and user privacy leakage concerns.\footnote{Google's autonomous driving vehicle can generate more than 750\;MB of data per second~\cite{va2016millimeter}.} To address these issues, a potential solution is to facilitate AI services over a large number of network nodes in a distributed manner~\cite{hard2018federated}, which requires new networking protocols to coordinate multiple network nodes;
    
    \item \emph{Constrained Resources} - Network nodes, such as end users, have limited resources, while state-of-the-art AI models (e.g., DNN models with dozens of neural network layers) are complex. As such, running a complex AI model on a single network node can exhaust its computing resource and energy.\footnote{The energy consumption of using AlexNet to process an image on a tailored energy-efficient Eyeriss chip is up to 0.28\;W~\cite{chen2016eyeriss}.} With advanced model partition techniques (e.g., DNN partition), a complex AI model can be partitioned into multiple sub-models and embedded into a network with data exchange among the sub-models~\cite{model_partition}. Executing sub-models consumes computing resources of network nodes, and exchanging data between sub-models also consumes communication resources. Hence, running AI models at multiple network nodes in a cost-effective manner requires innovative network embedding designs;
    
    \item \emph{Network Heterogeneity and Dynamics} - 6G networks will be highly heterogeneous, in which network nodes possess different amounts of communication, computing, and storage resources. As complex AI models need to be deployed at multiple network nodes, executing AI tasks requires judiciously allocating resources of these network nodes. Moreover, network dynamics, such as time-varying channel conditions among network nodes and spatiotemporal service demands, further complicate the resource allocation decision making problem. Hence, it is necessary to design tailored resource management algorithms to optimize AI performance, while adapting to network dynamics.
\end{itemize}



The \emph{scope} of networking for AI covers the entire lifecycle of AI services, which consists of three stages. The first stage is \emph{data collection} for model training via  communication links. For instance, real-time service load data from end users need to be collected to train an AI model for service demand prediction. The second stage is \emph{model training}, which is to achieve a certain objective based on the collected data. For instance, a large number of images are processed to train DNN-based object detection modules until  the target accuracy requirement is satisfied. The third stage is \emph{model inference}, which is to apply well-trained models to complete specific computation tasks. For instance, AI-based object recognition for autonomous driving detects and classifies nearby vehicles, pedestrians, and obstacles based on real-time images captured by on-board cameras~\cite{lin2018architectural}. 


\subsubsection{State-of-the-Art Approaches}\label{subsubsec: related_work}
The research on networking for AI is still at its infancy stage with only a few existing works. In this subsection, the existing studies are categorized into data collection, model training, and model inference according to the lifecycle of AI services. Representative related works are summarized in Table~\ref{Table:networking_for_AI}.

\begin{table*}
	\centering
	\scriptsize{
		\caption{Summary of Related Works on Networking for AI} \label{Table:networking_for_AI}
		\begin{tabularx}{0.92\textwidth}{|l|l|p{10cm}|X|}
			\hline
			\textbf{Topic}  & \begin{tabular}[c]{@{}l@{}} \textbf{Work}\end{tabular}  & \textbf{Contribution} & \textbf{Highlight}     \\ \hline
			\multirow{5}{*}{\begin{tabular}[c]{@{}l@{}} Data \\Collection \end{tabular}}  
			&\centering \makebox{ \cite{wen2019overview}}  
			&  Scheduling data transmission based on users' data importance levels and channel conditions           &  Data importance-aware spectrum allocation  \\ \cline{2-4} 
			&\centering \makebox{\cite{wang2020machine}}  &  Allocating users' transmission power that can adjust the amount of  collected data  samples for multiple AI models to enhance the overall model accuracy    & Data amount-aware power allocation  \\ \cline{2-4} 
			&\centering \makebox{\cite{liu2020wireless}}  &  Designing an importance-aware ARQ  protocol, in which users' data importance levels and channel conditions are jointly considered to trigger data retransmission   &  Data importance-aware  retransmission protocol  \\ \hline
			
			\multirow{11}{*}{\begin{tabular}[c]{@{}l@{}}Model\\ Training\end{tabular}} 
			&\centering \makebox{\cite{liu2020client}}  &  Proposing an edge-cloud assisted FL framework, in which the edge and cloud servers alternatively aggregate local models to reduce communication overhead  &  Two-tier FL framework     \\  \cline{2-4} 
			&\centering \makebox{\cite{yang2020federated}}  &  Proposing an over-the-air computation approach for model aggregation   &  Over-the-air model   aggregation  \\ \cline{2-4} 
			&\centering \makebox{\cite{wang2020optimizing}}  &  Selecting users with more contribution to  convergence for model aggregation based on users' data distribution  &  Data distribution-aware user selection    \\  \cline{2-4} 
			&\centering \makebox{\cite{nishio2019client}}  &  Selecting  users with low training delay considering heterogeneity among  users  & Training latency-aware user selection  \\  \cline{2-4} 
			&\centering \makebox{\cite{wang2019adaptive}}  &  Optimizing the number of local model updates given a  resource budget &  Local update frequency optimization   \\  \cline{2-4} 
			&\centering \makebox{\cite{ren2020scheduling}}  &  Scheduling model uploading based on end users' model importance levels and channel conditions    &   Model importance-aware  model uploading  \\\hline
			
				\multirow{8}{*}{\begin{tabular}[c]{@{}l@{}}Model\\ Inference\end{tabular}}  
			&\centering \makebox{\cite{liu2018edge}}  & Optimizing video frame rate and input image resolution to balance service latency and   detection accuracy for virtual reality users  &  Data resolution optimization  \\  \cline{2-4} 

			&\centering \makebox{\cite{wang2020joint}}  &  Selecting the optimal DNN model for real-time video analytics    &   DNN model selection  \\  \cline{2-4} 
			&\centering \makebox{\cite{zhang2021autodidactic}}  &  Selecting the optimal DNN model's cut layer  to minimize inference latency  for user-edge DNN synergy     &   User-edge DNN model partition  \\  \cline{2-4} 
			&\centering \makebox{\cite{teerapittayanon2017distributed}}  & Partitioning a complicated DNN model across end users, the network edge, and the  cloud to reduce communication overhead    &  User-edge-cloud DNN model partition \\ \cline{2-4} 
		    &\centering \makebox{\cite{wu2020accuracy}}  & Designing a collaborative DNN model inference scheme with light-weight models at IoT devices and an uncompressed model at the network edge   & Collaborative DNN model inference \\  \cline{2-4} \hline

	\end{tabularx}}
\end{table*}

\textbf{Data Collection} - The objective is to efficiently collect the data from end users for optimizing AI performance. Since data are distributed across end users in the network, transmission resource is scheduled to end users for uploading their data. For instance, the level-one digital twins require periodical data synchronization with the end users, and such data can be provided for AI services. Data collection is a classic research problem widely investigated in wireless sensor networks~\cite{li2017wireless} and UAV networks~\cite{zhan2017energy}, and these works focus on optimizing either the reliability of data collection or the amount of  collected data. {In AI services, the collected data are used to train AI models, and the data samples may have different importance levels for model training. Merely maximizing the reliability or the amount of the collected data is not optimal. Hence, novel data collection designs taking model training into account are required for performance optimization. }



Recently, AI-centric data collection is investigated in the following two research directions:
\begin{itemize}
    \item  \emph{Resource Allocation} -  Data importance-aware resource allocation schemes have been proposed to optimize AI model accuracy. The idea is to schedule data transmission while taking both end users' channel conditions and data importance levels into account~\cite{wen2019overview}. The data importance level can be captured via data uncertainty, i.e., higher uncertainty means higher importance. The data uncertainty can be measured by entropy~\cite{holub2008entropy}. {Power allocation for data collection is investigated in multi-model training scenarios~\cite{wang2020machine}. Since the number of  collected data samples impacts the model accuracy, a learning-centric power allocation scheme can adjust the users' transmission power to determine the amount of  collected data  for different AI models, thereby maximizing the overall model accuracy given a transmission power budget;}

    \item\emph{Protocols} - There are a few AI-centric data collection protocols. In a  network environment with poor channel conditions, data retransmission is applied to improve data collection reliability. Existing automatic repeat-request (ARQ) retransmission protocols, such as hybrid ARQ in long term evolution (LTE) networks, trigger data retransmissions for lost packets once the end user's signal-to-noise ratio (SNR) threshold is satisfied. The importance of data samples should be incorporated in transmission protocols to speed up the model training process. An importance-aware ARQ  protocol is proposed for CNN-based classification model training in~\cite{liu2020wireless}. In the protocol, both data importance levels and channel conditions are taken into account to determine the data retransmission threshold, which can enhance the model training performance. 
\end{itemize}




\textbf{Model Training} - Due to the distributed data and user privacy concerns, distributed training is  suitable for training AI models in a network~\cite{verbraeken2020survey}. FL is one of the most promising distributed training paradigms, which can be applied in various fields such as smart healthcare and financial services~\cite{yang2019federated, lim2020federated, bonawitz2019towards}. The FL operates as follows. Each end user iteratively trains a local model with its own data, and the local model is uploaded to an edge server. Then, the edge server aggregates the local models to obtain a global model. The  model training lasts multiple rounds until the global model achieves satisfactory accuracy.

Since the model is trained locally, FL is communication-efficient and can preserve data privacy of end users~\cite{hard2018federated, li2020federated}. However, with the increase of  data sizes in state-of-the-art AI models,\footnote{The data sizes of ResNet32~\cite{he2016deep}, Inception-v3~\cite{szegedy2016rethinking}, AlexNet~\cite{krizhevsky2012imagenet} and VGG16~\cite{simonyan2014very} models are 50\;MB, 108\;MB, 240\;MB,  and 552\;MB, respectively~\cite{chen2019round}.} uploading local AI models still places a growing strain on spectrum-constrained wireless networks. In addition, end users with powerful computing servers can conduct local model training with a low delay. As such, the model uploading delay due to limited radio resources can be the dominant component in the entire FL delay.  Hence, it is necessary to maximize FL performance in resource-constrained wireless networks.

Recent research works optimize FL performance from the following perspectives: 
\begin{itemize}
    \item \emph{FL Framework Design} - A line of works focus on designing innovative FL frameworks to reduce communication overhead. A novel two-tier hierarchical FL framework is proposed in~\cite{liu2020client}, which coordinates  end users, edge servers, and the cloud server to perform FL. Each edge server aggregates local models from end users in its coverage in every FL round, and the cloud server aggregates the models from edge servers in its coverage once in a few FL rounds. {The proposed two-tier framework can accommodate a large number of end users for model training due to its broad coverage and, at the same time, reduce the backhaul data traffic between the cloud server and edge servers due to a low model aggregation frequency.} Such framework is applied to industrial IoT networks with geographically distributed  data in~\cite{zhang2021dynamic};
   
    \item \emph{Model Aggregation} - Another line of works study radio spectrum-efficient model aggregation. Over-the-air computation based approaches are investigated in~\cite{yang2020federated, amiri2020federated, zhang2021federated}. The basic idea is to exploit the superposition property of wireless multiple-access channels to perform model aggregation, which can reduce radio resource consumption;
    
    \item \emph{Resource Management} - The FL performance can be optimized via efficient resource management. As FL performance depends on multiple factors, such as end user selection, the number of local model updates, and local model importance levels, different resource management schemes are developed as follows: (1) User selection - How to select participating end users in the FL process impacts model convergence and training delay and hence is crucial to FL performance. A few end user selection algorithms are proposed based on principles such as training data distribution~\cite{wang2020optimizing} and local training latency~\cite{nishio2019client}; (2) FL parameters  -  To alleviate communication overhead, end users conduct a few local model updates before model uploading. Given a resource budget, the optimal number of local model updates is studied in~\cite{wang2019adaptive}, which provides a theoretical guideline for selecting the number of local model update; (3) Local model importance level, which is a concept extended from the idea of data importance\footnote{The model importance can be measured by layer-wise gradient norm. Local models with larger gradient norm contribute more to global model convergence in FL~\cite{wang2019adaptive}.} -  An importance-aware model uploading strategy is proposed in~\cite{ren2020scheduling}, in which end users with high   model importance levels and good channel conditions are scheduled with high priority, to speed up the convergence of FL. 
\end{itemize}

\textbf{Model Inference} - For many AI services in the network, AI models are usually deployed at end users and edge servers to achieve low service latency. The model inference is computation-intensive, while end users and edge servers usually have limited computing capabilities and battery power. Executing model inference tasks usually results in long service latency and  high energy consumption. Hence, performing model inference should satisfy service latency under node energy constraints, thereby calling for innovative model inference schemes.

Existing studies on model inference can be categorized as follows: 
\begin{itemize}
    \item \emph{Data Resolution} - Raw data are offloaded to edge or cloud servers for model inference. The input data resolution influences the inference accuracy. For instance, the accuracy of object detection  is related to the input image resolution~\cite{liu2018edge}, which in turn affects the offloaded data volume since the data size of high-resolution images is usually large. Taking into account the trade-off between the inference accuracy and the amount of offloaded data, the input image resolution should be optimized to satisfy the target AI service requirements. The optimal video frame rate and input image resolution are investigated in~\cite{liu2018edge} to balance service latency and detection accuracy for virtual reality users;
    
    \item \emph{Model Selection} -  An appropriate AI model is selected to satisfy specific AI service requirements. In addition to the data resolution, the inference accuracy depends on the type of AI models. A DNN model with more hidden layers can usually  achieve a higher inference accuracy than a shallow DNN model. Considering multiple available DNN models deployed at the network edge, the optimal DNN model selection  for real-time video analytics is investigated in~\cite{wang2020joint};
    
    \item \emph{Model Partition} - With advanced model partition techniques, an AI model can be partitioned into multiple sub-models and then embedded into different network nodes to conduct model inference. For instance, leveraging the layered structure of DNNs, the entire DNN model can be partitioned into an end user-side model and a server-side model at a proper DNN layer (i.e., the cut layer). As such, the end users and the edge servers can conduct model inference in a collaborative manner. DNN models can be partitioned for achieving different goals. For instance, the optimal model partition for minimizing inference latency is studied in~\cite{zhang2021autodidactic}, in which an online learning algorithm can adaptively determine the optimal cut layer. To reduce communication overhead among network nodes, complicated DNNs models can be partitioned into sub-models for end users, edge servers, and the cloud as in~\cite{teerapittayanon2017distributed};
    
     \item \emph{Model Compression} - Light-weight models are used to facilitate prompt model inference at end users. Computation-efficient compressed models can be obtained via various model compression techniques, such as weight pruning~\cite{han2015deep}, knowledge distilling~\cite{chen2017learning} and fast exiting~\cite{teerapittayanon2016branchynet}. For instance, weight pruning techniques remove less important model weights to reduce the  computational complexity of model inference, while achieving inference accuracy close to that of the uncompressed models. To enhance service performance, a collaborative model inference scheme that deploys light-weight models at IoT devices and uncompressed models at the network edge is proposed in industrial IoT networks~\cite{wu2020accuracy}. The IoT devices dynamically make AI task offloading decisions according to time-varying channel conditions to minimize the service delay while guaranteeing the accuracy requirements of DNN-based fault diagnosis services.
\end{itemize}





\begin{figure*}[h]
	\centering
	\renewcommand{\figurename}{Fig.}
	\includegraphics[width=0.9\textwidth]{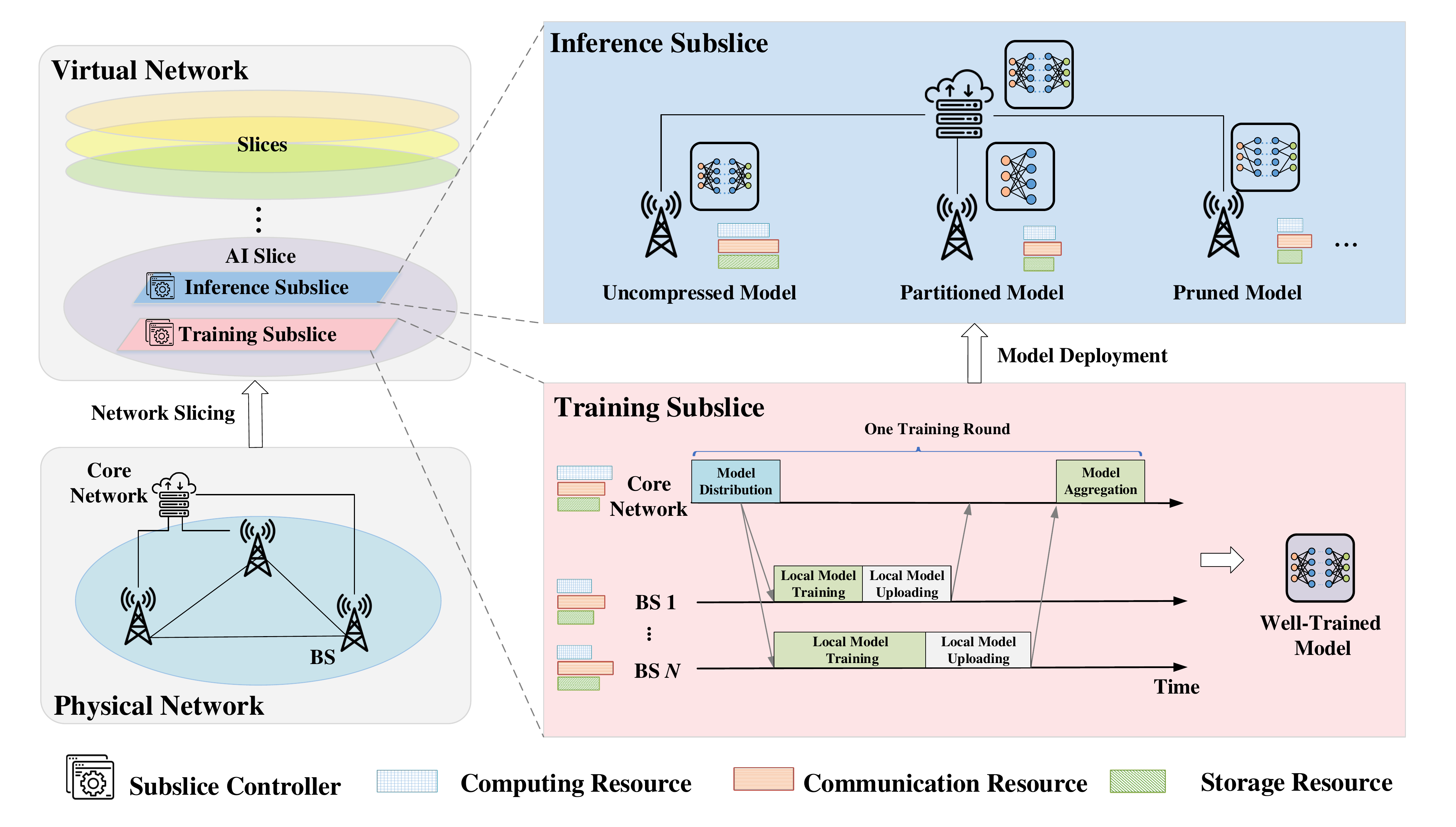}
	\caption{Conceptual AI slice consisting of a training subslice and an inference subslice.}
	\label{fig:slicing_for_AI}
\end{figure*}

\subsubsection{Research Challenges}\label{subsubsec: challenges}
 
Despite the aforementioned research efforts, facilitating AI services in a network faces various challenges, some of which are discussed in the following.


\emph{Complex Implementation Option Selection} - An AI service can be implemented by various options with different model structures, training procedures, and inference processes. For instance, a service of object detection can be implemented via different neural networks, such as AlexNet~\cite{krizhevsky2012imagenet} and SqueezeNet~\cite{iandola2016squeezenet}. Even if the model structure is the same, a model can be trained in different ways, such as centralized training, decentralized training (e.g., FL~\cite{yang2019federated}), and semi-centralized training (e.g., split training~\cite{thapa2020advancements}). In addition, model inference can be conducted in various manners, such as end user-only inference, edge-only inference, and collaborative inference. Different implementation options consume different amounts of computing, storage and communication resources. Hence, it is necessary to select an implementation solution for AI services that suits the service characteristics and network dynamics.    	 


\emph{Multi-Dimensional QoS Requirements} - The QoS requirements of AI services are multi-dimensional.  AI model accuracy is usually a key performance metric. In addition, AI services should be offered to end users with low latency in many use cases. For instance, the service latency of object detection in autonomous driving should be less than 100\;\emph{ms} for safety considerations~\cite{lin2018architectural}, whereas autonomous vehicles require an ultra-high accuracy in 3D object detection~\cite{chen2017multi}.  Moreover, these performance metrics are correlated. High-accuracy object detection usually requires high-resolution images as input and advanced AI models to process the input images, which can result in long service latency. How to satisfy multi-dimensional QoS requirements of AI services requires further investigation.

\subsubsection{AI Slice}\label{subsec:AI_slice}\label{subsubsec: AI_slice}
To better support AI services, we extend the network slice concept and propose an idea of \emph{AI slice} with two subslices. The basic idea is to construct a \emph{training subslice} for model training and an \emph{inference subslice} for model inference. The two subslices are logically isolated and use their own network resources. The rationale behind training and inference separation is that the two stages can have different goals.


{An illustration of an AI slice is given in Fig.~\ref{fig:slicing_for_AI}. In the AI slice, the training and inference subslices share the same resource pool and are coordinated to jointly support the AI service. First, the multi-dimensional QoS requirement of the AI slice is decoupled into two separate QoS requirements for the two subslices. For an object detection service in autonomous driving, both high detection accuracy (e.g., 99\%) and low service latency (e.g., 100\;\emph{ms}) are required. The training subslice should satisfy the detection accuracy requirement, while the inference subslice should satisfy the service latency requirement. Second, to satisfy the individual QoS requirements of the two subslices, the resources reserved for the AI slice are judiciously allocated between the two subslices, based on the performance of the two subslices and their QoS requirements. Then, given the allocated resources, the two subslices are configured to satisfy their individual QoS requirements, as described in the following: }
\begin{itemize}
    \item In the training subslice, based on the training data distribution in the network, a subslice controller determines training configurations (e.g., data collection schemes and model training methods) and schedules resources to network nodes to train a model given the target accuracy. In addition, since the training data vary over time in a dynamic network, the AI model may need to be retrained from time to time. Note that allocating dedicated resources for the training subslice can effectively mitigate the straggler effect that plagues distributed learning in large-scale networks, thereby speeding up the model training process;
    
    \item In the inference subslice, the subslice controller analyzes the service  demand pattern at each BS and  determines inference configurations (e.g., model inference and input data compression schemes) to satisfy the inference latency requirement. For instance, uncompressed models can be deployed at resource-abundant BSs, and partitioned and pruned models can be deployed at resource-limited BSs. This can achieve close inference service latency performance across different BSs.
\end{itemize}
{Overall, the two logically-isolated subslices focus on satisfying different QoS requirements and jointly support the AI service.}


To elaborate the idea of AI slices, we present the following example on real-time video analytics in vehicular networks~\cite{wang2020joint}. Smart cameras are deployed in  intersections to provide a video surveillance service such as vehicle plate recognition. In such service, a CNN model is trained using the video streams collected by smart cameras, and then the well-trained model is used to conduct video analytics tasks. Using the proposed AI slice framework, CNN model training is conducted in a training subslice, while real-time video analytics is conducted in an inference subslice. Specifically, in the training subslice, the CNN model can be trained via a FL framework for protecting data privacy. The corresponding computing resources at smart cameras and spectrum resources in the network are allocated to satisfy model training requirements, such as training accuracy. In the inference subslice,  different user-edge orchestration schemes (e.g., DNN model partition), input data compression schemes (e.g., frame rate reduction), and network resource management policies can be configured to satisfy the inference delay requirement in video analytics services based on time-varying service demands and network conditions due to vehicle mobility. With the AI slice for video analytics, both training accuracy and inference latency requirements can be satisfied in a dynamic network environment.

\subsection{{Summary}}
{In this section, we have reviewed some common AI techniques, explored the role of AI in 6G networks, and proposed a four-layer AI architecture for pervasive intelligence in 6G. Two perspectives of AI in wireless networks, i.e., AI for networking and networking for AI,  have been discussed, which correspond to using AI as a powerful tool for network management and optimizing networks to support AI applications, respectively. }

{Recent advancements in ML algorithms have accelerated the deployment of AI in wireless networks. In 5G, AI techniques are used to address particular networking problems, whereas, in 6G, AI will penetrate every corner of wireless networks from network management to network services. Therefore,  an architecture for AI is needed for identifying the role of AI and characterizing the functionalities of AI across a network.}

Appropriate AI techniques should be selected to tackle networking problems with different characteristics and on different decision time scales when it comes to AI for networking. Furthermore, the collaboration among intelligent modules is important to implement AI-driven networks efficiently and flexibly. The idea of connected AI is to enable cooperative decision making among intelligent modules for network control. In terms of networking for AI, a distributed architecture of AI algorithms connects AI models and network resources located at network edge. The study of networking for AI is still in its infancy but essential to supporting an expanding group of AI services. Network slicing will remain to be an enabler for delivering AI services, but slicing policies should be customized according to the features of AI algorithms and the training and inference stages of AI.



\section{A Potential Network Architecture for 6G}
\label{sec:PNA}
 \begin{figure*}[t]
	\centering
	\renewcommand{\figurename}{Fig.}
	\includegraphics[width=0.95\textwidth]{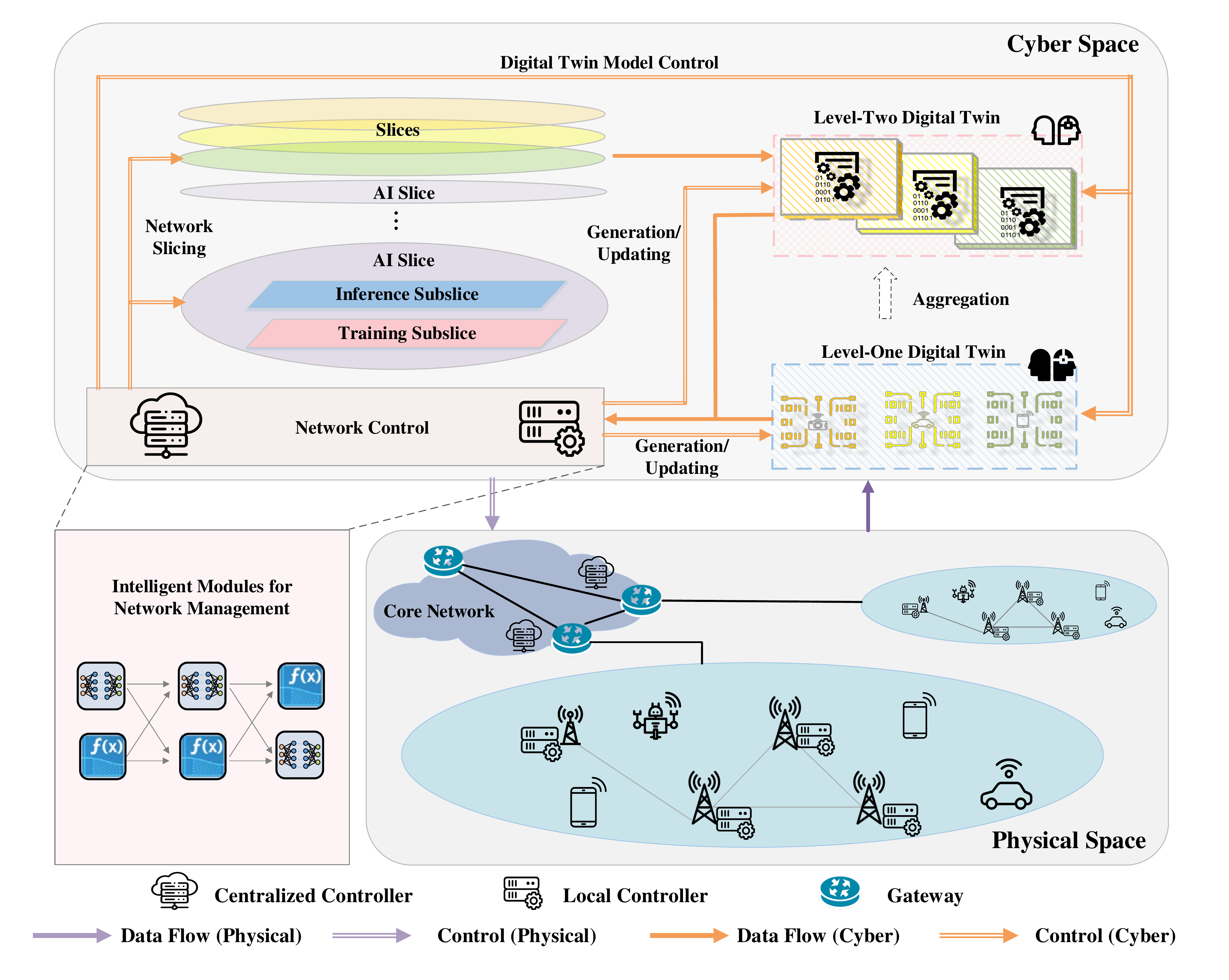}
	\caption{The proposed network architecture for 6G networks.}
	\label{fig:proposed_architecture}
\end{figure*}

In this section, we propose a conceptual network architecture for 6G, which integrates holistic network virtualization (including digital twins and network slicing) and pervasive network intelligence (including connected AI and AI slices). Then, we illustrate three types of interplay enabled by the proposed architecture, i.e., the interplay between digital twin paradigm and network slicing, model-driven methods and data-driven methods, and virtualization and AI, respectively.

\subsection{Related Studies on Architecture for 6G} \label{ss:related6Garc}
Several works have proposed architectures with various focuses for 6G networks, e.g., space-air-ground integrated networks for global coverage~\cite{M_NZhang}, cell-free massive multiple-input multiple-output (MIMO) architecture for inter-cell interference mitigation~\cite{ngo2015cell}, and multi-tier computing architecture for ubiquitous computing service provisioning~\cite{yang2019multi}. Pursuing the goal of advanced network management, most of the proposed architectures highlight AI techniques to optimize network architecture, control, and management~\cite{M_KLetaief2019, itu2019architectural}. For example, AI-based data analytics functions, which mine historical data for network operation troubleshooting, network resource optimization, and network traffic prediction, are incorporated in the network architecture in ~\cite{M_KLetaief2019}. The ITU specifies a high-level AI-based architectural framework for future networks, in which several novel components such as ML management and orchestration functionalities are incorporated for flexible AI-based function placement~\cite{itu2019architectural}. In addition to AI techniques, some recent conceptual network architectures start to embrace digital twin techniques~\cite{M_Q_Yu_2019, khan2021digital}. For example, a digital twin-based network architecture constructs a digital twin for each end user to serve as its communication assistant and data asset manager~\cite{M_Q_Yu_2019}. Another digital twin-enabled network architecture adopts three categories of digital twins, i.e., edge-based, cloud-based, and hybrid  digital twins, for supporting different types of services~\cite{khan2021digital}.

Different from the existing network architectures, our proposed network architecture features novel holistic network virtualization, which incorporates network slicing and digital twin paradigms, and pervasive network intelligence, which integrates AI for networking and networking for AI. Moreover, featuring the designs in Sections~\ref{sec:HNV}~and~\ref{sec:PI}, the proposed architecture enables various interplay among its key elements to empower 6G. In the following subsections, we present the details of the proposed architecture.

\subsection{Architecture Overview}\label{ss:ArcOv}


The overall network architecture is illustrated in Fig.~\ref{fig:proposed_architecture}, which consists of the physical space and the cyber space. The physical space includes end users and network infrastructure at the edge and the core networks. Data describing end users are collected from the physical network to create level-one digital twins as introduced in detail in Subsection~\ref{ss:HNV}, and network slices are created for various services. The slices are further abstracted into level-two digital twins, which are supplemented with service-specific information aggregated from level-one digital twins. The six-layer virtualization architecture in Fig.~\ref{fig:digital_twin} applies to the network slices and the digital twins, both of which reside in the cyber space in Fig.~\ref{fig:proposed_architecture}.

AI pervades the entire architecture, which supports both AI for networking and networking for AI. First, AI is used to manage network slices and digital twins, as shown in the logic network control section in Fig.~\ref{fig:proposed_architecture}. For network management, a connected AI solution discussed in Subsection~\ref{ss:AIforNet} is applied to enable intelligent modules, which in turn manage network slices and digital twins. The connected AI solution corresponds to AL 3 and 4 in Fig.~\ref{fig:1}. Second, the  architecture supports dedicated AI slices with training and inference separation for AI service provisioning, as mentioned in Subsection~\ref{ss:NetforAI}. AI slices provide services corresponding to AL 1 and 2 in Fig.~\ref{fig:1}, while the management of AI slices is conducted by intelligent modules.  

With the overall network architecture in Fig.~\ref{fig:proposed_architecture}, we integrate holistic network virtualization and pervasive network intelligence for 6G. Virtualization is supported from the aspects of both the network and the end users, while intelligence is reflected through both AI for networking and networking for AI. Taking advantage of digital twin paradigm and network slicing as well as those of virtualization and AI, the proposed architecture aims at exceeding flexibility, scalability, adaptivity, and intelligence. 

\subsection{Components and Subsystems}\label{ss:comp}

In the physical space, the proposed architecture includes both RANs and core networks. Specifically, the following components are involved:
\begin{itemize}
    \item Assorted APs: This component includes MBSs, SBSs, mobile APs (such as UAVs), satellites, and other non-cellular APs; 
    \item Network controllers: This component includes local controllers located at APs or servers on  network edge and the centralized controller located at servers in core networks or in the cloud. Each controller can consist of computing servers and affiliated network storage servers;
    \item General computing servers: This component includes computing servers for implementing network functions, such as routing and firewall, and hosting the VNFs;  
    \item Application servers: This component includes computing and network storage servers for supporting general edge computing and AI services. These servers are not used for  network management or implementing network functions;   
    \item Other network devices: This component includes specialized network hardware that are not general computing servers, such as baseband processing units and network switches;
    \item End users: This component includes human mobile users, sensors, vehicles, and various IoT devices, such as meters, actuators, and robots. 
\end{itemize}
In the cyber space, the proposed architecture includes three subsystems, i.e., network slices, digital twins, and connected AI, as follows:
\begin{itemize}
    \item Network slices: This subsystem includes all virtual networks created in network slicing, including AI slices. A network slice can involve a RAN, a core network, or both. General slices are inherited from existing networks, while AI slices are described in detail in Subsection~\ref{ss:NetforAI};   
    \item Digital twins: This subsystem includes level-one and level-two digital twins. The digital twin subsystem is described in detail in Subsection~\ref{ss:HNV};
    \item Connected AI: This subsystem includes intelligent modules deployed across a network at both the local controllers and the centralized controller. The connected AI subsystem is described in detail in Subsection~\ref{ss:AIforNet}.  
\end{itemize}
Interconnections between different components and subsystems of the proposed architecture are elaborated in Subsections~\ref{ss:IntDTNS}~to~\ref{ss:IntVTAI}, which highlight the interplay between digital twin paradigm and network slicing, between model-driven and data-driven methods, and between virtualization and AI, in the proposed architecture. Some open issues and challenges regarding the architecture are presented in Section~\ref{sec:Open}. 

Note that the proposed conceptual architecture can apply to various types of physical networks, such as vehicular networks and integrated terrestrial-satellite networks, although Fig.~\ref{fig:proposed_architecture} cannot illustrate every possible network scenario. In different physical networks, the implementation of holistic network virtualization and pervasive network intelligence can be different and require certain customization. For example, the deployment of intelligent modules and the data flow among the modules in a satellite network segment can be different from those in a terrestrial network segment. Furthermore, the migration of digital twins can be more important in a vehicular network than in a static IoT network. Related discussions can be found in Section~\ref{sec:Open}, where we present challenges and open issues. Nevertheless, the basic ideas in the proposed conceptual architecture, including the two-level digital twins, intelligent modules, and AI slices, are applicable in various physical networks.

\subsection{Implementation\label{ss:Implement}}
In this subsection, we provide a case study on a vehicular network to demonstrate the potential implementation of the proposed network architecture. Roadside BSs co-located with edge computing and caching servers facilitate autonomous driving services for vehicles on the road. To implement the proposed network architecture, the following steps are conducted.

\begin{itemize}
    \item \emph{Network Slice Establishment}: Multiple network slices are established  for autonomous driving services with different QoS requirements, achieving network virtualization. Conventional network slices are established for non-AI based services, e.g., high-definition map downloading, while AI slices consisting of training and inference subslices are established for AI based services, such as deep learning based cooperative sensing. The network slices are stored and managed by a centralized controller.
    
    \item \emph{Digital Twin Construction}: By collecting extensive data from  physical entities, digital twins are constructed for vehicle users, roadside BSs, and the established network slices, achieving the virtualization of end users and slices. Digital twins of vehicle users and roadside BSs are located at edge servers, while digital twins of network slices are located at a cloud server. Due to high vehicle mobility, digital twins of vehicle users should be migrated across edge servers to ensure service continuity. In addition to collected data, digital twins can include generated user and service specific data, such as predicted vehicle trajectory and spatial-temporal service demands, via mining historical data. The generated vehicle data will be used for network management and service provision.
    
    \item \emph{AI Module Deployment}: AI modules with different functionalities can be deployed at both the centralized and local network controllers, achieving intelligent network management. The AI modules at the centralized network controller are in charge of network planning. For guaranteeing QoS requirements of different slices, these AI modules can make resource reservation decisions based on the predicted service demands from the digital twins of roadside BSs and collected slice performance data from the digital twins of network slices. The AI modules at local network controllers are in charge of network operations. For enhancing the perceived performance of the vehicle users, the AI modules schedule on-demand network resources based on the collected data (e.g.,  vehicle users' channel conditions) and the generated data (e.g., predicted vehicle trajectory) from the digital twins of vehicle users. 
    
    
\end{itemize}

\subsection{Interplay between Digital Twin Paradigm and Network Slicing}\label{ss:IntDTNS}

As the two components of holistic network virtualization, digital twin paradigm and network slicing are connected in the following two aspects.

First, the digital twin paradigm for end user virtualization focuses on data management, while network slicing focuses on network management. Data may be viewed as a new type of resources in future networks, in addition to communication, computing, caching, and sensing resources. Meanwhile, as a resource, data has its unique features. First, data can be considered as an application-layer resource rather than a physical-layer resource. Second, different from computing or communication resources, the amount of data resources available to a network is not fixed but progressive. Last, the collection and processing of data, which is necessary for utilizing any data resource, consume other network resources. On one hand, effectual utilization of the data resource will benefit network management, and hence digital twin paradigm can enhance network slicing. On the other hand, network management should take into account the need and cost of allocating other network resources for utilizing the data resource. Hence, network slicing can facilitate digital twins.      

Second, digital twins will enable user-centric networking in future networks, while network slicing enables service-centric networking. Creating an isolated slice for each service and provisioning the service through managing the slice yield a service-centric focus in network management. Meanwhile, creating a digital copy of each end user and administrating data that characterize the end user provide a user-centric perspective of network management. Having a set of information, selected by the centralized controller through digital twin model control, to describe various characteristics of the end users, such as their location, service request profile, resource utilization, and channel information, creates the possibility of user-specific scheduling within each slice in the network operation stage. For instance, access control and resource allocation decisions for an end user may depend on the data profile from its digital twin, while different data profiles may lead to different scheduling policies.  Accordingly, future networks may feature service-centric network planning and user-centric network operations, which can improve the granularity of network management for handling highly diversified end users and dynamic network environments.

\subsection{Interplay between Model-Driven and Data-Driven Methods}\label{ss:codriven} \label{subsec:Model-data Co-driven}

The second interplay enabled by the proposed architecture is the interplay between model-driven and data-driven methods in network operation and service provision. This interplay applies to the intelligent modules for network management shown in Fig.~\ref{fig:proposed_architecture}.

Network management mostly relied on model-driven or heuristic methods before 5G. Prior to the prevalence of AI, mathematical tools such as optimization methods and game theory have been widely used for network management. Optimization methods formulate the objective and constraints in a closed form, and the corresponding network management problems are solved using optimization algorithms~\cite{Mushu_UAV, Leconte,Halabian}. Game-theoretic approaches analyze the interactions among network entities in either cooperative or non-cooperative scenarios to identify  the optimal strategy of each entity~\cite{Xiong1, Xiong2, Caballero}. Mechanism design, an analytical framework in game theory, has also been used to coordinate network entities with locally-held information to achieve desirable network-wide solutions in network utility maximization problems~\cite{J_Jie_MD_2018}.


Through characterizing the relations among several key variables, model-driven methods can lead to either closed-form solutions or algorithms for network management problems. Based on mathematical models, model-driven methods are usually explainable and generalize well for different specific problems~\cite{zappone2019wireless}.\footnote{For instance, the water-filling algorithm could be applied to various power allocation problems, and the Rayleigh fading model could characterize channels in various network environments.} However, when networks become complex (i.e., when there are a large number of variables and/or complicated correlation among them) or highly dynamic (e.g., when the network environment changes too rapidly for an optimization algorithm to converge or for a game to achieve an equilibrium), model-driven methods may no longer be accurate or applicable. 

The investigation of data-driven methods for network management has gained momentum since 5G. Through collecting and exploiting real-world data, data-driven methods implicitly characterize the relations among variables to generate and fine-tune policies for network management. Given sufficient data and a stationary network environment, data-driven methods can provide close-to-optimal solutions to problems that are too complicated for model-driven methods. However, when the network environment is non-stationary so that new and unknown situations occur from time to time, the performance of data-driven methods can be questionable~\cite{yang2020deep}. In addition, data-driven methods may not generalize well due to their strong dependence on data collected from a specific network environment.             

\begin{figure}[t]
	\centering
	\subfloat[Backup/switching]{
		\includegraphics[width=0.36\textwidth]{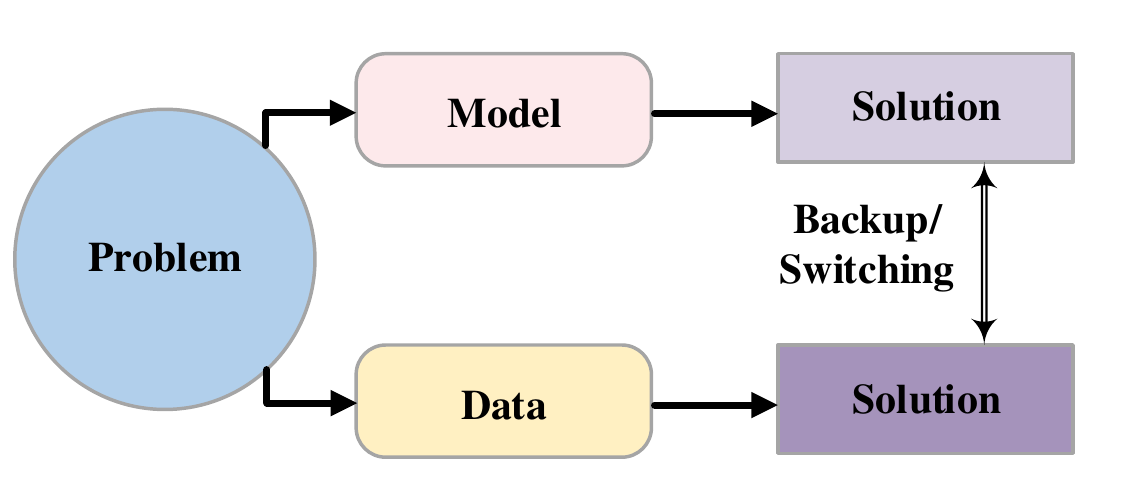}\label{fig:HDMD1}}\\
		\vspace{-3mm}
	\subfloat[Task division]{
		\includegraphics[width=0.36\textwidth]{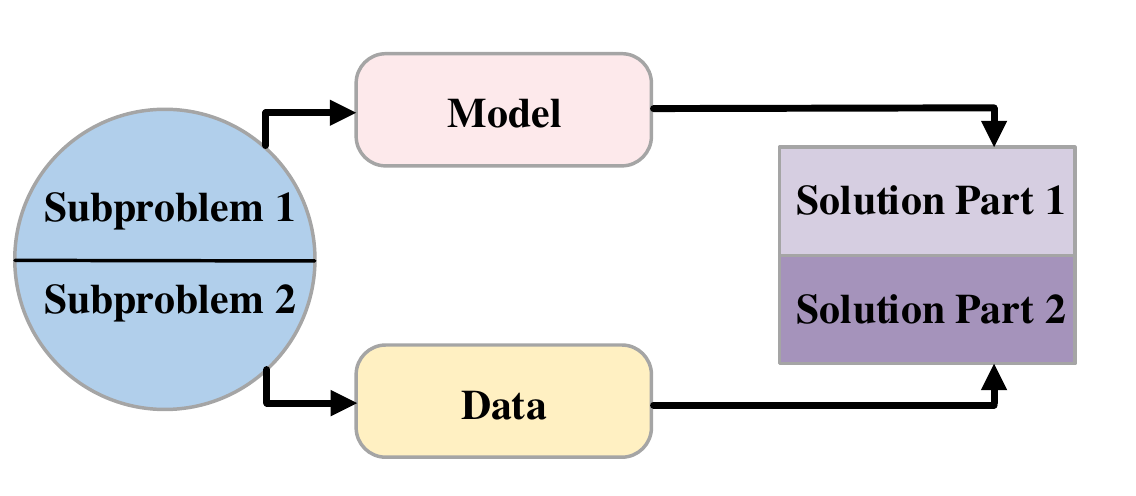}\label{fig:HDMD2}}\\
		\vspace{-3mm}
	\subfloat[Refinement]{
		\includegraphics[width=0.36\textwidth]{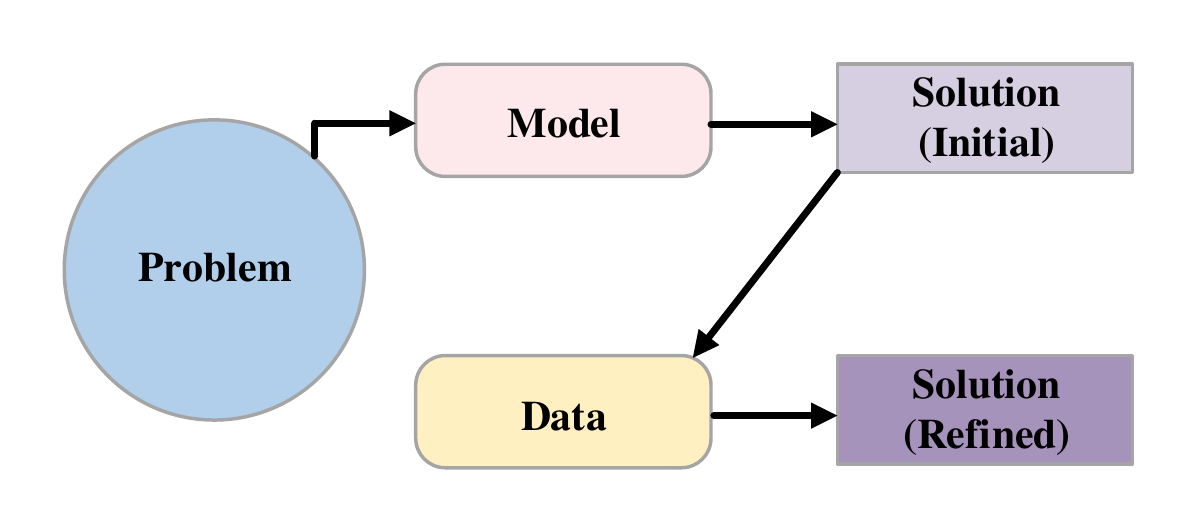}\label{fig:HDMD3}}\\
		\vspace{-5mm}
	\subfloat[Mixing]{
		\includegraphics[width=0.36\textwidth]{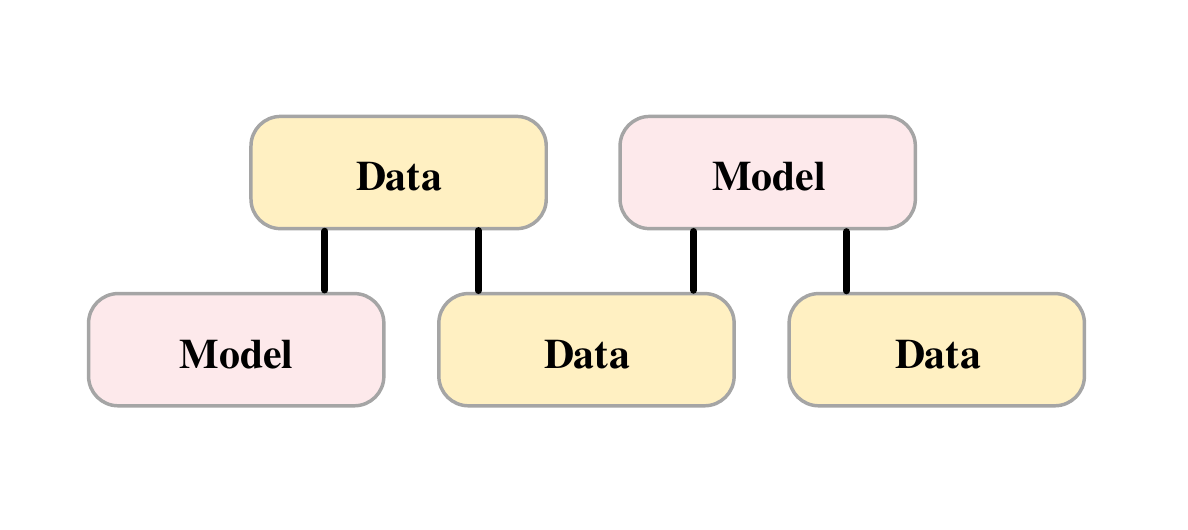}\label{fig:HDMD4}}
	\caption{Options for hybrid data-model driven methods. The ``Data'' and ``Model'' blocks represent ``data-driven methods'' and ``model-driven methods'', respectively. }
	\label{fig:HDMD}
\end{figure}

In 6G, data-driven and model-driven methods should work in synergy. The proposed architecture enables the interplay between data-driven and model-driven methods for creating advanced \emph{hybrid data-model driven} methods. There are different options of hybrid data-model driven methods, as illustrated in Fig.~\ref{fig:HDMD} and elaborated below. The first three options suit AI for networking, while the last option suits networking for AI.   

\begin{itemize}
    \item \emph{Backup/Switching} - Data-driven and model-driven methods can be the backup for each other. For instance, models can be selected to back up data-driven methods, for the case when unknown situations occur in the network environment and degradation in the performance of data-driven methods appears. Meanwhile, switching between data-driven and model-driven methods, e.g., based on the available resources, can potentially increase the adaptivity of network management. 
    \item \emph{Task Division} - Date-driven and model-driven methods can target different steps and solve different subproblems of network management. Specifically, data-driven methods can solve the subproblems with a large number of variables or complicated coupling relations among variables, while model-driven methods can solve relatively isolated subproblems with a few key variables. This would allow data-driven and model-driven methods to play to their respective strengths.   
    \item \emph{Refinement} - Model-driven methods can provide rough solutions based on general mathematical models, and then data-driven methods, taking the rough solutions as input and exploiting real-world data from the network, can refine the solutions for the specific network scenario. Having the initial solution generated from models may reduce either the amount of data or the amount of time needed by data-driven methods.      
    \item \emph{Mixing} - In networking for AI, while deploying a service function chain for an AI service, some of the function modules can use data-driven methods, while other function modules in the same service function chain can use model-driven methods. For example, in an AI-based image processing service, a model-driven module can be used for image resolution adjustment prior to a data-driven module for object detection. The idea is similar to task division, except that the scenario here is networking for AI instead of AI for networking  \cite{li2021slicing}.
\end{itemize}

\subsection{Interplay between Virtualization and AI}\label{ss:IntVTAI}
\begin{figure}[t]
	\centering
	\renewcommand{\figurename}{Fig.}
	\includegraphics[width=88mm]{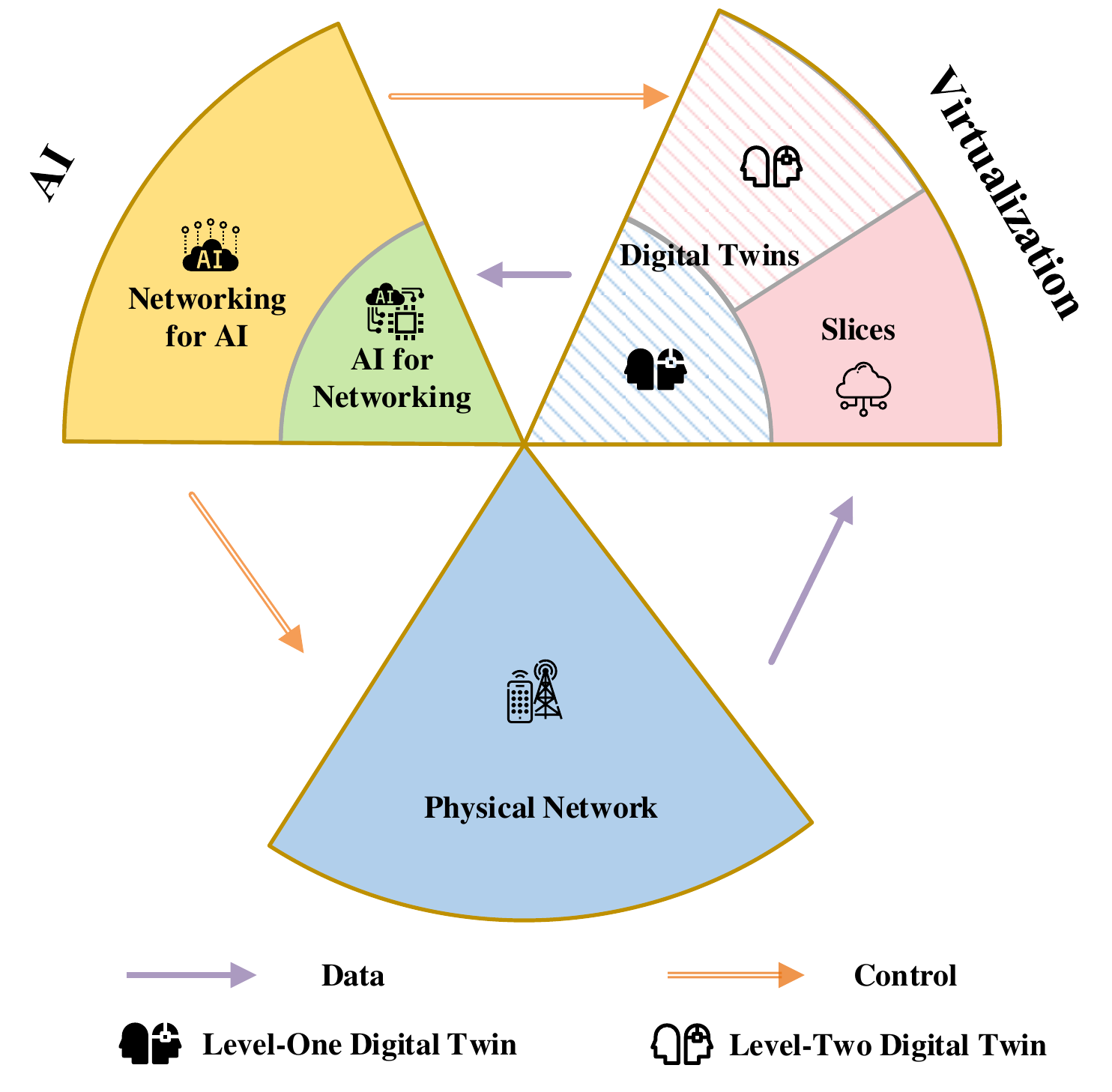}
	\caption{Interplay between virtualization and AI.}
	\label{fig:digital_twin_AI}
\end{figure}

The third interplay enabled by the proposed architecture, i.e., the interplay between virtualization and AI, is illustrated in Fig.~\ref{fig:digital_twin_AI}. 

First and foremost, virtualization and AI are coupled through data. With the introduction of digital twins, a vast amount of organized data regarding end users, i.e., level-one digital twins, and network services, i.e., level-two digital twins, become available. The data included in the digital twins can be provided to the intelligent modules, the training or inference subslice of an AI slice, or both. For instance, edge-hosted AI, possibly collaborating with end user-hosted AI, can perform user-specific data processing and prediction based on the data from digital twins. The results, such as prediction results, resource scheduling schemes, or slicing policies, can be fed back to the digital twins to record certain predicted status, e.g., location and mobility, of the end users. Correspondingly, data in the digital twins of end users, network infrastructure, and slices can be either the input or the output of AI modules, leading to a bidirectional interaction between virtualization and AI.\footnote{Interested readers are referred to~\cite{S_DC.Nguyen_ST_2021} for the relation between AI and data life cycle, although the discussions therein do not involve virtualization.}

The second connection between virtualization and AI is through control. Based on the data from digital twins, AI  functions hosted at the edge and core networks can make the network management and service provisioning decisions. The decisions may include network slice control, which are fed back to the physical network and network slices for execution and, at the same time, to the level-two digital twins for data update. In addition, the decisions may include digital twin model control for level-one and level-two digital twins. Digital twin model control may include the determination of the type and the amount of data to be included in digital twins, the frequency and the method of data collection, the format and the precision of stored data, and so on. The digital twin models affect the availability and quality of data available for network control, especially AI-driven network control, and thereby impact the network performance. Therefore, from the perspective of network control, the interaction between virtualization and AI is also bi-directional.\footnote{The interaction between digital twin and AI for intelligent network control is discussed in~\cite{O_Zhou_IETF21}. Note that the definition of digital twins therein is different from ours.}        

The third and implicit connection between virtualization and AI is through resources. Holistic network virtualization requires extensive resources, including computing resource for virtual network functions, caching resource for storing digital twins, and communication resource for the synchronization between end users and their digital twins. Similarly, pervasive network intelligence also requires extensive computing resource and possibly other resources, e.g., communication resource for distributed training as mentioned in Section~\ref{sec:PI}. Therefore, the network resources need to be shared and coordinated between virtualization and AI functions. However, this does not mean that virtualization and AI functions simply compete for resources. Instead, they can help each other improve resource utilization efficiency. 
Digital twin paradigm may reduce the resource consumption of AI functions by providing high-importance data only. This can be achieved by the aforementioned digital twin model control. Meanwhile, creating a digital twin for every end user may be too resource-demanding for networks in the near future. Using AI to select representative end users for generating digital twins and optimizing digital twin models may reduce the resource consumption of maintaining and updating digital twins. One potential implementation is using AI to categorize end users and select a portion of users from each category for creating digital twins. Alternatively, since it may be more challenging to provide QoE guarantee for some end users than others, using AI to select such end users for creating digital twins can potentially reduce the resource consumption on digital twins for 6G networks.

\subsection{Potential Network Architecture for 6G: A Summary}
This section has provided a potential network architecture for 6G, which integrates two key  elements, i.e., pervasive network intelligence and holistic network virtualization. In the proposed network architecture, detailed network components, subsystems, and potential implementation have been discussed. Moreover, three types of interplay in the architecture are provided to characterize the proposed network architecture.

The proposed network architecture holds great potential for achieving advanced network management schemes and supporting AI services in 6G networks. Firstly, integrating  digital twins and network slicing facilitates user-centric networking and improves the granularity of network management. Secondly, integrating data-driven methods and model-driven methods enables novel hybrid data-model driven methods, which has the potential to outperform existing network management methods in terms of adaptivity, granularity, and so on. Thirdly, leveraging the network slicing concept in AI services facilitates AI services targeting QoS performance guarantees.

\section{Challenges and Open Issues}
\label{sec:Open}

Many challenges and open issues are yet to be addressed for holistic network virtualization and pervasive network intelligence in 6G. In the following, we present some key challenges and open issues.


\subsection{Digital Twin}

The six-layer architecture in Subsection~\ref{ss:HNV} provides a high-level design for integrating the digital twin paradigm into network virtualization. Open issues to be investigated for practical implementation of this architecture include quantitative performance characterization of digital twins, the optimal digital twin model, digital twin migration, and data security. 

First, it is necessary to quantitatively characterize the network performance improvement from introducing digital twins, either from the perspective of QoS/QoE satisfaction or from the perspective of resource utilization. 
Second, level-one digital twin models configured by the centralized controller may be different for different edge networks to account for network heterogeneity, and how to determine effective digital twin models is a challenge. Third, the mobility of end users such as vehicles creates a need for updating and migrating digital twins across different edge networks, which requires further study. 
Last, ensuring the security of user data in the digital twin paradigm is yet another challenge. As the local and centralized network controllers have access to a vast amount of user data, developing proper security mechanisms for data collection, aggregation, and migration becomes essential. Readers are referred to \cite{J_RMinerva_ProcIEEE_2020,J_YLu_IoT_2021,J_WJiang_OJCS_2021,J_PBellavista_TII_2021} for discussions on some of the aforementioned challenges, such as the heterogeneity and migration of digital twins, and more open issues related to digital twins in 6G.
       
\subsection{Network Management Oriented Data Abstraction and Processing}

While digital twins provide data to enable AI for networking, including automated network slicing and AI-empowered network control, efficient data management can be challenging. First, it is necessary to develop data abstraction methods to aggregate the data with different levels of granularity for making different network management decisions. For instance, in network slicing, high-granularity data are required for determining the optimal network operation strategies and low-granularity data are sufficient for determining the optimal network planning strategies~\cite{J_XShen_2020,Mei}. How to determine the appropriate data granularity for different network management decisions is an open issue. 
A potential solution is to empirically adjust data granularity and the time scale for decision-making \cite{Marquez}. Meanwhile, as the number of variables and data types in network management can be huge, more scalable and efficient solutions are required. 
Second, while applying the connected AI solution for network management, the settings of intelligent modules, such as the selection of  algorithms, the input and output attributes, and the connections among intelligent modules, should be configured to maximize the utilization of data with low communication and processing overhead, yet finding the optimal settings is challenging. The cooperation between model-driven and data-driven methods in intelligent modules can be a potential approach to address the challenge, yet how to support such cooperation among different types of intelligent modules requires further investigation. Third, as data can be generated, transmitted, and processed at different network stakeholders, configurable and regulation-compliant data management is also a challenge. The integration of the blockchain and privacy-enhancing technologies can be a potential solution, while the trade-offs between privacy preservation and processing efficiency need in-depth investigation. Readers are referred to \cite{MNet_XShen_2020, zhou2019edge, wu2021ai, S_DC.Nguyen_ST_2021} for discussions on the aforementioned challenges, such as privacy preservation, AI model selection, intelligent modules, and more open issues about data abstraction and processing.

\subsection{Model and Resource Orchestration}
Networking for AI in Subsection~\ref{ss:NetforAI} can facilitate AI services in a network. One key issue is to optimize AI service performance, which requires judicious configuration of the network, including AI algorithm selection, data collection, and network resource allocation. The main challenge lies in modeling the relationship between AI performance and these network configurations. Establishing an accurate mathematical or empirical model requires extensive measurements in real-world networks. Even if establishing a model is viable, the  model may be suitable only for a chosen AI algorithm. In addition, to adapt to network dynamics (e.g., rapidly fluctuating service demands), an online network configuration scheme is desirable. Since reinforcement learning algorithms are able to make online decisions in a dynamic environment, developing cost-effective reinforcement learning algorithms for high-dimensional network configuration problems can be a promising approach. For example, a reinforcement learning algorithm is developed for joint AI model selection and resource allocation in industrial IoT~\cite{wu2020accuracy}. For more discussions on the above challenges, interested readers are referred to~\cite{partition1, partition2, he2020optimizing}. 


%
\subsection{Training and Inference Coordination} 
The concept of AI slice is proposed to meet specific QoS requirements of AI services in Subsection~\ref{ss:NetforAI}. The training and inference stages for an AI service consume multi-dimensional network resources~\cite{lin2019artificial, zhou2019edge}. In an AI slice, two subslices share the virtualized network resource pool, and hence resource reservation decisions for the two subslices are closely correlated. On the one hand, reserving abundant resources for the training subslice may help achieve a high training accuracy but potentially render resource insufficiency in the inference subslice, which can result in a long service latency. On the other hand, insufficient resource provisioning for the training subslice may yield a model with low accuracy and consequently create a bottleneck for inference accuracy. To optimize the performance of the AI service, resource reservation for training and inference subslices should be coordinated. Developing an accurate mathematical model to characterize the interplay between training and inference stages is difficult, since a large number of system factors should be taken into account. Hence, it is necessary to study efficient model-free approaches to characterize the interplay. 


\subsection{Energy Efficiency of AI}
With hundreds of neural network layers, thousands of neurons, and millions of parameters, state-of-the-art AI models usually consume extensive energy and incur substantial environmental costs.\footnote{The estimated carbon footprint of training a state-of-the-art natural language processing model is about five times the life emissions of an average car~\cite{strubell2019energy}.} Improving energy efficiency has become a major issue for  wide deployment of AI services.  In addition, recent research shows that improving the accuracy of an AI model may come at an exponential increase in the computation, environmental and economic costs~\cite{thompson2020computational}.\footnote{It is estimated that reducing the classification error probability from 11.5\% to 5\% over the ImageNet dataset needs to increase computation from 10$^{14}$ to 10$^{19}$\;Gflops, carbon emissions from 10$^6$ to 10$^{10}$\;lbs, and economic costs from 10$^6$ to 10$^{11}$\;USD~\cite{thompson2020computational}, respectively.} Hence, deploying energy-efficient AI services in a network is necessary for reducing costs for the network operator and meeting environmental standards. Several model compression techniques, such as weight pruning~\cite{han2015deep}, parameter quantization~\cite{zhu2016trained}, and model compression\cite{hinton2015distilling}, can be applied to alleviate the problem. 
In addition, hybrid data-model driven methods can train AI models with a reduced amount of data, which can also decrease energy consumption.

\subsection{Hybrid Data-Model Driven Methods}

The four options listed in Subsection~\ref{ss:codriven} provide our initial ideas for hybrid data-model driven methods. Related open issues to be investigated include the following. First, it is necessary to study how to determine which option to use and how to switch among options. Designing mechanisms for choosing and switching among options will allow networks to flexibly and adaptively integrate data-driven and model-driven methods. Second, for a chosen option, it is important to understand how much the data-driven and model-driven components affect the overall performance and how much impact they have on each other. For instance, in the mixing option, the AI service performance may depend on the combined choices of data-driven and model-driven methods, and finding a proper combination can be a challenge. Third, in addition to the four options as introduced, there should be other potential options for hybrid data-model driven methods, and identifying other promising options is an open issue of great importance. Last, due to the lack of explainability in existing data-driven methods, careful investigations and analysis should be directed to the management of critical network operations. The role of hybrid data-model driven methods in enhancing system robustness is an open issue that deserves further investigation. For more discussions on challenges in hybrid data-model driven methods for networks, interested readers are referred to \cite{zappone2019wireless, M_CWang_WCom_2020, M_TWang_ChinaCom_2019}.\footnote{ An application of a hybrid approach in vehicular network simulation can be found in \cite{C_BSliwa_GC_2020}.  }





\section{Conclusion}
\label{sec:conclusions}
Designing an architecture for future networks is challenging, especially when the use cases and defining techniques are still beneath the surface. Nevertheless, the evolution of networks through the previous generations demonstrates a necessity to support increasingly heterogeneous networks, diverse services, and stringent QoS/QoE requirements. This has been driving the trend of virtualization and generating significant interest in AI-driven networking. Recognizing the insufficiency of the existing scope and level of virtualization and AI for future 6G networks, we have presented a conceptual architecture design that integrates holistic network virtualization and pervasive network intelligence. To complement and solidify our overall network architecture, we have proposed several specific designs, including the six-layer holistic network virtualization based on digital twins, the connected AI solution for network management, as well as ideas, including AI slices and hybrid data-model driven methods. As a result, the proposed network architecture has the potential to achieve unprecedented scalability and flexibility due to the holistic network virtualization as well as exceeding adaptivity and intelligence due to the pervasive network intelligence. At last, we have identified some challenges and open issues related to the proposed architecture. We hope this study will lead to further discussions and developments on the architecture of 6G networks.

\section*{Acknowledgement}
The authors would like to thank Dr. Dongxiao Liu for helpful discussions on open issues related to data privacy and security.


%

\bibliographystyle{IEEEtran}
\bibliography{references_Conghao} 

\begin{thebibliography}{100}
\providecommand{\url}[1]{#1}
\csname url@samestyle\endcsname
\providecommand{\newblock}{\relax}
\providecommand{\bibinfo}[2]{#2}
\providecommand{\BIBentrySTDinterwordspacing}{\spaceskip=0pt\relax}
\providecommand{\BIBentryALTinterwordstretchfactor}{4}
\providecommand{\BIBentryALTinterwordspacing}{\spaceskip=\fontdimen2\font plus
\BIBentryALTinterwordstretchfactor\fontdimen3\font minus
  \fontdimen4\font\relax}
\providecommand{\BIBforeignlanguage}[2]{{%
\expandafter\ifx\csname l@#1\endcsname\relax
\typeout{** WARNING: IEEEtran.bst: No hyphenation pattern has been}%
\typeout{** loaded for the language `#1'. Using the pattern for}%
\typeout{** the default language instead.}%
\else
\language=\csname l@#1\endcsname
\fi
#2}}
\providecommand{\BIBdecl}{\relax}
\BIBdecl

\bibitem{J_You2021towards}
X.~You \emph{et~al.}, ``Towards {6G} wireless communication networks: Vision,
  enabling technologies, and new paradigm shifts,'' \emph{Sci. China Inf.
  Sci.}, vol.~64, no.~1, pp. 1--74, Nov. 2021.

\bibitem{M_WSaad_2020}
W.~{Saad}, M.~{Bennis}, and M.~{Chen}, ``A vision of {6G} wireless systems:
  Applications, trends, technologies, and open research problems,'' \emph{IEEE
  Network}, vol.~34, no.~3, pp. 134--142, May/June 2020.

\bibitem{M_MGiordani_2020}
M.~{Giordani}, M.~{Polese}, M.~{Mezzavilla}, S.~{Rangan}, and M.~{Zorzi},
  ``Toward {6G} networks: Use cases and technologies,'' \emph{IEEE Commun.
  Mag.}, vol.~58, no.~3, pp. 55--61, Mar. 2020.

\bibitem{MNet_XShen_2020}
X.~{Shen}, C.~{Huang}, D.~{Liu}, L.~{Xue}, W.~{Zhuang}, R.~{Sun}, and
  B.~{Ying}, ``Data management for future wireless networks: Architecture,
  privacy preservation, and regulation,'' \emph{IEEE Network}, vol.~35, no.~1,
  pp. 8--15, Mar./Apr. 2021.

\bibitem{J_ASodhro_2020}
A.~{Sodhro}, S.~{Pirbhulal}, L.~{Zongwei}, K.~{Muhammad}, and N.~{Zahid},
  ``Towards {6G} architecture for energy efficient communication in
  {IoT}-enabled smart automation systems,'' \emph{IEEE Internet Things J.},
  vol.~8, no.~7, pp. 5141--5148, Apr. 2021.

\bibitem{M_ZZhang_2019}
Z.~{Zhang}, Y.~{Xiao}, Z.~{Ma}, M.~{Xiao}, Z.~{Ding}, X.~{Lei},
  G.~{Karagiannidis}, and P.~{Fan}, ``{6G} wireless networks: Vision,
  requirements, architecture, and key technologies,'' \emph{IEEE Veh. Technol.
  Mag.}, vol.~14, no.~3, pp. 28--41, Sep. 2019.

\bibitem{M_BZong2019}
B.~{Zong}, C.~{Fan}, X.~{Wang}, X.~{Duan}, B.~{Wang}, and J.~{Wang}, ``{6G}
  technologies: Key drivers, core requirements, system architectures, and
  enabling technologies,'' \emph{IEEE Veh. Technol. Mag.}, vol.~14, no.~3, pp.
  18--27, Sep. 2019.

\bibitem{M_NZhang}
N.~{Zhang}, S.~{Zhang}, P.~{Yang}, O.~{Alhussein}, W.~{Zhuang}, and X.~{Shen},
  ``Software defined space-air-ground integrated vehicular networks: Challenges
  and solutions,'' \emph{IEEE Commun. Mag.}, vol.~55, no.~7, pp. 101--109, July
  2017.

\bibitem{M_SChen_2020}
S.~{Chen}, S.~{Sun}, and S.~{Kang}, ``System integration of terrestrial mobile
  communication and satellite communication — the trends, challenges and key
  technologies in {B5G} and {6G},'' \emph{China Commun.}, vol.~17, no.~12, pp.
  156--171, Dec. 2020.

\bibitem{M_ECStrinati2019}
E.~{Calvanese Strinati}, S.~{Barbarossa}, J.~{Gonzalez-Jimenez}, D.~{Ktenas},
  N.~{Cassiau}, L.~{Maret}, and C.~{Dehos}, ``{6G}: The next frontier: From
  holographic messaging to artificial intelligence using subterahertz and
  visible light communication,'' \emph{IEEE Veh. Technol. Mag.}, vol.~14,
  no.~3, pp. 42--50, Sep. 2019.

\bibitem{J_XShen_2020}
X.~{Shen}, J.~{Gao}, W.~{Wu}, K.~{Lyu}, M.~{Li}, W.~{Zhuang}, X.~{Li}, and
  J.~{Rao}, ``{AI}-assisted network-slicing based next-generation wireless
  networks,'' \emph{IEEE Open J. Veh. Technol.}, vol.~1, pp. 45--66, Jan. 2020.

\bibitem{M_KLetaief2019}
K.~{Letaief}, W.~{Chen}, Y.~{Shi}, J.~{Zhang}, and Y.~{Zhang}, ``The roadmap to
  {6G}: {AI} empowered wireless networks,'' \emph{IEEE Commun. Mag.}, vol.~57,
  no.~8, pp. 84--90, Aug. 2019.

\bibitem{yang2020artificial}
H.~Yang, A.~Alphones, Z.~Xiong, D.~Niyato, J.~Zhao, and K.~Wu,
  ``{Artificial-intelligence-enabled intelligent {6G} networks},'' \emph{IEEE
  Network}, vol.~34, no.~6, pp. 272--280, Nov. 2020.

\bibitem{M_MSayed_2002}
M.~{El-Sayed} and J.~{Jaffe}, ``A view of telecommunications network
  evolution,'' \emph{IEEE Commun. Mag.}, vol.~40, no.~12, pp. 74--81, Dec.
  2002.

\bibitem{M_BBjerke_2011}
B.~{Bjerke}, ``{LTE}-advanced and the evolution of {LTE} deployments,''
  \emph{IEEE Wireless Commun.}, vol.~18, no.~5, pp. 4--5, Oct. 2011.

\bibitem{S_DKreutz_2015}
D.~{Kreutz}, F.~{Ramos}, P.~{Veríssimo}, C.~{Rothenberg}, S.~{Azodolmolky},
  and S.~{Uhlig}, ``Software-defined networking: A comprehensive survey,''
  \emph{Proc. IEEE}, vol. 103, no.~1, pp. 14--76, Jan. 2015.

\bibitem{S_AChecko_2015}
A.~{Checko}, H.~{Christiansen}, Y.~{Yan}, L.~{Scolari}, G.~{Kardaras},
  M.~{Berger}, and L.~{Dittmann}, ``Cloud {RAN} for mobile networks—a
  technology overview,'' \emph{IEEE Commun. Surveys Tuts.}, vol.~17, no.~1, pp.
  405--426, 4th Quart. 2015.

\bibitem{J_MBagaa2018}
M.~{Bagaa}, T.~{Taleb}, A.~{Laghrissi}, A.~{Ksentini}, and H.~{Flinck},
  ``Coalitional game for the creation of efficient virtual core network slices
  in {5G} mobile systems,'' \emph{IEEE J. Sel. Areas Commun.}, vol.~36, no.~3,
  pp. 469--484, Mar. 2018.

\bibitem{J_WWu2020}
W.~{Wu}, N.~{Chen}, C.~{Zhou}, M.~{Li}, X.~{Shen}, W.~{Zhuang}, and X.~{Li},
  ``Dynamic {RAN} slicing for service-oriented vehicular networks via
  constrained learning,'' \emph{IEEE J. Sel. Areas Commun.}, vol.~39, no.~7,
  pp. 2076--2089, July 2021.

\bibitem{M_KQu_2020}
K.~{Qu}, W.~{Zhuang}, Q.~{Ye}, X.~{Shen}, X.~{Li}, and J.~{Rao}, ``Traffic
  engineering for service-oriented {5G} networks with {SDN-NFV} integration,''
  \emph{IEEE Network}, vol.~34, no.~4, pp. 234--241, July 2020.

\bibitem{S_IAfolabi2018}
I.~{Afolabi}, T.~{Taleb}, K.~{Samdanis}, A.~{Ksentini}, and H.~{Flinck},
  ``Network slicing and softwarization: A survey on principles, enabling
  technologies, and solutions,'' \emph{IEEE Commun. Surveys Tuts.}, vol.~20,
  no.~3, pp. 2429--2453, 3rd Quart. 2018.

\bibitem{O_Ali6gWhite_2020}
S.~{Ali} \emph{et~al.}, ``{6G} white paper on machine learning in wireless
  communication networks,'' \emph{arXiv:2004.13875}, 2020, [Online]. Available:
  http://arxiv.org/abs/2004.13875.

\bibitem{J_HZhang2016}
H.~Zhang, ``Future wireless network: {MyNET} platform and end-to-end network
  slicing,'' \emph{arXiv:1611.07601}, 2016, [Online]. Available:
  https://arxiv.org/abs/1611.07601.

\bibitem{S_Fadlullah2017}
Z.~{Fadlullah}, F.~{Tang}, B.~{Mao}, N.~{Kato}, O.~{Akashi}, T.~{Inoue}, and
  K.~{Mizutani}, ``State-of-the-art deep learning: Evolving machine
  intelligence toward tomorrow’s intelligent network traffic control
  systems,'' \emph{IEEE Commun. Surveys Tuts.}, vol.~19, no.~4, pp. 2432--2455,
  4th Quart. 2017.

\bibitem{J_AToma2020}
A.~{Toma}, A.~{Krayani}, M.~{Farrukh}, H.~{Qi}, L.~{Marcenaro}, Y.~{Gao}, and
  C.~{Regazzoni}, ``{AI}-based abnormality detection at the {PHY}-layer of
  cognitive radio by learning generative models,'' \emph{IEEE Trans. Cogn.
  Commun. Netw.}, vol.~6, no.~1, pp. 21--34, Mar. 2020.

\bibitem{M_SHan2020}
S.~{Han}, T.~{Xie}, C.~{I}, L.~{Chai}, Z.~{Liu}, Y.~{Yuan}, and C.~{Cui},
  ``Artificial-intelligence-enabled air interface for {6G}: Solutions,
  challenges, and standardization impacts,'' \emph{IEEE Commun. Mag.}, vol.~58,
  no.~10, pp. 73--79, Oct. 2020.

\bibitem{J_MLi2021}
M.~{Li}, J.~{Gao}, L.~{Zhao}, and X.~{Shen}, ``Adaptive computing scheduling
  for edge-assisted autonomous driving,'' \emph{IEEE Trans. Veh. Technol.},
  vol.~70, no.~6, pp. 5318--5331, June 2021.

\bibitem{li2021slicing}
M.~Li, J.~Gao, C.~Zhou, W.~Zhuang, and X.~Shen, ``Slicing-based {AI} service
  provisioning on network edge,'' \emph{IEEE Veh. Technol. Mag.}, to be
  published, doi:10.1109/MVT.2021.3114655.

\bibitem{O_PChemouil2020}
P.~{Chemouil}, P.~{Hui}, W.~{Kellerer}, N.~{Limam}, R.~{Stadler}, and Y.~{Wen},
  ``Guest editorial special issue on advances in artificial intelligence and
  machine learning for networking,'' \emph{IEEE J. Sel. Areas Commun.},
  vol.~38, no.~10, pp. 2229--2233, Oct. 2020.

\bibitem{3gpp.29.520}
3GPP, ``Technical specification group core network and terminals; {5G} system;
  {N}etwork data analytics services; {S}tage 3 ({R}elease 17),'' Tech. Rep.
  3GPP TS29.520 V0.0.0, Oct. 2017.

\bibitem{3gpp.21.916}
------, ``Technical specification group services and system aspects; {R}elease
  16 description; summary of {R}el-16 work items ({R}elease 16),'' Tech. Rep.
  3GPP TR21.916 V1.0.0, Dec. 2020.

\bibitem{J_RMinerva_ProcIEEE_2020}
R.~Minerva, G.~M. Lee, and N.~Crespi, ``{Digital twin in the {IoT} context: A
  survey on technical features, scenarios, and architectural models},''
  \emph{Proc. IEEE}, vol. 108, no.~10, pp. 1785--1824, Oct. 2020.

\bibitem{boutaba2018comprehensive}
R.~Boutaba, M.~A. Salahuddin, N.~Limam, S.~Ayoubi, N.~Shahriar,
  F.~Estrada-Solano, and O.~M. Caicedo, ``{A comprehensive survey on machine
  learning for networking: {Evolution}, applications and research
  opportunities},'' \emph{J. Internet Serv. Appl.}, vol.~9, no.~1, pp. 1--99,
  June 2018.

\bibitem{zhang2019deep}
C.~Zhang, P.~Patras, and H.~Haddadi, ``{Deep learning in mobile and wireless
  networking: A survey},'' \emph{IEEE Commun. Surveys Tuts.}, vol.~21, no.~3,
  pp. 2224--2287, 3rd Quart. 2019.

\bibitem{M_Mosharaf_COMM_2009}
N.~{Chowdhury} and R.~{Boutaba}, ``Network virtualization: State of the art and
  research challenges,'' \emph{IEEE Commun. Mag.}, vol.~47, no.~7, pp. 20--26,
  July 2009.

\bibitem{rossi1986proposal}
G.~Rossi and C.~Garavaglia, ``A proposal for an improved network layer of an
  {LAN},'' \emph{ACM SIGCOMM Computer Communication Review}, vol.~16, no.~1,
  pp. 1--5, Feb. 1986.

\bibitem{J_KSato_TCOM_1990}
K.~{Sato}, S.~{Ohta}, and I.~{Tokizawa}, ``Broad-band {ATM} network
  architecture based on virtual paths,'' \emph{IEEE Trans. Commun.}, vol.~38,
  no.~8, pp. 1212--1222, Aug. 1990.

\bibitem{J_KQu_TCom_2020}
K.~{Qu}, W.~{Zhuang}, Q.~{Ye}, X.~{Shen}, X.~{Li}, and J.~{Rao}, ``Dynamic flow
  migration for embedded services in {SDN/NFV}-enabled {5G} core networks,''
  \emph{IEEE Trans. Commun.}, vol.~68, no.~4, pp. 2394--2408, Apr. 2020.

\bibitem{C_chioC_2012}
M.~Chiosi \emph{et~al.}, ``Network functions virtualisation: An introduction,
  benefits, enablers, challenges and call for action,'' in \emph{Proc. SDN and
  OpenFlow World Congress}, Darmstadt, Germany, Oct. 2012.

\bibitem{J_AGDalla-Costa_JSAC_2020}
A.~{Dalla-Costa}, L.~{Bondan}, J.~{Wickboldt}, C.~{Both}, and L.~{Granville},
  ``Orchestra: A customizable split-aware {NFV} orchestrator for dynamic cloud
  radio access networks,'' \emph{IEEE J. Sel. Areas Commun.}, vol.~38, no.~6,
  pp. 1014--1024, June 2020.

\bibitem{J_GZhang2007}
G.~{Zhang}, J.~{Shu}, W.~{Xue}, and W.~{Zheng}, ``Design and implementation of
  an out-of-band virtualization system for large {SANs},'' \emph{IEEE Trans.
  Comput.}, vol.~56, no.~12, pp. 1654--1665, Dec. 2007.

\bibitem{J_FXu_2014}
F.~{Xu}, F.~{Liu}, H.~{Jin}, and A.~{Vasilakos}, ``Managing performance
  overhead of virtual machines in cloud computing: A survey, state of the art,
  and future directions,'' \emph{Proc. IEEE}, vol. 102, no.~1, pp. 11--31, Jan.
  2014.

\bibitem{J_PBellavista_2019}
P.~{Bellavista}, A.~{Corradi}, L.~{Foschini}, S.~{Luciano}, and M.~{Solimando},
  ``A simulation framework for virtualized resources in cloud data center
  networks,'' \emph{IEEE J. Sel. Areas Commun.}, vol.~37, no.~8, pp.
  1808--1819, Aug. 2019.

\bibitem{J_XYuan2021}
X.~{Yuan}, M.~{Sun}, and W.~{Lou}, ``A dynamic deep-learning-based virtual edge
  node placement scheme for edge cloud systems in mobile environment,''
  \emph{IEEE Trans. Cloud Comput.}, to be published, doi:
  10.1109/TCC.2020.2974948.

\bibitem{J_SYang2017}
S.~{Yang}, P.~{Wieder}, R.~{Yahyapour}, S.~{Trajanovski}, and X.~{Fu},
  ``Reliable virtual machine placement and routing in clouds,'' \emph{IEEE
  Trans. Parallel Distrib. Syst.}, vol.~28, no.~10, pp. 2965--2978, Oct. 2017.

\bibitem{J_MNagy_2018}
M.~{Nagy}, J.~{Tapolcai}, and G.~{Rétvári}, ``Node virtualization for {IP}
  level resilience,'' \emph{IEEE/ACM Trans. Netw.}, vol.~26, no.~3, pp.
  1250--1263, June 2018.

\bibitem{M_Khan2015}
I.~{Khan}, F.~{Belqasmi}, R.~{Glitho}, N.~{Crespi}, M.~{Morrow}, and
  P.~{Polakos}, ``Wireless sensor network virtualization: Early architecture
  and research perspectives,'' \emph{IEEE Network}, vol.~29, no.~3, pp.
  104--112, May 2015.

\bibitem{J_SZaidi_2019}
S.~{Zaidi}, O.~{Ben Smida}, S.~{Affes}, U.~{Vilaipornsawai}, L.~{Zhang}, and
  P.~{Zhu}, ``User-centric base-station wireless access virtualization for
  future {5G} networks,'' \emph{IEEE Trans. Commun.}, vol.~67, no.~7, pp.
  5190--5202, July 2019.

\bibitem{J_HDu2013}
H.~{Du}, W.~{Wu}, Q.~{Ye}, D.~{Li}, W.~{Lee}, and X.~{Xu}, ``{CDS}-based
  virtual backbone construction with guaranteed routing cost in wireless sensor
  networks,'' \emph{IEEE Trans. Parallel Distrib. Syst.}, vol.~24, no.~4, pp.
  652--661, Apr. 2013.

\bibitem{J_FHosseini_2019}
F.~{Hosseini}, A.~{James}, and M.~{Ghaderi}, ``Probabilistic virtual link
  embedding under demand uncertainty,'' \emph{IEEE Trans. Netw. Service
  Manag.}, vol.~16, no.~4, pp. 1552--1566, Dec. 2019.

\bibitem{J_STomovic2019}
S.~{Tomovic} and I.~{Radusinovic}, ``Toward a scalable, robust, and {QoS}-aware
  virtual-link provisioning in {SDN}-based {ISP} networks,'' \emph{IEEE Trans.
  Netw. Service Manag.}, vol.~16, no.~3, pp. 1032--1045, Sep. 2019.

\bibitem{J_CPapagianni_2013}
C.~Papagianni, A.~Leivadeas, S.~Papavassiliou, V.~Maglaris, C.~Cervello-Pastor,
  and A.~Monje, ``On the optimal allocation of virtual resources in cloud
  computing networks,'' \emph{IEEE Trans. Comput.}, vol.~62, no.~6, pp.
  1060--1071, June 2013.

\bibitem{J_TWood2015}
T.~{Wood}, K.~{Ramakrishnan}, P.~{Shenoy}, J.~{Van der Merwe}, J.~{Hwang},
  G.~{Liu}, and L.~{Chaufournier}, ``Cloudnet: Dynamic pooling of cloud
  resources by live {WAN} migration of virtual machines,'' \emph{IEEE/ACM
  Trans. Netw.}, vol.~23, no.~5, pp. 1568--1583, Oct. 2015.

\bibitem{L_MKali_2016}
M.~{Kalil}, A.~{Moubayed}, A.~{Shami}, and A.~{Al-Dweik}, ``Efficient
  low-complexity scheduler for wireless resource virtualization,'' \emph{IEEE
  Wireless Commun. Lett.}, vol.~5, no.~1, pp. 56--59, Feb. 2016.

\bibitem{J_XLu_2019}
X.~{Lu}, Q.~{Ni}, D.~{Zhao}, W.~{Cheng}, and H.~{Zhang}, ``Resource
  virtualization for customized delay-bounded {QoS} provisioning in uplink
  {VMIMO-SC-FDMA} systems,'' \emph{IEEE Trans. Commun.}, vol.~67, no.~4, pp.
  2951--2967, Apr. 2019.

\bibitem{J_SZhang2020}
S.~{Zhang}, H.~{Luo}, J.~{Li}, W.~{Shi}, and X.~{Shen}, ``Hierarchical soft
  slicing to meet multi-dimensional {QoS} demand in cache-enabled vehicular
  networks,'' \emph{IEEE Trans. Wireless Commun.}, vol.~19, no.~3, pp.
  2150--2162, Mar. 2020.

\bibitem{J_OAlhussein_2020}
O.~{Alhussein}, P.~{Do}, Q.~{Ye}, J.~{Li}, W.~{Shi}, W.~{Zhuang}, X.~{Shen},
  X.~{Li}, and J.~{Rao}, ``A virtual network customization framework for
  multicast services in {NFV}-enabled core networks,'' \emph{IEEE J. Sel. Areas
  Commun.}, vol.~38, no.~6, pp. 1025--1039, June 2020.

\bibitem{J_NZhang_2017}
N.~{Zhang}, Y.~{Liu}, H.~{Farmanbar}, T.~{Chang}, M.~{Hong}, and Z.~{Luo},
  ``Network slicing for service-oriented networks under resource constraints,''
  \emph{IEEE J. Sel. Areas Commun.}, vol.~35, no.~11, pp. 2512--2521, Nov.
  2017.

\bibitem{J_JTang2019}
J.~{Tang}, B.~{Shim}, and T.~Q.~S. {Quek}, ``Service multiplexing and revenue
  maximization in sliced {C-RAN} incorporated with {URLLC} and multicast
  {eMBB},'' \emph{IEEE J. Sel. Areas Commun.}, vol.~37, no.~4, pp. 881--895,
  Apr. 2019.

\bibitem{J_QYe2018}
Q.~{Ye}, W.~{Zhuang}, S.~{Zhang}, A.~{Jin}, X.~{Shen}, and X.~{Li}, ``Dynamic
  radio resource slicing for a two-tier heterogeneous wireless network,''
  \emph{IEEE Trans. Veh. Technol.}, vol.~67, no.~10, pp. 9896--9910, Oct. 2018.

\bibitem{J_TGuo2019}
T.~{Guo} and A.~{Suárez}, ``Enabling {5G} {RAN} slicing with {EDF} slice
  scheduling,'' \emph{IEEE Trans. Veh. Technol.}, vol.~68, no.~3, pp.
  2865--2877, Mar. 2019.

\bibitem{J_JNi2018}
J.~{Ni}, X.~{Lin}, and X.~{Shen}, ``Efficient and secure service-oriented
  authentication supporting network slicing for {5G}-enabled {IoT},''
  \emph{IEEE J. Sel. Areas Commun.}, vol.~36, no.~3, pp. 644--657, Mar. 2018.

\bibitem{J_MKessler_2008}
M.~{Kessler}, A.~{Reifert}, D.~{Lamp}, and T.~{Voith}, ``A service-oriented
  infrastructure for providing virtualized networks,'' \emph{Bell Labs
  Technical Journal}, vol.~13, no.~3, pp. 111--127, Fall 2008.

\bibitem{S_JVBelt_2017}
J.~{van de Belt}, H.~{Ahmadi}, and L.~{Doyle}, ``Defining and surveying
  wireless link virtualization and wireless network virtualization,''
  \emph{IEEE Commun. Surveys Tuts.}, vol.~19, no.~3, pp. 1603--1627, 3rd Quart.
  2017.

\bibitem{mckeown2008openflow}
N.~McKeown, T.~Anderson, H.~Balakrishnan, G.~Parulkar, L.~Peterson, J.~Rexford,
  S.~Shenker, and J.~Turner, ``{OpenFlow: Enabling innovation in campus
  networks},'' \emph{ACM SIGCOMM Comput. Commun. Review}, vol.~38, no.~2, pp.
  69--74, Apr. 2008.

\bibitem{gudipati2013softran}
A.~Gudipati, D.~Perry, L.~E. Li, and S.~Katti, ``{SoftRAN: Software defined
  radio access network},'' in \emph{Proc. ACM SIGCOMM Workshop HotSDN}, Hong
  Kong, China, Aug. 2013.

\bibitem{foukas2016flexran}
X.~Foukas, N.~Nikaein, M.~M. Kassem, M.~K. Marina, and K.~Kontovasilis,
  ``{FlexRAN: A flexible and programmable platform for software-defined radio
  access networks},'' in \emph{Proc. ACM CoNEXT}, Irvine, CA, USA, Dec. 2016,
  pp. 427--441.

\bibitem{ORAN}
{O-RAN Alliance}, 2021, [Online]. Available: https://www.o-ran.org/software.

\bibitem{breen2021powder}
J.~Breen \emph{et~al.}, ``{POWDER: Platform for open wireless data-driven
  experimental research},'' \emph{Computer Networks}, vol. 197, p. 108281, Oct.
  2021.

\bibitem{johnson2021open}
D.~Johnson, D.~Maas, and J.~Van Der~Merwe, ``{Open source RAN slicing on
  POWDER: A top-to-bottom O-RAN use case},'' in \emph{Proc. ACM MobiSys},
  Virtual Conference, June 2021.

\bibitem{foukas2021concordia}
X.~Foukas and B.~Radunovic, ``{Concordia: Teaching the 5G vRAN to share
  compute},'' in \emph{Proc. ACM SIGCOMM}, Virtual Conference, Aug. 2021.

\bibitem{J_AConte_2008}
A.~{Conte}, S.~{Kerboeuf}, and L.~{Thomas}, ``Network-hosted avatar:
  User-terminal virtualization in the network,'' \emph{Bell Labs Technical
  Journal}, vol.~13, no.~2, pp. 117--126, Summer 2008.

\bibitem{S_MNitti_2016}
M.~{Nitti}, V.~{Pilloni}, G.~{Colistra}, and L.~{Atzori}, ``The virtual object
  as a major element of the internet of things: A survey,'' \emph{IEEE Commun.
  Surveys Tuts.}, vol.~18, no.~2, pp. 1228--1240, 2nd Quart. 2016.

\bibitem{W_MGrieves_2014}
M.~Grieves, ``Digital twin: Manufacturing excellence through virtual factory
  replication,'' \emph{White paper}, vol.~1, pp. 1--7, Mar. 2015.

\bibitem{khan2021digital}
L.~U. Khan, W.~Saad, D.~Niyato, Z.~Han, and C.~S. Hong, ``Digital-twin-enabled
  {6G: Vision,} architectural trends, and future directions,''
  \emph{arXiv:2102.12169}, 2021, [Online]. Available:
  https://arxiv.org/abs/2102.12169.

\bibitem{glaessgen2012digital}
E.~Glaessgen and D.~Stargel, ``{The digital twin paradigm for future NASA and
  US Air Force vehicles},'' in \emph{Proc. AIAA structures, structural dynamics
  and materials conference}, Honolulu, HI, USA, Apr. 2012.

\bibitem{TII_Gao}
Z.~Gao, A.~Paul, and X.~Wang, ``{Guest editorial: Digital twinning: Integrating
  AI-ML and big data analytics for virtual representation},'' \emph{IEEE Trans.
  Ind. Informat.}, vol.~18, no.~2, pp. 1355--1358, Feb. 2022.

\bibitem{madni2019leveraging}
A.~M. Madni, C.~C. Madni, and S.~D. Lucero, ``{Leveraging digital twin
  technology in model-based systems engineering},'' \emph{Systems}, vol.~7,
  no.~1, p.~7, Jan. 2019.

\bibitem{J_PJia_2020}
P.~{Jia}, X.~{Wang}, and X.~{Shen}, ``Digital twin enabled intelligent
  distributed clock synchronization in industrial {IoT} systems,'' \emph{IEEE
  Internet Things J.}, vol.~8, no.~6, pp. 4548--4559, Mar. 2021.

\bibitem{J_YDai_2020}
Y.~{Dai}, K.~{Zhang}, S.~{Maharjan}, and Y.~{Zhang}, ``Deep reinforcement
  learning for stochastic computation offloading in digital twin networks,''
  \emph{IEEE Trans. Ind. Informat.}, vol.~17, no.~7, pp. 4968--4977, July 2021.

\bibitem{M_LZhao_2020}
L.~{Zhao}, G.~{Han}, Z.~{Li}, and L.~{Shu}, ``Intelligent digital twin-based
  software-defined vehicular networks,'' \emph{IEEE Network}, vol.~34, no.~5,
  pp. 178--184, Oct. 2020.

\bibitem{C_JTaylor_2020}
J.~{Taylor} and H.~{Sharif}, ``Leveraging digital twins to enhance performance
  of {IoT} in disadvantaged networks,'' in \emph{Proc. IEEE IWCMC}, Limassol,
  Cyprus, June 2020.

\bibitem{M_Q_Yu_2019}
Q.~{Yu}, J.~{Ren}, Y.~{Fu}, Y.~{Li}, and W.~{Zhang}, ``Cybertwin: An origin of
  next generation network architecture,'' \emph{IEEE Wireless Commun.},
  vol.~26, no.~6, pp. 111--117, Dec. 2019.

\bibitem{J_ABarbie_2021}
A.~{Barbie}, N.~{Pech}, W.~{Hasselbring}, S.~{Flogel}, F.~{Wenzhofer},
  M.~{Walter}, E.~{Shchekinova}, M.~{Busse}, M.~{Turk}, M.~{Hofbauer}, and
  S.~{Sommer}, ``Developing an underwater network of ocean observation systems
  with digital twin prototypes - a field report from the baltic sea,''
  \emph{IEEE Internet Comput.}, to be published, doi:10.1109/MIC.2021.3065245.

\bibitem{J_XXu2020}
X.~{Xu}, B.~{Shen}, S.~{Ding}, G.~{Srivastava}, M.~{Bilal}, M.~{Khosravi},
  V.~{Menon}, M.~{Jan}, and W.~{Maoli}, ``Service offloading with deep
  {Q}-network for digital twinning empowered internet of vehicles in edge
  computing,'' \emph{IEEE Trans. Ind. Informat.}, vol.~18, no.~2, pp.
  1414--1423, Feb. 2022.

\bibitem{J_HWang2020}
H.~{Wang}, Y.~{Wu}, G.~{Min}, and W.~{Miao}, ``A graph neural network-based
  digital twin for network slicing management,'' \emph{IEEE Trans. Ind.
  Informat.}, vol.~18, no.~2, pp. 1367--1376, Feb. 2022.

\bibitem{J_TWang2020}
T.~{Wang}, J.~{Cheng}, Y.~{Yang}, C.~{Esposito}, H.~{Snoussi}, and F.~{Tao},
  ``Adaptive optimization method in digital twin conveyor systems via
  range-inspection control,'' \emph{IEEE Trans. Autom. Sci. Eng.}, to be
  published, doi:10.1109/TASE.2020.3043393.

\bibitem{M_OEMarai2020}
O.~E. {Marai}, T.~{Taleb}, and J.~{Song}, ``Roads infrastructure digital twin:
  A step toward smarter cities realization,'' \emph{IEEE Network}, vol.~35,
  no.~2, pp. 136--143, Mar./Apr. 2021.

\bibitem{J_RMinerval_2021}
R.~{Minerva}, F.~{Awan}, and N.~{Crespi}, ``Exploiting digital twin as enablers
  for synthetic sensing,'' \emph{IEEE Internet Comput.}, to be published,
  doi:10.1109/MIC.2021.3051674.

\bibitem{J_ACastellani2020}
A.~{Castellani}, S.~{Schmitt}, and S.~{Squartini}, ``Real-world anomaly
  detection by using digital twin systems and weakly-supervised learning,''
  \emph{IEEE Trans. Ind. Informat.}, vol.~17, no.~7, pp. 4733--4742, July 2021.

\bibitem{J_HElayan2021}
H.~{Elayan}, M.~{Aloqaily}, and M.~{Guizani}, ``Digital twin for intelligent
  context-aware {IoT} healthcare systems,'' \emph{IEEE Internet Things J.},
  vol.~8, no.~23, pp. 16\,749--16\,757, Dec. 2021.

\bibitem{J_MSchluse2018}
M.~{Schluse}, M.~{Priggemeyer}, L.~{Atorf}, and J.~{Rossmann}, ``Experimentable
  digital twins—streamlining simulation-based systems engineering for
  industry 4.0,'' \emph{IEEE Trans. Ind. Informat.}, vol.~14, no.~4, pp.
  1722--1731, Apr. 2018.

\bibitem{J_RDong2019}
R.~{Dong}, C.~{She}, W.~{Hardjawana}, Y.~{Li}, and B.~{Vucetic}, ``Deep
  learning for hybrid {5G} services in mobile edge computing systems: {L}earn
  from a digital twin,'' \emph{IEEE Trans. Wireless Commun.}, vol.~18, no.~10,
  pp. 4692--4707, Oct. 2019.

\bibitem{J_CGehrmann2020}
C.~{Gehrmann} and M.~{Gunnarsson}, ``A digital twin based industrial automation
  and control system security architecture,'' \emph{IEEE Trans. Ind.
  Informat.}, vol.~16, no.~1, pp. 669--680, Jan. 2020.

\bibitem{J_KMAlam2017}
K.~{Alam} and A.~{El Saddik}, ``{C2PS}: A digital twin architecture reference
  model for the cloud-based cyber-physical systems,'' \emph{IEEE Access},
  vol.~5, pp. 2050--2062, Jan. 2017.

\bibitem{J_LFRivera2021}
L.~{Rivera}, M.~{Jimenez}, N.~{Villegas}, G.~{Tamura}, and H.~{Muller}, ``The
  forging of autonomic and cooperating digital twins,'' \emph{IEEE Internet
  Comput.}, pp. 1--10, to be published, doi:10.1109/MIC.2021.3051902.

\bibitem{zhang2020manufacturing}
C.~Zhang, G.~Zhou, H.~Li, and Y.~Cao, ``{Manufacturing blockchain of things for
  the configuration of a data-and knowledge-driven digital twin manufacturing
  cell},'' \emph{IEEE Internet Things J.}, vol.~7, no.~12, pp.
  11\,884--11\,894, Dec. 2020.

\bibitem{fang2019digital}
Y.~Fang, C.~Peng, P.~Lou, Z.~Zhou, J.~Hu, and J.~Yan, ``{Digital-twin-based job
  shop scheduling toward smart manufacturing},'' \emph{IEEE Trans. Ind.
  Informat.}, vol.~15, no.~12, pp. 6425--6435, Dec. 2019.

\bibitem{xu2019digital}
Y.~Xu, Y.~Sun, X.~Liu, and Y.~Zheng, ``{A digital-twin-assisted fault diagnosis
  using deep transfer learning},'' \emph{IEEE Access}, vol.~7, pp.
  19\,990--19\,999, Jan. 2019.

\bibitem{jain2019digital}
P.~Jain, J.~Poon, J.~P. Singh, C.~Spanos, S.~R. Sanders, and S.~K. Panda, ``{A
  digital twin approach for fault diagnosis in distributed photovoltaic
  systems},'' \emph{IEEE Trans. Power Electron.}, vol.~35, no.~1, pp. 940--956,
  Jan. 2019.

\bibitem{IOT_Sun_2021}
W.~Sun, P.~Wang, N.~Xu, G.~Wang, and Y.~Zhang, ``{Dynamic digital twin and
  distributed incentives for resource allocation in aerial-assisted {Internet}
  of vehicles},'' \emph{IEEE Internet Things J.}, pp. 1--14, 2021, to be
  published, doi:10.1109/JIOT.2021.3058213.

\bibitem{TCSS_Zhang_2021}
K.~Zhang, J.~Cao, S.~Maharjan, and Y.~Zhang, ``{Digital twin empowered content
  caching in social-aware vehicular edge networks},'' \emph{IEEE Trans. Comput.
  Soc. Syst.}, pp. 1--13, 2021, to be published, doi:10.1109/TCSS.2021.3068369.

\bibitem{hartigan1979algorithm}
J.~A. Hartigan and M.~A. Wong, ``{Algorithm AS 136: A k-means clustering
  algorithm},'' \emph{J. Roy. Stat. Soc. C (Appl. Stat.)}, vol.~28, no.~1, pp.
  100--108, 1979.

\bibitem{Parwez}
M.~Parwez, D.~Rawat, and M.~Garuba, ``{Big data analytics for user-activity
  analysis and user-anomaly detection in mobile wireless network},'' \emph{IEEE
  Trans. Ind. Informat.}, vol.~13, no.~4, pp. 2058--2065, Aug. 2017.

\bibitem{topchy2004mixture}
A.~Topchy, A.~Jain, and W.~Punch, ``{A mixture model for clustering
  ensembles},'' in \emph{Proc. SIAM Int. Conf. Data Mining}, Lake Buena Vista,
  FL, USA, 2004.

\bibitem{Shih}
M.~Shih and A.~Hero, ``{Unicast-based inference of network link delay
  distributions with finite mixture models},'' \emph{IEEE Trans. Signal
  Process.}, vol.~51, no.~8, pp. 2219--2228, Aug. 2003.

\bibitem{Bega}
D.~{Bega}, M.~{Gramaglia}, M.~{Fiore}, A.~{Banchs}, and X.~{Costa-Perez},
  ``{DeepCog}: Optimizing resource provisioning in network slicing with
  {AI}-based capacity forecasting,'' \emph{IEEE J. Sel. Areas Commun.},
  vol.~38, no.~2, pp. 361--376, Feb. 2020.

\bibitem{ledig2017photo}
C.~Ledig, L.~Theis, F.~Husz{\'a}r, J.~Caballero, A.~Cunningham, A.~Acosta,
  A.~Aitken, A.~Tejani, J.~Totz, Z.~Wang \emph{et~al.}, ``{Photo-realistic
  single image super-resolution using a generative adversarial network},'' in
  \emph{Proc. IEEE CVPR}, Honolulu, HA, USA, 2017.

\bibitem{Erpek}
T.~Erpek, Y.~Sagduyu, and Y.~Shi, ``{Deep learning for launching and mitigating
  wireless jamming attacks},'' \emph{IEEE Trans. Cogn. Commun. Netw.}, vol.~5,
  no.~1, pp. 2--14, Mar. 2019.

\bibitem{SVM_MAC}
S.~Hu, Y.~Yao, and Z.~Yang, ``{MAC protocol identification using support vector
  machines for cognitive radio networks},'' \emph{IEEE Wireless Commun.},
  vol.~21, no.~1, pp. 52--60, Feb. 2014.

\bibitem{You_LR}
C.~You and R.~Zhang, ``{3D trajectory optimization in Rician fading for
  UAV-enabled data harvesting},'' \emph{IEEE Trans. Wireless Commun.}, vol.~18,
  no.~6, pp. 3192--3207, June 2019.

\bibitem{Peng_CNN}
S.~Peng, H.~Jiang, H.~Wang, H.~Alwageed, Y.~Zhou, M.~Sebdani, and Y.~Yao,
  ``{Modulation classification based on signal constellation diagrams and deep
  learning},'' \emph{IEEE Trans. Neural Netw. Learn. Syst.}, vol.~30, no.~3,
  pp. 718--727, Mar. 2019.

\bibitem{zhao2017lstm}
Z.~Zhao, W.~Chen, X.~Wu, P.~Chen, and J.~Liu, ``{LSTM network: A deep learning
  approach for short-term traffic forecast},'' \emph{IET Intell. Transp.
  Syst.}, vol.~11, no.~2, pp. 68--75, Feb. 2017.

\bibitem{DQN2}
S.~Gu, T.~Lillicrap, I.~Sutskever, and S.~Levine, ``{Continuous deep Q-learning
  with model-based acceleration},'' in \emph{Proc. ICML}, New York City, NY,
  USA, 2016.

\bibitem{DQN1}
J.~Zhu, Y.~Song, D.~Jiang, and H.~Song, ``{A new deep-Q-learning-based
  transmission scheduling mechanism for the cognitive Internet of things},''
  \emph{IEEE Internet Things J.}, vol.~5, no.~4, pp. 2375--2385, Aug. 2018.

\bibitem{Yaohua}
Y.~Sun, M.~Peng, and S.~Mao, ``{Deep reinforcement learning-based mode
  selection and resource management for green fog radio access networks},''
  \emph{IEEE Internet Things J.}, vol.~6, no.~2, pp. 1960--1971, Apr. 2019.

\bibitem{silver2014deterministic}
D.~Silver, G.~Lever, N.~Heess, T.~Degris, D.~Wierstra, and M.~Riedmiller,
  ``{Deterministic policy gradient algorithms},'' in \emph{Proc. ICML},
  Beijing, China, 2014.

\bibitem{Somuyiwa}
S.~O. Somuyiwa, A.~Gyorgy, and D.~Gunduz, ``{A reinforcement-learning approach
  to proactive caching in wireless networks},'' \emph{IEEE J. Sel. Areas
  Commun.}, vol.~36, no.~6, pp. 1331--1344, June 2018.

\bibitem{Hoang}
D.~T. Hoang, D.~Niyato, P.~Wang, and D.~I. Kim, ``{Performance optimization for
  cooperative multiuser cognitive radio networks with RF energy harvesting
  capability},'' \emph{IEEE Trans. Wireless Commun.}, vol.~14, no.~7, pp.
  3614--3629, July 2015.

\bibitem{konda2000actor}
V.~R. Konda and J.~N. Tsitsiklis, ``{Actor-critic algorithms},'' in \emph{Proc.
  NIPS}, Denver, CO, USA, 2000.

\bibitem{Cheng_AC}
N.~Cheng, F.~Lyu, W.~Quan, C.~Zhou, H.~He, W.~Shi, and X.~Shen,
  ``{Space/aerial-assisted computing offloading for IoT applications: A
  learning-based approach},'' \emph{IEEE J. Sel. Areas Commun.}, vol.~37,
  no.~5, pp. 1117--1129, May 2019.

\bibitem{lillicrap2015continuous}
T.~P. Lillicrap, J.~J. Hunt, A.~Pritzel, N.~Heess, T.~Erez, Y.~Tassa,
  D.~Silver, and D.~Wierstra, ``{Continuous control with deep reinforcement
  learning},'' \emph{arXiv:1509.02971}, 2015, [Online]. Available:
  https://arxiv.org/abs/1509.02971.

\bibitem{TCCN_MUSHU}
M.~{Li}, J.~{Gao}, L.~{Zhao}, and X.~{Shen}, ``Deep reinforcement learning for
  collaborative edge computing in vehicular networks,'' \emph{IEEE Trans. Cogn.
  Commun. Netw.}, vol.~6, no.~4, pp. 1122--1135, Dec. 2020.

\bibitem{konevcny2016federated}
J.~Kone{\v{c}}n{\`y}, H.~B. McMahan, F.~X. Yu, P.~Richt{\'a}rik, A.~T. Suresh,
  and D.~Bacon, ``{Federated learning: Strategies for improving communication
  efficiency},'' \emph{arXiv:1610.05492}, 2016, [Online]. Available:
  https://arxiv.org/abs/1610.05492.

\bibitem{Zhaohui}
Z.~Yang, M.~Chen, W.~Saad, C.~S. Hong, and M.~Shikh-Bahaei, ``{Energy efficient
  federated learning over wireless communication networks},'' \emph{IEEE Trans.
  Wireless Commun.}, vol.~20, no.~3, pp. 1935--1949, Mar. 2021.

\bibitem{thapa2020advancements}
C.~Thapa, M.~Chamikara, and S.~Camtepe, ``Advancements of federated learning
  towards privacy preservation: from federated learning to split learning,''
  \emph{arXiv:2011.14818}, 2020, [Online]. Available:
  https://arxiv.org/abs/2011.14818.

\bibitem{Weisen}
W.~Shi, J.~Li, H.~Wu, C.~Zhou, N.~Cheng, and X.~Shen, ``{Drone-cell trajectory
  planning and resource allocation for highly mobile networks: A hierarchical
  DRL approach},'' \emph{IEEE Internet Things J.}, vol.~8, no.~12, pp.
  9800--9813, June 2021.

\bibitem{Sana}
M.~{Sana}, A.~{De Domenico}, W.~{Yu}, Y.~{Lostanlen}, and E.~{Calvanese
  Strinati}, ``Multi-agent reinforcement learning for adaptive user association
  in dynamic {mmWave} networks,'' \emph{IEEE Trans. Wireless Commun.}, vol.~19,
  no.~10, pp. 6520--6534, Oct. 2020.

\bibitem{Ruijin}
R.~Ding, Y.~Xu, F.~Gao, and X.~Shen, ``{Trajectory design and access control
  for air-ground coordinated communications system with multi-agent deep
  reinforcement learning},'' \emph{IEEE Internet Things J.}, pp. 1--14, 2021,
  to be published, doi:10.1109/JIOT.2021.3062091.

\bibitem{Yunting}
Y.~Xu, H.~Zhou, T.~Ma, J.~Zhao, B.~Qian, and X.~Shen, ``{Leveraging multiagent
  learning for automated vehicles scheduling at nonsignalized intersections},''
  \emph{IEEE Internet Things J.}, vol.~8, no.~14, pp. 11\,427--11\,439, July
  2021.

\bibitem{zhou2019edge}
Z.~Zhou, X.~Chen, E.~Li, L.~Zeng, K.~Luo, and J.~Zhang, ``Edge intelligence:
  Paving the last mile of artificial intelligence with edge computing,''
  \emph{Proc. IEEE}, vol. 107, no.~8, pp. 1738--1762, Aug. 2019.

\bibitem{Samarakoon}
S.~Samarakoon, M.~Bennis, W.~Saad, and M.~Latva-aho, ``{Dynamic clustering and
  on/off strategies for wireless small cell networks},'' \emph{IEEE Trans.
  Wireless Commun.}, vol.~15, no.~3, pp. 2164--2178, Aug. 2016.

\bibitem{Zhu_RNN}
Y.~Zhu, X.~Dong, and T.~Lu, ``{An adaptive and parameter-free recurrent neural
  structure for wireless channel prediction},'' \emph{IEEE Trans. Commun.},
  vol.~67, no.~11, pp. 8086--8096, Nov. 2019.

\bibitem{Yang_INFOCOM}
K.~Yang, C.~Shen, and T.~Liu, ``{Deep reinforcement learning based wireless
  network optimization: A comparative study},'' in \emph{Proc. IEEE INFOCOM
  Workshops}, Virtual Conference, 2020.

\bibitem{DRL4}
M.~{Alsenwi}, N.~{Tran}, M.~{Bennis}, S.~{Pandey}, A.~{Bairagi}, and C.~{Hong},
  ``Intelligent resource slicing for {eMBB} and {URLLC} coexistence in {5G} and
  beyond: A deep reinforcement learning based approach,'' \emph{IEEE Trans.
  Wireless Commun.}, vol.~20, no.~7, pp. 4585--4600, July 2021.

\bibitem{MEC_survey}
Y.~Mao, C.~You, J.~Zhang, K.~Huang, and K.~B. Letaief, ``A survey on mobile
  edge computing: The communication perspective,'' \emph{IEEE Commun. Surveys
  Tuts.}, vol.~19, no.~4, pp. 2322--2358, 4th Quart. 2017.

\bibitem{Niknam}
S.~Niknam, H.~S. Dhillon, and J.~H. Reed, ``{Federated learning for wireless
  communications: Motivation, opportunities, and challenges},'' \emph{IEEE
  Commun. Mag.}, vol.~58, no.~6, pp. 46--51, June 2020.

\bibitem{lim2020federated}
W.~{Lim}, N.~{Luong}, D.~{Hoang}, Y.~{Jiao}, Y.~{Liang}, Q.~{Yang},
  D.~{Niyato}, and C.~{Miao}, ``Federated learning in mobile edge networks: A
  comprehensive survey,'' \emph{IEEE Commun. Surveys Tuts.}, vol.~22, no.~3,
  pp. 2031--2063, 3rd Quart. 2020.

\bibitem{sun2018learning}
H.~Sun, X.~Chen, Q.~Shi, M.~Hong, X.~Fu, and N.~Sidiropoulos, ``Learning to
  optimize: {Training} deep neural networks for interference management,''
  \emph{IEEE Trans. Signal Process.}, vol.~66, no.~20, pp. 5438--5453, Oct.
  2018.

\bibitem{zhu2017overview}
H.~Zhu, K.~Yuen, L.~Mihaylova, and H.~Leung, ``Overview of environment
  perception for intelligent vehicles,'' \emph{IEEE Trans. Intell. Transp.
  Syst.}, vol.~18, no.~10, pp. 2584--2601, Oct. 2017.

\bibitem{gunduz2019machine}
D.~Gunduz, P.~de~Kerret, N.~Sidiropoulos, D.~Gesbert, C.~Murthy, and M.~van~der
  Schaar, ``Machine learning in the air,'' \emph{IEEE J. Sel. Areas in
  Commun.}, vol.~37, no.~10, pp. 2184--2199, Oct. 2019.

\bibitem{chen2019deep}
J.~Chen and X.~Ran, ``Deep learning with edge computing: {A} review,''
  \emph{Proc. IEEE}, vol. 107, no.~8, pp. 1655--1674, Aug. 2019.

\bibitem{lin2018architectural}
S.~Lin, Y.~Zhang, C.~Hsu, M.~Skach, M.~Haque, L.~Tang, and J.~Mars, ``The
  architectural implications of autonomous driving: {C}onstraints and
  acceleration,'' in \emph{Proc. ASPLOS}, Williamsburg, VA, USA, Mar. 2018.

\bibitem{liu2018edge}
Q.~Liu, S.~Huang, J.~Opadere, and T.~Han, ``An edge network orchestrator for
  mobile augmented reality,'' in \emph{Proc. IEEE INFOCOM}, Honolulu, HI, USA,
  Apr. 2018.

\bibitem{huang2008labeled}
G.~Huang, M.~Mattar, T.~Berg, and E.~Learned-Miller, ``Labeled faces in the
  wild: {A} database forstudying face recognition in unconstrained
  environments,'' in \emph{Workshop on faces in `Real-Life' Images: detection,
  alignment, and recognition}, Amherst, MA, USA, Oct. 2008.

\bibitem{deng2020edge}
S.~Deng, H.~Zhao, W.~Fang, J.~Yin, S.~Dustdar, and A.~Zomaya, ``{Edge
  intelligence: The confluence of edge computing and artificial
  intelligence},'' \emph{IEEE Internet Things J.}, vol.~7, no.~8, pp.
  7457--7469, Aug. 2020.

\bibitem{peltonen20206g}
E.~Peltonen \emph{et~al.}, ``{6G} white paper on edge intelligence,''
  \emph{arXiv:2004.14850}, 2020, [Online]. Available:
  http://arxiv.org/abs/2004.14850.

\bibitem{hard2018federated}
A.~Hard, K.~Rao, R.~Mathews, S.~Ramaswamy, F.~Beaufays, S.~Augenstein,
  H.~Eichner, C.~Kiddon, and D.~Ramage, ``Federated learning for mobile
  keyboard prediction,'' in \emph{Proc. IEEE IJCNN}, Budapest, Hungary, July
  2019.

\bibitem{wang2020cellular}
W.~Wang, C.~Zhou, H.~He, W.~Wu, W.~Zhuang, and X.~Shen, ``Cellular traffic load
  prediction with {LSTM and Gaussian} process regression,'' in \emph{Proc. IEEE
  ICC}, Dublin, Ireland, June 2020.

\bibitem{deo2018convolutional}
N.~Deo and M.~Trivedi, ``Convolutional social pooling for vehicle trajectory
  prediction,'' in \emph{Proc. IEEE CVPR Workshops}, Salt Lake City, UT, USA,
  June 2020.

\bibitem{cocskun2017face}
M.~Co{\c{s}}kun, A.~U{\c{c}}ar, {\"O}.~Yildirim, and Y.~Demir, ``Face
  recognition based on convolutional neural network,'' in \emph{Proc. IEEE
  MEES}, Kremenchuk, Ukraine, 2017.

\bibitem{erhan2014scalable}
D.~Erhan, C.~Szegedy, A.~Toshev, and D.~Anguelov, ``Scalable object detection
  using deep neural networks,'' in \emph{Proc. IEEE CVPR}, Columbus, OH, USA,
  June 2014.

\bibitem{muller2016context}
S.~M{\"u}ller, O.~Atan, M.~van~der Schaar, and A.~Klein, ``Context-aware
  proactive content caching with service differentiation in wireless
  networks,'' \emph{IEEE Trans. Wireless Commun.}, vol.~16, no.~2, pp.
  1024--1036, Feb. 2017.

\bibitem{zhou2020deep}
C.~{Zhou}, W.~{Wu}, H.~{He}, P.~{Yang}, F.~{Lyu}, N.~{Cheng}, and X.~{Shen},
  ``Deep reinforcement learning for delay-oriented {IoT} task scheduling in
  {SAGIN},'' \emph{IEEE Trans. Wireless Commun.}, vol.~20, no.~2, pp. 911--925,
  Feb. 2021.

\bibitem{sciancalepore2019rl}
V.~Sciancalepore, X.~Costa-Perez, and A.~Banchs, ``{RL-NSB}: Reinforcement
  learning-based {5G} network slice broker,'' \emph{IEEE/ACM Trans. Netw.},
  vol.~27, no.~4, pp. 1543--1557, Aug. 2019.

\bibitem{qu2020dynamic}
K.~Qu, W.~Zhuang, X.~Shen, X.~Li, and J.~Rao, ``Dynamic resource scaling for
  {VNF} over nonstationary traffic: A learning approach,'' \emph{IEEE Trans.
  Cogn. Commun. Netw.}, vol.~7, no.~2, pp. 648--662, June 2021.

\bibitem{Huynh}
N.~{Van Huynh}, D.~{Thai Hoang}, D.~{Nguyen}, and E.~{Dutkiewicz}, ``Optimal
  and fast real-time resource slicing with deep dueling neural networks,''
  \emph{IEEE J. Sel. Areas Commun.}, vol.~37, no.~6, pp. 1455--1470, June 2019.

\bibitem{gnn2}
Y.~{Hua}, R.~{Li}, Z.~{Zhao}, X.~{Chen}, and H.~{Zhang}, ``{GAN}-powered deep
  distributional reinforcement learning for resource management in network
  slicing,'' \emph{IEEE J. Sel. Areas Commun.}, vol.~38, no.~2, pp. 334--349,
  Feb. 2020.

\bibitem{Li_CL}
R.~{Li}, C.~{Wang}, Z.~{Zhao}, R.~{Guo}, and H.~{Zhang}, ``The {LSTM}-based
  advantage actor-critic learning for resource management in network slicing
  with user mobility,'' \emph{IEEE Commun. Lett.}, vol.~24, no.~9, pp.
  2005--2009, Sep. 2020.

\bibitem{Chergui}
H.~{Chergui} and C.~{Verikoukis}, ``Offline {SLA}-constrained deep learning for
  {5G} networks reliable and dynamic end-to-end slicing,'' \emph{IEEE J. Sel.
  Areas Commun.}, vol.~38, no.~2, pp. 350--360, Feb. 2020.

\bibitem{drl2}
X.~{Chen}, H.~{Zhang}, C.~{Wu}, S.~{Mao}, Y.~{Ji}, and M.~{Bennis}, ``Optimized
  computation offloading performance in virtual edge computing systems via deep
  reinforcement learning,'' \emph{IEEE Internet Things J.}, vol.~6, no.~3, pp.
  4005--4018, June 2019.

\bibitem{Xiang_TVT}
H.~{Xiang}, M.~{Peng}, Y.~{Sun}, and S.~{Yan}, ``Mode selection and resource
  allocation in sliced fog radio access networks: A reinforcement learning
  approach,'' \emph{IEEE Trans. Veh. Technol.}, vol.~69, no.~4, pp. 4271--4284,
  Apr. 2020.

\bibitem{Xiang_TWC}
H.~{Xiang}, S.~{Yan}, and M.~{Peng}, ``A realization of fog-{RAN} slicing via
  deep reinforcement learning,'' \emph{IEEE Trans. Wireless Commun.}, vol.~19,
  no.~4, pp. 2515--2527, Apr. 2020.

\bibitem{Chen_JSAC}
X.~{Chen}, Z.~{Zhao}, C.~{Wu}, M.~{Bennis}, H.~{Liu}, Y.~{Ji}, and H.~{Zhang},
  ``Multi-tenant cross-slice resource orchestration: A deep reinforcement
  learning approach,'' \emph{IEEE J. Sel. Areas Commun.}, vol.~37, no.~10, pp.
  2377--2392, Oct. 2019.

\bibitem{Messaoud}
S.~{Messaoud}, A.~{Bradai}, O.~{Ben Ahmed}, P.~{Quang}, M.~{Atri}, and
  M.~{Hossain}, ``Deep federated {Q}-learning-based network slicing for
  industrial {IoT},'' \emph{IEEE Trans. Ind. Informat.}, vol.~17, no.~8, pp.
  5572--5582, Aug. 2021.

\bibitem{Dandachi}
G.~{Dandachi}, A.~{De Domenico}, D.~{Hoang}, and D.~{Niyato}, ``An artificial
  intelligence framework for slice deployment and orchestration in {5G}
  networks,'' \emph{IEEE Trans. Cogn. Commun. Netw.}, vol.~6, no.~2, pp.
  858--871, June 2020.

\bibitem{kmean}
H.~{Xiao}, Y.~{Chen}, Q.~{Zhang}, A.~{Chronopoulos}, Z.~{Zhang}, and
  S.~{Ouyang}, ``Joint clustering and power allocation for the cross roads
  congestion scenarios in cooperative vehicular networks,'' \emph{IEEE Trans.
  Intell. Transp. Syst.}, vol.~20, no.~6, pp. 2267--2277, June 2019.

\bibitem{drl1}
Y.~{Yu}, T.~{Wang}, and S.~{Liew}, ``Deep-reinforcement learning multiple
  access for heterogeneous wireless networks,'' \emph{IEEE J. Sel. Areas
  Commun.}, vol.~37, no.~6, pp. 1277--1290, June 2019.

\bibitem{HDL1}
Y.~{Zhang}, Z.~{Mou}, F.~{Gao}, L.~{Xing}, J.~{Jiang}, and Z.~{Han},
  ``Hierarchical deep reinforcement learning for backscattering data collection
  with multiple {UAVs},'' \emph{IEEE Internet Things J.}, vol.~8, no.~5, pp.
  3786--3800, Mar. 2021.

\bibitem{Qian}
Y.~{Qian}, R.~{Wang}, J.~{Wu}, B.~{Tan}, and H.~{Ren}, ``Reinforcement
  learning-based optimal computing and caching in mobile edge network,''
  \emph{IEEE J. Sel. Areas Commun.}, vol.~38, no.~10, pp. 2343--2355, Oct.
  2020.

\bibitem{Dorner}
S.~{Dörner}, S.~{Cammerer}, J.~{Hoydis}, and S.~{Brink}, ``Deep learning based
  communication over the air,'' \emph{IEEE J. Sel. Topics Signal Process.},
  vol.~12, no.~1, pp. 132--143, Feb. 2018.

\bibitem{JieG_mac1}
J.~{Gao}, W.~{Zhuang}, M.~{Li}, X.~{Shen}, and X.~{Li}, ``{MAC} for
  machine-type communications in industrial {IoT}—{Part I}: Protocol design
  and analysis,'' \emph{IEEE Internet Things J.}, vol.~8, no.~12, pp.
  9945--9957, June 2021.

\bibitem{Jie_mac2}
J.~{Gao}, M.~{Li}, W.~{Zhuang}, X.~{Shen}, and X.~{Li}, ``{MAC} for machine
  type communications in industrial {IoT} – {Part II}: Scheduling and
  numerical results,'' \emph{IEEE Internet Things J.}, vol.~8, no.~12, pp.
  9958--9969, June 2021.

\bibitem{Yiping}
Y.~Kang, J.~Hauswald, C.~Gao, A.~Rovinski, T.~Mudge, J.~Mars, and L.~Tang,
  ``Neurosurgeon: Collaborative intelligence between the cloud and mobile
  edge,'' in \emph{Proc. ACM ASPLOS}, Xi'an, China, Apr. 2017.

\bibitem{partition1}
E.~{Li}, L.~{Zeng}, Z.~{Zhou}, and X.~{Chen}, ``Edge {AI}: On-demand
  accelerating deep neural network inference via edge computing,'' \emph{IEEE
  Trans. Wireless Commun.}, vol.~19, no.~1, pp. 447--457, Jan. 2020.

\bibitem{partition3}
W.~{He}, S.~{Guo}, S.~{Guo}, X.~{Qiu}, and F.~{Qi}, ``Joint {DNN} partition
  deployment and resource allocation for delay-sensitive deep learning
  inference in {IoT},'' \emph{IEEE Internet Things J.}, vol.~7, no.~10, pp.
  9241--9254, Oct. 2020.

\bibitem{cui2021fully}
Y.~Cui, Z.~Liu, W.~Yao, Q.~Li, A.~B. Chan, T.~Kuo, and C.~Xue, ``Fully nested
  neural network for adaptive compression and quantization,'' in \emph{Proc.
  IJCAI}, Yokohama, Japan, Jan. 2021.

\bibitem{nest1}
K.~{Lin}, Y.~{Li}, Q.~{Zhang}, and G.~{Fortino}, ``{AI}-driven collaborative
  resource allocation for task execution in {6G}-enabled massive {IoT},''
  \emph{IEEE Internet Things J.}, vol.~8, no.~7, pp. 5264--5273, Apr. 2021.

\bibitem{Haykin}
S.~{Haykin}, ``Cognitive radio: brain-empowered wireless communications,''
  \emph{IEEE J. Sel. Areas Commun.}, vol.~23, no.~2, pp. 201--220, Feb. 2005.

\bibitem{chemouil2019special}
P.~Chemouil, P.~Hui, W.~Kellerer, Y.~Li, R.~Stadler, D.~Tao, Y.~Wen, and
  Y.~Zhang, ``Special issue on artificial intelligence and machine learning for
  networking and communications,'' \emph{IEEE J. Sel. Areas Commun.}, vol.~37,
  no.~6, pp. 1185--1191, June 2019.

\bibitem{sorour2020returning}
S.~Sorour, U.~Mohammad, A.~Abutuleb, and H.~Hassanein, ``Returning the favor:
  {What} wireless networking can offer to {AI} and edge learning,''
  \emph{arXiv:2006.07453}, 2020, [Online]. Available:
  https://arxiv.org/abs/2006.07453.

\bibitem{wu2021ai}
W.~Wu, C.~Zhou, M.~Li, H.~Wu, H.~Zhou, N.~Zhang, W.~Zhuang, and X.~Shen,
  ``{AI-native network slicing for 6G networks},'' \emph{arXiv:2105.08576},
  2021, [Online]. Available: https://arxiv.org/abs/2105.08576.

\bibitem{itu2019architectural}
ITU-T, ``Architectural framework for machine learning in future networks
  including {IMT}-2020,'' 2019, [Online]. Available:
  https://https://www.itu.int/rec/T-REC-Y.3172/en.

\bibitem{itu2019Unified}
ITU, ``Unified architecture for machine learning in {5G} and future networks,''
  Jan. 2019.

\bibitem{3GPP2019Study}
3GPP, ``Study of enablers for network automation for {5G},'' no. 3GPP TR 23.791
  V16.2.0, June 2019.

\bibitem{wilhelmi2020flexible}
F.~Wilhelmi, S.~Barrachina-Mu{\~n}oz, B.~Bellalta, C.~Cano, A.~Jonsson, and
  V.~Ram, ``A flexible machine-learning-aware architecture for future
  {WLANs},'' \emph{IEEE Commun. Mag.}, vol.~58, no.~3, pp. 25--31, Mar. 2020.

\bibitem{va2016millimeter}
V.~Va, T.~Shimizu, G.~Bansal, and R.~W. Heath~Jr, ``Millimeter wave vehicular
  communications: A survey,'' \emph{Found. Trends Netw.}, vol.~10, no.~1, pp.
  1--126, June 2016.

\bibitem{chen2016eyeriss}
Y.~Chen, T.~Krishna, J.~Emer, and V.~Sze, ``Eyeriss: {An} energy-efficient
  reconfigurable accelerator for deep convolutional neural networks,''
  \emph{IEEE J. Solid-State Circuits}, vol.~52, no.~1, pp. 127--138, Jan. 2017.

\bibitem{model_partition}
C.~{Hu}, W.~{Bao}, D.~{Wang}, and F.~{Liu}, ``Dynamic adaptive {DNN} surgery
  for inference acceleration on the edge,'' in \emph{Proc. IEEE INFOCOM},
  Paris, France, Apr. 2019.

\bibitem{wen2019overview}
D.~Wen, X.~Li, Q.~Zeng, J.~Ren, and K.~Huang, ``An overview of data-importance
  aware radio resource management for edge machine learning,'' \emph{J. Commun.
  Netw.}, vol.~4, no.~4, pp. 1--14, Dec. 2019.

\bibitem{wang2020machine}
S.~Wang, Y.~C. Wu, M.~Xia, R.~Wang, and V.~Poor, ``Machine intelligence at the
  edge with learning centric power allocation,'' \emph{IEEE Trans. Wireless
  Commun.}, vol.~19, no.~11, pp. 7293--7308, Nov. 2020.

\bibitem{liu2020wireless}
D.~Liu, G.~Zhu, Q.~Zeng, J.~Zhang, and K.~Huang, ``Wireless data acquisition
  for edge learning: {Data-importance} aware retransmission,'' \emph{IEEE
  Trans. Wireless Commun.}, vol.~20, no.~1, pp. 406--420, Jan. 2021.

\bibitem{liu2020client}
L.~Liu, J.~Zhang, S.~Song, and K.~Letaief, ``Client-edge-cloud hierarchical
  federated learning,'' in \emph{Proc. IEEE ICC}, Virtual Conference, June
  2020.

\bibitem{yang2020federated}
K.~Yang, T.~Jiang, Y.~Shi, and Z.~Ding, ``Federated learning via over-the-air
  computation,'' \emph{IEEE Trans. Wireless Commun.}, vol.~19, no.~3, pp.
  2022--2035, Mar. 2020.

\bibitem{wang2020optimizing}
H.~Wang, Z.~Kaplan, D.~Niu, and B.~Li, ``Optimizing federated learning on
  non-iid data with reinforcement learning,'' in \emph{Proc. IEEE INFOCOM},
  Toronto, ON, Canada, July 2020.

\bibitem{nishio2019client}
T.~Nishio and R.~Yonetani, ``Client selection for federated learning with
  heterogeneous resources in mobile edge,'' in \emph{Proc. IEEE ICC}, Shanghai,
  China, May 2019.

\bibitem{wang2019adaptive}
S.~Wang, T.~Tuor, T.~Salonidis, K.~Leung, C.~Makaya, T.~He, and K.~Chan,
  ``Adaptive federated learning in resource constrained edge computing
  systems,'' \emph{IEEE J. Sel. Areas Commun.}, vol.~37, no.~6, pp. 1205--1221,
  June 2019.

\bibitem{ren2020scheduling}
J.~Ren, Y.~He, D.~Wen, G.~Yu, K.~Huang, and D.~Guo, ``Scheduling for cellular
  federated edge learning with importance and channel awareness,'' \emph{IEEE
  Trans. Wireless Commun.}, vol.~19, no.~11, pp. 7690--7703, Nov. 2020.

\bibitem{wang2020joint}
C.~Wang, S.~Zhang, Y.~Chen, Z.~Qian, J.~Wu, and M.~Xiao, ``Joint configuration
  adaptation and bandwidth allocation for edge-based real-time video
  analytics,'' in \emph{Proc. IEEE INFOCOM}, Toronto, ON, Canada, July 2020.

\bibitem{zhang2021autodidactic}
L.~Zhang, L.~Chen, and J.~Xu, ``Autodidactic neurosurgeon: {C}ollaborative deep
  inference for mobile edge intelligence via online learning,''
  \emph{arXiv:2102.02638}, 2021, [Online]. Available:
  https://arxiv.org/abs/2102.02638.

\bibitem{teerapittayanon2017distributed}
S.~Teerapittayanon, B.~McDanel, and H.-T. Kung, ``Distributed deep neural
  networks over the cloud, the edge and end devices,'' in \emph{Proc. IEEE
  ICDCS}, Atlanta, GA, USA, June 2017.

\bibitem{wu2020accuracy}
W.~Wu, P.~Yang, W.~Zhang, C.~Zhou, and X.~Shen, ``Accuracy-guaranteed
  collaborative {DNN} inference in industrial {IoT} via deep reinforcement
  learning,'' \emph{IEEE Trans. Ind. Informat.}, vol.~17, no.~7, pp.
  4988--4998, July 2020.

\bibitem{li2017wireless}
K.~Li, W.~Ni, L.~Duan, M.~Abolhasan, and J.~Niu, ``Wireless power transfer and
  data collection in wireless sensor networks,'' \emph{IEEE Trans. Veh.
  Technol.}, vol.~67, no.~3, pp. 2686--2697, Mar. 2017.

\bibitem{zhan2017energy}
C.~Zhan, Y.~Zeng, and R.~Zhang, ``Energy-efficient data collection in {UAV}
  enabled wireless sensor network,'' \emph{IEEE Commun. Lett.}, vol.~7, no.~3,
  pp. 328--331, June 2017.

\bibitem{holub2008entropy}
A.~Holub, P.~Perona, and M.~Burl, ``Entropy-based active learning for object
  recognition,'' in \emph{Proc. IEEE CVPR Workshops}, Anchorage, Alaska, USA,
  June 2020.

\bibitem{verbraeken2020survey}
J.~Verbraeken, M.~Wolting, J.~Katzy, J.~Kloppenburg, T.~Verbelen, and
  J.~Rellermeyer, ``A survey on distributed machine learning,'' \emph{ACM
  Computing Surveys}, vol.~53, no.~2, pp. 1--33, Mar. 2020.

\bibitem{yang2019federated}
Q.~Yang, Y.~Liu, T.~Chen, and Y.~Tong, ``{Federated machine learning: Concept
  and applications},'' \emph{ACM Trans. Intell. Syst. Technol.}, vol.~10,
  no.~2, pp. 1--19, Jan. 2019.

\bibitem{bonawitz2019towards}
K.~Bonawitz, H.~Eichner, W.~Grieskamp, D.~Huba, A.~Ingerman, V.~Ivanov,
  C.~Kiddon, J.~Kone{\v{c}}n{\`y}, S.~Mazzocchi, B.~McMahan \emph{et~al.},
  ``Towards federated learning at scale: System design,''
  \emph{arXiv:1902.01046}, 2019, [Online]. Available:
  https://arxiv.org/abs/1902.01046.

\bibitem{li2020federated}
T.~Li, A.~Sahu, A.~Talwalkar, and V.~Smith, ``{Federated learning: Challenges,
  methods, and future directions},'' \emph{IEEE Signal Process. Mag.}, vol.~37,
  no.~3, pp. 50--60, May 2020.

\bibitem{he2016deep}
K.~He, X.~Zhang, S.~Ren, and J.~Sun, ``Deep residual learning for image
  recognition,'' in \emph{Proc. IEEE CVPR}, Las Vegas, NV, USA, June 2016.

\bibitem{szegedy2016rethinking}
C.~Szegedy, V.~Vanhoucke, S.~Ioffe, J.~Shlens, and Z.~Wojna, ``Rethinking the
  inception architecture for computer vision,'' in \emph{Proc. IEEE CVPR}, Las
  Vegas, NV, USA, June 2016.

\bibitem{krizhevsky2012imagenet}
A.~Krizhevsky, I.~Sutskever, and G.~Hinton, ``{ImageNet} classification with
  deep convolutional neural networks,'' in \emph{Proc. IEEE NIPS}, Lake Tahoe,
  Nevada, USA, Dec. 2012.

\bibitem{simonyan2014very}
K.~Simonyan and A.~Zisserman, ``Very deep convolutional networks for
  large-scale image recognition,'' \emph{arXiv:1409.1556}, 2014, [Online].
  Available: https://arxiv.org/abs/1409.1556.

\bibitem{chen2019round}
C.~Chen, W.~Wang, and B.~Li, ``{Round-robin synchronization: Mitigating
  communication bottlenecks in parameter servers},'' in \emph{Proc. IEEE
  INFOCOM}, Paris, France, Apr. 2019.

\bibitem{zhang2021dynamic}
W.~Zhang, D.~Yang, W.~Wu, H.~Peng, N.~Zhang, H.~Zhang, and X.~Shen,
  ``{Optimizing federated learning in distributed industrial {IoT: A}
  multi-agent approach},'' \emph{IEEE J. Sel. Areas Commun.}, vol.~39, no.~12,
  pp. 3688--3703, Dec. 2021.

\bibitem{amiri2020federated}
M.~Amiri and D.~G{\"u}nd{\"u}z, ``Federated learning over wireless fading
  channels,'' \emph{IEEE Trans. Wireless Commun.}, vol.~19, no.~5, pp.
  3546--3557, May 2020.

\bibitem{zhang2021federated}
J.~Zhang, N.~Li, and M.~Dedeoglu, ``Federated learning over wireless networks:
  {A} band-limited coordinated descent approach,'' \emph{arXiv preprint
  arXiv:2102.07972}, 2021, [Online]. Available:
  https://arxiv.org/abs/2102.07972.

\bibitem{han2015deep}
S.~Han, H.~Mao, and W.~Dally, ``Deep compression: {Compressing} deep neural
  networks with pruning, trained quantization and {Huffman} coding,'' in
  \emph{Proc. ICLR}, San Juan, Puerto Rico, May 2016.

\bibitem{chen2017learning}
G.~Chen, W.~Choi, X.~Yu, T.~Han, and M.~Chandraker, ``Learning efficient object
  detection models with knowledge distillation,'' in \emph{Proc. IEEE NIPS},
  Long Beach, CA, USA, Dec. 2017.

\bibitem{teerapittayanon2016branchynet}
S.~Teerapittayanon, B.~McDanel, and H.~Kung, ``{Branchynet: Fast} inference via
  early exiting from deep neural networks,'' in \emph{Proc. IEEE ICPR}, Cancun,
  Mexico, Dec. 2016.

\bibitem{iandola2016squeezenet}
F.~Iandola, S.~Han, M.~Moskewicz, K.~Ashraf, W.~Dally, and K.~Keutzer,
  ``{SqueezeNet: AlexNet-level accuracy with 50x fewer parameters and $<$0.5 MB
  model size},'' \emph{arXiv:1602.07360}, 2016, [Online]. Available:
  http://arxiv.org/abs/1602.07360.

\bibitem{chen2017multi}
X.~Chen, H.~Ma, J.~Wan, B.~Li, and T.~Xia, ``Multi-view {3D} object detection
  network for autonomous driving,'' in \emph{Proc. IEEE CVPR}, Honolulu, HI,
  USA, July 2017.

\bibitem{ngo2015cell}
H.~Q. Ngo, A.~Ashikhmin, H.~Yang, E.~G. Larsson, and T.~L. Marzetta,
  ``{Cell-free massive MIMO: Uniformly great service for everyone},'' in
  \emph{Proc. IEEE SPAWC}, Stockholm, Sweden, July 2015.

\bibitem{yang2019multi}
Y.~Yang, ``{Multi-tier computing networks for intelligent {IoT}},''
  \emph{Nature Electronics}, vol.~2, no.~1, pp. 4--5, Jan. 2019.

\bibitem{Mushu_UAV}
M.~Li, N.~Cheng, J.~Gao, Y.~Wang, L.~Zhao, and X.~Shen, ``{Energy-efficient
  {UAV}-assisted mobile edge computing: {R}esource allocation and trajectory
  optimization},'' \emph{IEEE Trans. Veh. Technol.}, vol.~69, no.~3, pp.
  3424--3438, Mar. 2020.

\bibitem{Leconte}
M.~Leconte, G.~S. Paschos, P.~Mertikopoulos, and U.~C. Kozat, ``{A resource
  allocation framework for network slicing},'' in \emph{Proc. IEEE INFCOM},
  Honolulu, HI, USA, Apr. 2018.

\bibitem{Halabian}
H.~Halabian, ``{Distributed resource allocation optimization in {5G}
  virtualized networks},'' \emph{IEEE J. Sel. Areas Commun.}, vol.~37, no.~3,
  pp. 627--642, Mar. 2019.

\bibitem{Xiong1}
Z.~Xiong, S.~Feng, W.~Wang, D.~Niyato, P.~Wang, and Z.~Han, ``{Cloud/Fog
  computing resource management and pricing for blockchain networks},''
  \emph{IEEE Internet Things J.}, vol.~6, no.~3, pp. 4585--4600, June 2019.

\bibitem{Xiong2}
J.~Nie, J.~Luo, Z.~Xiong, D.~Niyato, and P.~Wang, ``{A Stackelberg game
  spproach toward socially-aware incentive mechanisms for mobile
  crowdsensing},'' \emph{IEEE Trans. Wireless Commun.}, vol.~18, no.~1, pp.
  724--738, Jan. 2019.

\bibitem{Caballero}
P.~Caballero, A.~Banchs, G.~De~Veciana, and X.~Costa-Pérez, ``{Network slicing
  games: Enabling customization in multi-tenant mobile networks},''
  \emph{IEEE/ACM Trans. Netw.}, vol.~27, no.~2, pp. 662--675, Apr. 2019.

\bibitem{J_Jie_MD_2018}
J.~Gao, L.~Zhao, and X.~Shen, ``{Network utility maximization based on an
  incentive mechanism for truthful reporting of local information},''
  \emph{IEEE Trans. Veh. Technol.}, vol.~67, no.~8, pp. 7523--7537, Aug. 2018.

\bibitem{zappone2019wireless}
A.~Zappone, M.~Di~Renzo, and M.~Debbah, ``Wireless networks design in the era
  of deep learning: Model-based, ai-based, or both{?}'' \emph{IEEE Trans.
  Commun.}, vol.~67, no.~10, pp. 7331--7376, Oct. 2019.

\bibitem{yang2020deep}
Y.~Yang, F.~Gao, Z.~Zhong, B.~Ai, and A.~Alkhateeb, ``Deep transfer
  learning-based downlink channel prediction for {FDD} massive {MIMO}
  systems,'' \emph{IEEE Trans. Commun.}, vol.~68, no.~12, pp. 7485--7497, Dec.
  2020.

\bibitem{S_DC.Nguyen_ST_2021}
D.~C. Nguyen, P.~Cheng, M.~Ding, D.~Lopez-Perez, P.~N. Pathirana, J.~Li,
  A.~Seneviratne, Y.~Li, and H.~V. Poor, ``{Enabling AI in future wireless
  networks: A data life cycle perspective},'' \emph{IEEE Commun. Surveys
  Tuts.}, vol.~23, no.~1, pp. 553--595, 1st Quart. 2021.

\bibitem{O_Zhou_IETF21}
\BIBentryALTinterwordspacing
C.~Zhou, H.~Yang, X.~Duan, D.~Lopez, A.~Pastor, Q.~Wu, M.~Boucadair, and
  C.~Jacquenet, ``{Digital twin network: Concepts and reference
  architecture},'' Internet Engineering Task Force, Internet-Draft, July 2021.
  [Online]. Available:
  \url{https://datatracker.ietf.org/doc/html/draft-zhou-nmrg-digitaltwin-network-concepts-04}
\BIBentrySTDinterwordspacing

\bibitem{J_YLu_IoT_2021}
Y.~Lu, S.~Maharjan, and Y.~Zhang, ``{Adaptive edge association for wireless
  digital twin networks in {6G}},'' \emph{IEEE Internet Things J.}, vol.~8,
  no.~22, pp. 16\,219--16\,230, July 2021.

\bibitem{J_WJiang_OJCS_2021}
W.~Jiang, B.~Han, M.~A. Habibi, and H.~D. Schotten, ``{The road towards 6{G}: A
  comprehensive survey},'' \emph{IEEE Open J. Commun. Society}, vol.~2, pp.
  334--366, Feb. 2021.

\bibitem{J_PBellavista_TII_2021}
P.~Bellavista, C.~Giannelli, M.~Mamei, M.~Mendula, and M.~Picone,
  ``{Application-driven network-aware digital twin management in industrial
  edge environments},'' \emph{IEEE Trans. Ind. Informat.}, vol.~17, no.~11, pp.
  7791--7801, Nov. 2021.

\bibitem{Mei}
J.~Mei, X.~Wang, K.~Zheng, G.~Boudreau, A.~B. Sediq, and H.~Abou-zeid,
  ``{Intelligent radio access network slicing for service provisioning in 6G: A
  hierarchical deep reinforcement learning approach},'' \emph{IEEE Trans.
  Commun.}, vol.~69, no.~9, pp. 6063--6078, Sep. 2021.

\bibitem{Marquez}
C.~Marquez, M.~Gramaglia, M.~Fiore, A.~Banchs, and X.~Costa-Perez, ``{How
  should I slice my network? A multi-service empirical evaluation of resource
  sharing efficiency},'' in \emph{Proc. MobiCom}, New Delhi, India, Oct. 2020.

\bibitem{partition2}
T.~{Mohammed}, C.~{Joe-Wong}, R.~{Babbar}, and M.~{Francesco}, ``Distributed
  inference acceleration with adaptive {DNN} partitioning and offloading,'' in
  \emph{Proc. IEEE INFOCOM}, Virtual Conference, July 2020.

\bibitem{he2020optimizing}
Y.~He, J.~Ren, G.~Yu, and Y.~Cai, ``{Optimizing the learning performance in
  mobile augmented reality systems with {CNN}},'' \emph{IEEE Trans. Wireless
  Commun.}, vol.~19, no.~8, pp. 5333--5344, Aug. 2020.

\bibitem{lin2019artificial}
W.~Lin, G.~Wu, X.~Wang, and K.~Li, ``{An artificial neural network approach to
  power consumption model construction for servers in cloud data centers},''
  \emph{IEEE Trans. Sustain. Comput.}, vol.~5, no.~3, pp. 329--340, July 2019.

\bibitem{strubell2019energy}
E.~Strubell, A.~Ganesh, and A.~McCallum, ``Energy and policy considerations for
  deep learning in {NLP},'' \emph{arXiv:1906.02243}, 2019, [Online]. Available:
  http://arxiv.org/abs/1906.02243.

\bibitem{thompson2020computational}
N.~Thompson, K.~Greenewald, K.~Lee, and G.~Manso, ``The computational limits of
  deep learning,'' \emph{arXiv:2007.05558}, 2020, [Online]. Available:
  http://arxiv.org/abs/2007.05558.

\bibitem{zhu2016trained}
C.~Zhu, S.~Han, H.~Mao, and W.~Dally, ``Trained ternary quantization,'' in
  \emph{Proc. ICLR}, Toulon, France, Apr. 2017.

\bibitem{hinton2015distilling}
G.~Hinton, O.~Vinyals, and J.~Dean, ``Distilling the knowledge in a neural
  network,'' in \emph{Proc. IEEE NIPS Workshops}, Montreal, Canada, Dec. 2014.

\bibitem{M_CWang_WCom_2020}
C.-X. Wang, M.~D. Renzo, S.~Stanczak, S.~Wang, and E.~G. Larsson, ``{Artificial
  intelligence enabled wireless networking for {5G} and beyond: recent advances
  and future challenges},'' \emph{IEEE Wireless Commun.}, vol.~27, no.~1, pp.
  16--23, Feb. 2020.

\bibitem{M_TWang_ChinaCom_2019}
T.~Wang, S.~Wang, and Z.-H. Zhou, ``{Machine learning for {5G} and beyond: From
  model-based to data-driven mobile wireless networks},'' \emph{China Commun.},
  vol.~16, no.~1, pp. 165--175, Jan. 2019.

\bibitem{C_BSliwa_GC_2020}
B.~Sliwa, M.~Patchou, and C.~Wietfeld, ``{The Best of both worlds: Hybrid
  data-driven and model-based vehicular network simulation},'' in \emph{Proc.
  IEEE GLOBECOM}, Virtual Conference, Dec. 2020.

\end{thebibliography}

\end{document}